\documentclass[%
aps, 
pre, 
twocolumn, 
nofootinbib, 
floatfix,
tightenlines, 
superscriptaddress, 
amsmath, 
amssymb, 
final, 
10pt
]{revtex4-1}

\usepackage{comment}
\usepackage{adjustbox}
\usepackage[dvipsnames]{xcolor}
\usepackage{hyperref}
\usepackage{hhline}
\usepackage{placeins}
\usepackage{graphicx}
\usepackage{dcolumn}
\usepackage{bm}

\hypersetup{
    colorlinks=true, 
    linkcolor=BrickRed, 
    citecolor=MidnightBlue, 
    filecolor=RedViolet,      
    urlcolor=MidnightBlue
    }

\begin{document}

\newcommand{\model}{\texttt{nODE}}


\title{When do neural ordinary differential equations generalize on complex networks?}

\author{Moritz Laber}
\email{laber.m@northeastern.edu}
\affiliation{Network Science Institute, Northeastern University, Boston, Massachusetts, USA}
\affiliation{Complexity Science Hub Vienna, Vienna, Austria}

\author{Tina Eliassi-Rad}
\affiliation{Network Science Institute, Northeastern University, Boston, Massachusetts, USA}
\affiliation{Khoury College of Computer Sciences, Northeastern University, Boston, Massachusetts, USA}
\affiliation{Santa Fe Institute, Santa Fe, New Mexico, USA}

\author{Brennan Klein}
\affiliation{Network Science Institute, Northeastern University, Boston, Massachusetts, USA}
\affiliation{Institute for Experiential AI, Northeastern University, Boston, Massachusetts, USA}
\affiliation{Department of Physics, Northeastern University, Boston, Massachusetts, USA}

\date{\today}

\begin{abstract}
Neural ordinary differential equations (neural ODEs) can effectively learn dynamical systems from time series data, but their behavior on graph-structured data remains poorly understood, especially when applied to graphs with different size or structure than encountered during training.
We study neural ODEs (\model{}s) with vector fields following the Barab\'asi-Barzel form, trained on synthetic data from five common dynamical systems on graphs. 
Using the $\mathbb{S}^1$-model to generate graphs with realistic and tunable structure, we find that degree heterogeneity and the type of dynamical system are the primary factors in determining \model{}s' ability to generalize across graph sizes and properties. This extends to \model{}s' ability to capture fixed points and maintain performance amid missing data. Average clustering plays a secondary role in determining \model{} performance.
Our findings highlight \model{}s as a powerful approach to understanding complex systems but underscore challenges emerging from degree heterogeneity and clustering in realistic graphs. 
\end{abstract}

\maketitle

\section{Introduction}\label{sec:introduction}

Complex systems in the natural and social sciences are often modeled as dynamical processes on graphs: The system's structure is represented as a graph, and its time evolution is specified using ordinary differential equations (ODEs) that describe how each node's state changes due to local interactions along the edges of the graph~\cite{barrat2008_dynamicalprocessescomplex, porter2016_dynamicalsystemsnetworks, lehmann2018_complexspreadingphenomen}. This modeling framework has yielded insight into various forms of collective behavior, revealing how systems respond to perturbations and enabling the construction of effective dynamics~\cite{barzel2013_universalitynetworkdynamics, gao2016_universalresiliencepatterns, hens2019_spatiotemporalsignalpropagation, meena2023_emergentstabilitycomplex, thibeault2024_lowrankhypothesiscomplex, thibeault2025_kuramotomeetskoopman}.

Traditionally, such models have been handcrafted and studied analytically or through numerical methods~\cite{ralston2001_firstcoursenumerical, iserles2008_firstcoursenumerical, teschl2012_ordinarydifferentialequations, strogatz2018_nonlineardynamicschaos}. In recent years, research into scientific machine learning has extended this toolbox~\cite{brunton2019_datadrivenscienceengineering, cicirello2024_physicsenhancedmachinelearning, vinuesa2025_decodingcomplexityhow}: Neural operators encode flexible mappings between function spaces, creating learnable solution operators for differential equations~\cite{li2020_fourierneuraloperator, lu2021_learningnonlinearoperators, kovachki2023_neuraloperatorlearning}. Physics-Informed Neural Networks incorporate constraints into neural network loss functions to solve differential equations or inverse problems~\cite{raissi2019_physicsinformedneuralnetworks, cuomo2022_scientificmachinelearning}. Symbolic regression aims to discover interpretable governing equations in closed form directly from data~\cite{brunton2016_discoveringgoverningequations, cranmer2020_discoveringsymbolicmodels, guimera2020_bayesianmachinescientist, yu2025_discoveringnetworkdynamics}. Techniques from generative modeling hold the promise of creating foundation models for scientific machine learning, attempting to emulate their success in the language domain~\cite{subramanian2023_foundationmodels, mccabe2024_multiplephysicspretraining}.

Our focus is on neural ordinary differential equations (neural ODEs)~\cite{chen2018_neuralordinarydifferential, rackauckas2020_universaldifferentialequations, kidger2022_neuraldifferentialequations, oh2025_comprehensivereviewneural}: By replacing the vector field of an ODE with a neural network and employing differentiable numerical solvers, they enable learning of data-driven dynamical systems directly from time series. This basic idea can be extended by incorporating suitable inductive biases into the neural network architecture. For example, combining graph neural networks with neural ODEs leads to models well suited for predicting and uncovering node-wise dynamics~\cite{poli2022_graphneuralordinary, liu2025_GraphODEsComprehensive, zang2020_neuraldynamicscomplex, vasiliauskaite2024_generalizationneuralnetwork, gao2024_learninginterpretabledynamics}.

While neural ODEs have become firmly established as a part of the scientific machine learning ecosystem, we are only beginning to understand their capabilities and limitations when applied to complex systems, whose structure is described by sparse graphs with heavy-tailed degree distributions, high average clustering, and small average shortest path length~\cite{albert2002_statisticalmechanicscomplex, newman2010_networksintroduction, dorogovtsev2022_naturecomplexnetworks}.

Prior work by Vasiliauskaite \& Antulov-Fantulin~\cite{vasiliauskaite2024_generalizationneuralnetwork} has established the ability of neural ODEs to generalize to graphs larger than the training graph but not to initial conditions outside the domain covered by the training data, creating a nuanced perspective on their performance on out-of-distribution data~\cite{liu2023_OutOfDistributionGeneralizationSurvey, li2025_OutofDistributionGeneralizationGraphs}. However, this study focuses on the Erd\H{o}s-R\'enyi random graph model, which is degree homogeneous and possesses low clustering at reasonable sparsity. Zang et al.~\cite{zang2020_neuraldynamicscomplex}, on the other hand, apply neural ODEs to time series tasks on graphs from different random graph models, but do not discuss the influence of graph topology on performance. And while there are mathematically rigorous results, e.g., generalization bounds~\cite{marion2023_generalization}, they are not focused on the graph setting.

Here, we examine how graph structure and the type of data generating dynamical system impact the performance and robustness of neural ODEs applied to complex systems. To this end, we use the hypercanonical $\mathbb{S}^1$-model, a flexible random graph model capable of generating training and test graphs with realistic but precisely controlled properties~\cite{serrano2008_selfsimilaritycomplexnetworks, krioukov2010_hyperbolicgeometrycomplex, vanderhoorn2018_sparsemaximumentropyrandom, boguna2020_smallworldsclustering}, and we train neural ODEs on synthetic data from five commonly studied dynamical systems, covering domains from epidemiology to neuroscience~\cite{hens2019_spatiotemporalsignalpropagation, vasiliauskaite2024_generalizationneuralnetwork}. By mirroring the structure of these systems in our neural ODE architecture, which we refer to as \model{}, we create a best-case scenario for applying our four evaluation strategies:
First, we show that \model{}s can generalize to graphs much larger than the training graph, an ability known as size generalization~\cite{xu2020_howneuralnetworks, yehudai2021_localstructuressize, anil2022_exploringlengthgeneralization}, but that degree heterogeneity impairs this capability for some dynamical systems.
Second, by exploring the space of graph topologies we reveal that \model{}s can generalize to less degree heterogeneous and less clustered graphs but struggle if degree heterogeneity and clustering are higher than in the training graph.
Third, we test whether \model{}s encode fixed points and their stability accurately, finding that \model{}s approach stable fixed points but that these fixed points might differ from those of the data generating dynamical system.
Fourth, motivated by classic assessments of the robustness of complex systems~\cite{artime2024_robustnessresiliencecomplex} and the literature on missing data~\cite{little2021_missingdataassumptions}, we show that depending on the graph structure and dynamical system, only a few unobserved nodes are enough to degrade the predictive performance of \model{}s.

Together, these findings show that \model{} performance and robustness depend on a complex interplay of data generating dynamical system and graph structure, thus highlighting the need for diverse assessment frameworks that allow careful exploration of relevant structural parameters.

\section{Results}

\subsection{Overview}

\begin{figure*}[t!]
    \centering
    \includegraphics[width=0.965\textwidth]{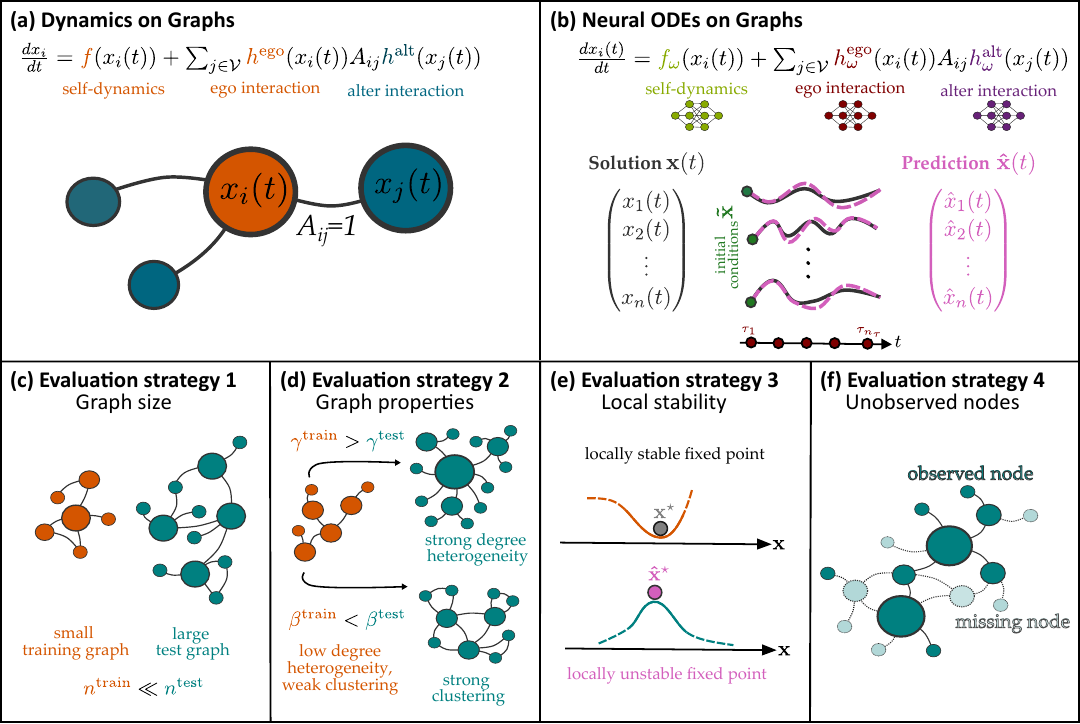}
    \caption{\textbf{Schematic depiction of (neural) ODEs on graphs and our four evaluation strategies.} \textbf{(a)} In dynamical systems of BB form the vector field at a specific node $i$ (orange) consists of the self-dynamics $f(x_i)$, describing the time evolution of node $i$'s state $x_i(t)$ in the absence of any interactions, and a factorized interaction term $h^\mathrm{ego}(x_i)h^\mathrm{alt}(x_j)$ describing how node $i$ and $j$ (blue) interact if they are connected, i.e., $A_{ij}=1$. \textbf{(b)} We investigate neural ODEs that mirror this structure (\model{}s) but replace $f, \, h^\mathrm{ego}, \, h^\mathrm{alt}$ with neural networks $f_\omega$ (light green), $h_\omega^\mathrm{ego}$ (red), and $h_\omega^\mathrm{alt}$ (purple), each with its own parameters. We evaluate these models by comparing their predictions $\mathbf{\hat{x}}(t)$ (pink) at discrete time points $\{\tau_r\}_{r=1}^{n_\tau}$ (dark red) with the ground truth time evolution $\mathbf{x}(t)$ (gray) starting from the same initial conditions $\mathbf{\tilde{x}}$ (dark green). We evaluate \model{}s in terms of their ability \textbf{(c)} to generalize to graphs much larger than the training graph, \textbf{(d)} to generalize to graphs that differ in their parameters from the training graph, \textbf{(e)} to faithfully capture fixed points and their stability, and \textbf{(f)} to make accurate predictions if only a subset of nodes is observed at test time.}
    \label{fig:schema}
\end{figure*}

Our evaluation of neural ODEs is concerned with dynamical systems of Barab\'asi-Barzel (BB) form (Fig.~\ref{fig:schema}\textbf{(a)}). Each node $i\in\{1, \dots, n\}$ is characterized by a scalar node state $x_i(t)\in\mathbb{R}$ at time $t$, that evolves according to a system of coupled ODEs, 
\begin{widetext}
\begin{equation}\label{eq:bb-dynamics}
    \frac{\mathrm{d}x_i(t)}{\mathrm{d}t}=f(x_i(t))+\sum_{j=1}^{n}h^{\mathrm{ego}}(x_i(t))A_{ij}h^{\mathrm{alt}}(x_j(t)), 
\end{equation}
\end{widetext}
where $f(x_i)$ is called the self-dynamics, as it describes the change of a node $i$'s state in absence of interactions, and interactions take place along the edges of a graph, encoded in its adjacency matrix $A$, and factorize into $h^\mathrm{ego}(x_i)$, only depending on node $i$'s state, and $h^\mathrm{alt}(x_j)$, only depending on the state of the neighbor $j$. This factorized interaction term in Eq.~\eqref{eq:bb-dynamics} is the defining feature of the BB form. 
%
The BB form specifies an entire class of dynamical systems. We investigate five specific instances covering various scientific domains: the Susceptible-Infected-Susceptible (SIS) model from epidemiology, the Mass-Action Kinetics (MAK) model from chemistry, the Michaelis-Menten (MM) model describing gene regulation, a Birth-Death (BD) process model arising in population dynamics, and a model of Neuronal Dynamics (ND) describing the activity of interacting brain regions. We detail their governing equations in Tab.~\ref{tab:dynamics}, and provide further detail on these systems and how we solve them to generate training data in Section~\ref{sec:methods:network_dynamics}.

We study neural ODEs modeled after Eq.~\eqref{eq:bb-dynamics} and refer to them as \model{}s. We replace the functions $f$, $h^\mathrm{ego}$, and $h^\mathrm{alt}$ with neural networks $f_\omega$, $h^\mathrm{ego}_\omega$, and $h^\mathrm{alt}_\omega$, each with its own parameters $\omega$ (Fig.~\ref{fig:schema}\textbf{(b)}), thereby creating suitable inductive biases for learning dynamical systems in BB form. These \model{}s can make predictions $\mathbf{\hat{x}}(t)=(\hat{x}_1(t), \dots, \hat{x}_n(t))^\mathsf{T}$ at arbitrary time $t$ given an initial condition $\mathbf{\tilde{x}}=(\tilde{x}_1, \dots, \tilde{x}_n)^\mathsf{T}$ at $t=0$ and a graph's adjacency matrix $A$. During training and testing we compare these predictions to the ground truth time evolution $\mathbf{x}(t)=(x_1(t), \dots, x_n(t))^\mathsf{T}$ at $n_\tau$ discrete time points $\{\tau_r\}_{r=1}^{n_\tau}$. We describe the exact architecture and training procedure in Section~\ref{sec:methods:graph_neural_odes} and Section~\ref{sec:methods:training} respectively.

In our four evaluation strategies, we make use of the hypercanonical $\mathbb{S}^1$-model~\cite{serrano2008_selfsimilaritycomplexnetworks, krioukov2010_hyperbolicgeometrycomplex, vanderhoorn2018_sparsemaximumentropyrandom, boguna2020_smallworldsclustering}, a flexible random graph model that allows us to control the properties of training and test graphs using only four input parameters: The number of nodes $n$, the average degree $\bar{k}$, the exponent $\gamma$ of the degree distribution, and inverse temperature $\beta$. The exponent $\gamma$ determines how degree heterogeneous graphs sampled from the model are, i.e., whether some nodes, referred to as hubs, have many more neighbors than an average node. The smaller $\gamma$ the larger the degree heterogeneity. The inverse temperature $\beta$ controls the average clustering coefficient $\bar{c}$ of graphs sampled from the model, i.e., the probability that two neighbors of a random node are connected. The clustering coefficient increases as $\beta$ increases. With our training graphs we cover five different regimes, very degree heterogeneous graphs with weak clustering, $(\gamma^\mathrm{train}, \, \beta^\mathrm{train})=(2.1, \, 0.1)$, or strong clustering, $(\gamma^\mathrm{train}, \, \beta^\mathrm{train})=(2.1, \, 4.1)$, less degree heterogeneous graphs with weak clustering,  $(\gamma^\mathrm{train}, \, \beta^\mathrm{train})=(3.9, \, 0.1)$, or strong clustering, $(\gamma^\mathrm{train}, \, \beta^\mathrm{train})=(3.9, \, 4.1)$, and finally an intermediate case, $(\gamma^\mathrm{train}, \, \beta^\mathrm{train})=(3.0, \, 1.1)$. We provide more background on the $\mathbb{S}^1$-model in Section~\ref{sec:methods:s1-model}, where we also describe how we sample training and test graphs.

Our first evaluation strategy (Fig.~\ref{fig:schema}\textbf{(c)}) focuses on size generalization: Can \model{}s trained on time series from a specific dynamical system on small graph, $n^\mathrm{train}=64$, make predictions on graphs of equal or larger size, $64\leq n^\mathrm{test}\leq 8192$,  with the same parameters, $(\gamma^\mathrm{test}, \, \beta^\mathrm{test})=(\gamma^\mathrm{train}, \, \beta^\mathrm{train})$? We study this question in Section~\ref{sec:results:size_generalization_global}.

Our second strategy (Fig.~\ref{fig:schema}\textbf{(d)}) assesses the ability of \model{}s to generalize to graphs with different parameter values, $(\gamma^\mathrm{test}, \, \beta^\mathrm{test})\neq(\gamma^\mathrm{train}, \, \beta^\mathrm{train})$, and of the same or larger size, $n^\mathrm{test}\in\{64, \, 8192\}$, than the training graph. We present the relevant results in Section~\ref{sec:results:property_generalization}.

Third, we investigate whether \model{}s can capture fixed-points $\mathbf{x}^\star$ and their local stability properties (Fig.~\ref{fig:schema}\textbf{(e)}). In this evaluation strategy, presented in Section~\ref{sec:results:stability}, we focus on graphs with the same parameters as the training graph, $(\gamma^\mathrm{test}, \, \beta^\mathrm{test})=(\gamma^\mathrm{train}, \, \beta^\mathrm{train})$, but vary the test graph size, $n^\mathrm{test}\in\{64, \, 8192\}$.

Finally, we determine whether \model{}s employed on graphs much larger than the training graph, $n^\mathrm{test}=8192$, but with the same parameters, $(\gamma^\mathrm{test}, \, \beta^\mathrm{test})=(\gamma^\mathrm{train}, \, \beta^\mathrm{train})$, are robust to the presence of unobserved nodes (Fig.~\ref{fig:schema}\textbf{(f)}). Can a \model{} make accurate predictions when only the initial conditions of $n^\mathrm{obs}<n^\mathrm{test}$ nodes and their induced subgraph are available? We answer this question in Section~\ref{sec:results:missingness}.

We provide further details on the experimental setup for each of these evaluation strategies in Section~\ref{sec:methods:evaluation}.

\subsection{Generalizing to larger graphs}\label{sec:results:size_generalization_global}

\begin{figure*}[tb]
    \centering
    \includegraphics[width=0.965\textwidth]{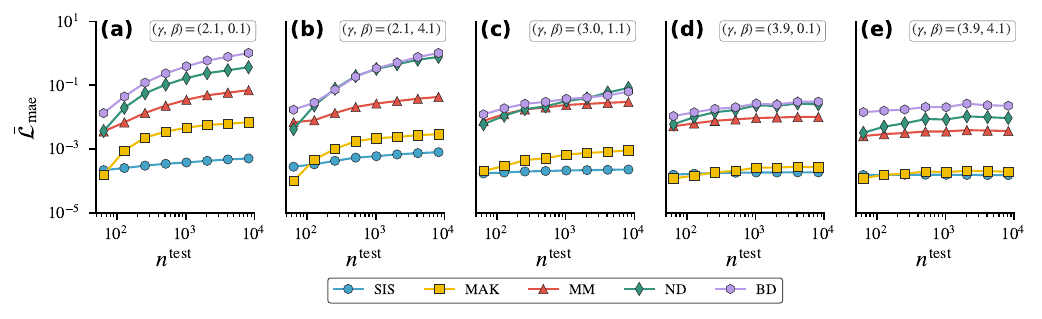}
    \caption{\textbf{Generalization across graph sizes.} The ability of \model{}s trained to approximate different dynamical systems (SIS (blue circles), MAK (yellow squares), MM (red triangles), ND (green diamonds), BD (purple hexagons)) on small graphs, $n^\mathrm{train}=64$, to make accurate predictions on larger graphs, $n^\mathrm{test}$, with the same properties as the training graph, $(\gamma^\mathrm{test},\,\beta^\mathrm{test})=(\gamma^\mathrm{train}, \, \beta^\mathrm{train})=(\gamma, \, \beta)$, differs between \textbf{(a)} very degree heterogeneous graphs with weak clustering, $(\gamma, \, \beta)=(2.1, \, 0.1)$, \textbf{(b)} very degree heterogeneous graphs with strong clustering, $(\gamma, \, \beta)=(2.1, \, 4.1)$, \textbf{(c)} graphs with moderate degree heterogeneity and clustering, $(\gamma, \, \beta)=(3.0, \, 1.1)$, \textbf{(d)} less degree heterogeneous graphs with weak clustering $(\gamma, \, \beta)=(3.9, \, 0.1)$, and \textbf{(e)} less degree heterogeneous graphs with strong clustering, $(\gamma, \, \beta)=(3.9, \, 4.1)$. The mean node-wise MAE, $\bar{\mathcal{L}}_\mathrm{mae}$, over $n_G^\mathrm{test}=100$ test graphs, stays constant or increases slowly on less degree heterogeneous graphs independent of clustering but increases noticeably on more degree heterogeneous graphs for most \model{}s. Those trained on the SIS model show the smallest increase in mean MAE. This means degree heterogeneity is a limiting factor for size generalization of \model{}s predicting dynamical systems on graphs.}
    \label{fig:size_generalization_mean_mae_noiseless}
\end{figure*}

In this section, we investigate the ability of \model{}s trained on small graphs, $n^\mathrm{train}=64$, to generalize to much larger graphs, $64 \leq n^\mathrm{test}\leq 8192$, with the same parameters as the training graph, $(\gamma^\mathrm{test}, \, \beta^\mathrm{test})=(\gamma^\mathrm{train}, \, \beta^\mathrm{train})=(\gamma, \, \beta)$.

The ability of \model{}s to generalize across graph sizes (Fig.~\ref{fig:size_generalization_mean_mae_noiseless}) depends considerably on the dynamical system under consideration as well as the graph properties. 

On very degree heterogeneous graphs with weak clustering ($(\gamma, \, \beta)=(2.1, \, 0.1)$, Fig.~\ref{fig:size_generalization_mean_mae_noiseless}\textbf{(a)}), and with strong clustering ($(\gamma, \, \beta)=(2.1, \, 4.1)$, Fig.~\ref{fig:size_generalization_mean_mae_noiseless}\textbf{(b)}) the mean $\bar{\mathcal{L}}_\mathrm{mae}$ of the node-wise mean absolute error (MAE) increases with test graph size. Only in the case of the SIS model (blue circles) does it remain low. This means size generalization is hard to achieve on degree heterogeneous graphs.

At intermediate degree heterogeneity and clustering ($(\gamma, \, \beta)=(3.0, \, 1.1)$, Fig.~\ref{fig:size_generalization_mean_mae_noiseless}\textbf{(c)}) mean MAE increases less with graph size independent of the dynamical system. However, for \model{}s trained on the MM (red triangles), ND (green diamonds), and BD (purple hexagons) model the increase is still substantial. Thus, size generalization is possible for \model{}s trained on the SIS and MAK (yellow squares) model but not for \model{}s trained on other systems.

Finally, for less degree heterogeneous graphs with weak clustering ($(\gamma, \, \beta)=(3.9, \, 0.1)$, Fig.~\ref{fig:size_generalization_mean_mae_noiseless}\textbf{(d)}) and with strong clustering ($(\gamma, \, \beta)=(3.9, \, 4.1)$), Fig.~\ref{fig:size_generalization_mean_mae_noiseless}\textbf{(e)}), mean MAE does not increase substantially, independent of the dynamical system. In the less degree heterogeneous case, size generalization is thus achievable.

The differences between \model{}s trained on different dynamical systems and the influence of degree heterogeneity can be explained by the scaling of a node's state with its degree. As we show in Appendix~\ref{appendix:scaling}, the magnitude of a node's state grows with its degree in the MM, ND, and BD models. This means if the graph is degree heterogeneous and large, the \model{} will encounter regions of the state space not covered by the training data when making predictions for high degree nodes. This problem is less severe for more degree homogeneous graphs or in the SIS model or MAK model, for which the node state's magnitude does not grow with graph size.

In Appendix~\ref{appendix:size_generalization}, we show that our findings hold for other measures of predictive performance and for when the training data has been corrupted by Gaussian noise. As graph-level evaluations can be misleading, we provide a node-level perspective on size generalization in Appendix~\ref{appendix:size_generalization_local}, which further underscores the role of hubs in shaping performance.

\subsection{Generalizing across graph properties}\label{sec:results:property_generalization}

\begin{figure*}[tb]
    \centering
    \includegraphics[width=0.965\textwidth]{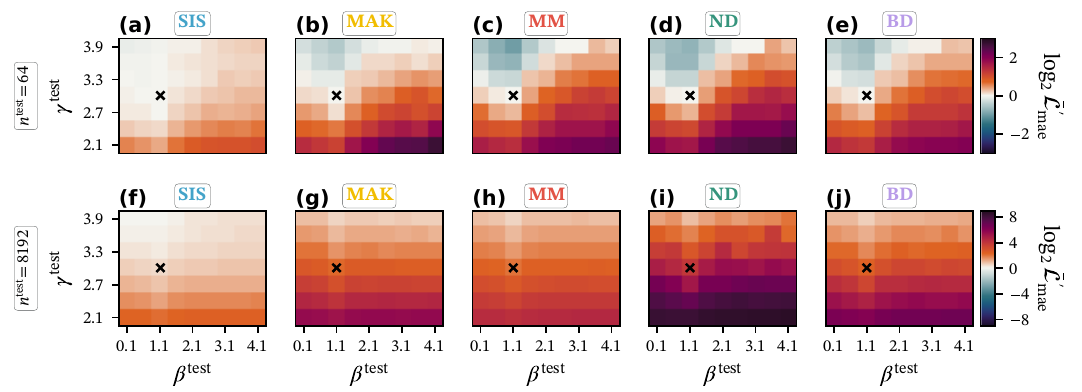}
    \caption{\textbf{Generalizing to graphs with different properties.} The ability of \model{}s trained on small graphs, $n^\mathrm{train}=64$, with moderate degree heterogeneity and clustering, $(\gamma^\mathrm{train}, \, \beta^\mathrm{train})=(3.0, \, 1.1)$ (black cross), to generalize to graphs with different properties, $(\gamma^\mathrm{test}, \, \beta^\mathrm{test})$, on $n_G^\mathrm{test}=100$ test graphs of the same size, $n^\mathrm{test}=64$ (upper row), or larger size, $n^\mathrm{test}=8192$ (lower row), is measured as the change in mean node-wise MAE (color coded) relative to its value on graphs with the same properties and size as the training graph, denoted $\bar{\mathcal{L}}^\prime_\mathrm{mae}$. On small graphs, the normalized MAE decreases towards lower degree heterogeneity and less clustering, and increases towards more degree heterogeneity and more clustering for all dynamical systems (\textbf{(a)} SIS, \textbf{(b)} MAK, \textbf{(c)} MM, \textbf{(d)} ND, \textbf{(e)} BD model). On larger graphs, normalized MAE increases throughout the $\mathbb{S}^1$-model parameter range independent of the dynamical system (\textbf{(f)} SIS, \textbf{(g)} MAK, \textbf{(h)} MM, \textbf{(i)} ND, \textbf{(j)} BD), and most strongly towards higher degree heterogeneity. The increase is moderate for the \model{} trained on the SIS model but substantial for \model{}s trained on the ND model. This means \model{}s are to some extent robust to changes in graph properties, especially if deployed on graphs with low degree heterogeneity and clustering and similar size to the training graph.}
\label{fig:property_generalization_mae_3011}
\end{figure*}
%
Complex systems in the real world can be hard to measure and often evolve over time, leading to a potential mismatch between the properties of the graph structure encountered during training and the one relevant during deployment of a \model{}. This motivates us to examine whether \model{}s trained on small graphs, $n^\mathrm{train}=64$, with a given set of parameters can generalize to graphs of the same or larger size with different parameters. Here, we focus on the case of intermediate degree heterogeneity and clustering, $(\gamma^\mathrm{train}, \, \beta^\mathrm{train})=(3.0, \, 1.1)$, and evaluate performance in terms of mean node-wise MAE normalized to its values on graphs of the same size and properties as the training graph, denoted $\bar{\mathcal{L}}_\mathrm{mae}^\prime$. In Appendix~\ref{appendix:property_generalization}, we discuss other settings and show that our findings also hold when accounting for the magnitude of node states by using MRE instead of MAE.
%
On graphs of the same size as the training graph ($n^\mathrm{test}=64$, Fig.~\ref{fig:property_generalization_mae_3011}\textbf{(a)}-\textbf{(e)}), we find that \model{}s show equivalent or even lower mean MAE on graphs that are less degree heterogeneous, $\gamma^\mathrm{test}>3.0$, and have lower clustering, $\beta^\mathrm{test}<1.1$, whereas MAE increases as degree heterogeneity, $\gamma^\mathrm{test}<3.0$, or clustering, $\beta^\mathrm{test}>1.1$, increase. These trends are present irrespective of the dynamical system the \model{} was trained on. The \model~trained on the SIS model maintains lowest normalized MAE, while the \model~trained on the ND model shows the strongest increase in MAE towards graphs with more degree heterogeneity and clustering.

Considering generalization across properties on graphs larger ($n^\mathrm{test}=8192$, Fig.~\ref{fig:property_generalization_mae_3011}\textbf{(f)}-\textbf{(j)}) than the training graph results differ by dynamical systems: For \model{}s trained on the SIS model, mean MAE remains low on larger graphs that are equally or less degree heterogeneous than the training graph, and increases moderately on more degree heterogeneous graphs. For \model{}s trained on other dynamical systems,  mean MAE increases on larger graphs for all parameter settings, but to varying extent. This increase is most noticeable for the ND model. Across dynamical systems, the increase in MAE is most benign on less degree heterogeneous graphs, $\gamma^\mathrm{test}>3.0$, and more severe on more degree heterogeneous graphs, $\gamma^\mathrm{test}<3.0$. The average clustering coefficient of the test graphs, tuned via the inverse temperature $\beta^\mathrm{test}$, shows limited effect on normalized MAE.

Overall, this means that \model{}s have some capability to generalize across graph properties, especially towards less degree heterogeneous graphs or graphs with lower average clustering. However, this capability declines on graphs that are larger than the training graph, and generalization towards more degree heterogeneous graphs can fail drastically in this case. This is in line with our findings in Section~\ref{sec:results:size_generalization_global}.

\subsection{Fixed point approximation and local stability}\label{sec:results:stability}

\begin{figure*}[tb]
    \centering
    \includegraphics[width=0.965\textwidth]{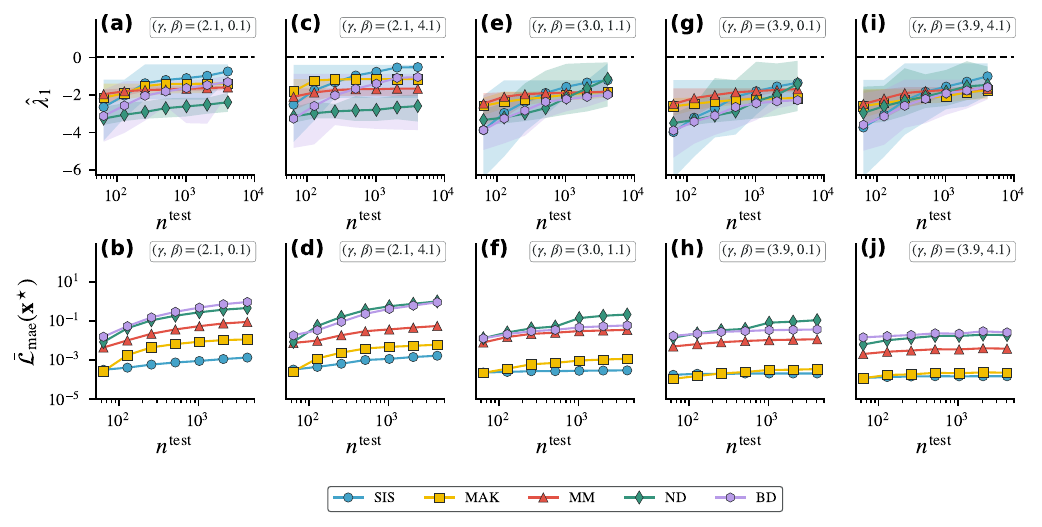}
    \caption{\textbf{Local stability and fixed point approximation.} Across dynamical systems (SIS (blue circles), MAK (yellow squares), MM (red triangles), ND (green diamonds), BD (purple hexagons)) and graph parameters (\textbf{(a)}, \textbf{(b)} $(\gamma, \, \beta)=(2.1, \, 0.1)$, \textbf{(c)}, \textbf{(d)} $(\gamma, \, \beta)=(2.1, \, 4.1)$, \textbf{(e)}, \textbf{(f)} $(\gamma, \, \beta)=(3.0, \, 1.1)$, \textbf{(g)}, \textbf{(h)} $(\gamma, \, \beta)=(3.9, \, 0.1)$, \textbf{(i)}, \textbf{(j)} $(\gamma, \, \beta)=(3.9, \, 4.1)$) the largest eigenvalue $\hat{\lambda}_1$ of the Jacobian (upper row) remains negative (shaded band) for \model{}s trained on small graphs, $n^\mathrm{train}=64$, with the same parameters as the test graph, $(\gamma^\mathrm{train}, \,  \beta^\mathrm{train})=(\gamma^\mathrm{test}, \, \beta^\mathrm{test})=(\gamma, \, \beta)$, even though it increases on average (solid line) with the size of the test graph $n^\mathrm{test}$. The average MAE (lower row) of the fixed point approximation, $\bar{\mathcal{L}}_\mathrm{mae}^\star$, increases noticeably  across dynamical systems on very degree heterogeneous graphs independent of clustering but stays constant or increases only slightly in the less degree heterogeneous case. This means that even though \model{}s approach stable fixed points, these fixed points can differ from the dynamical system's true fixed point,  especially on very degree heterogeneous graphs.}
    \label{fig:jac_mean_mae_noiseless}
\end{figure*}

Fixed points and their local stability are fundamental to understanding dynamical systems, as they offer insights into long term behavior and response to perturbations. If the system is poised at a stable fixed point perturbations decay, whereas they are amplified at an unstable fixed point. These two cases can be distinguished by the sign of the largest eigenvalue $\lambda_1$ of the Jacobian $J(\mathbf{x}^\star)$ at the fixed point $\mathbf{x}^\star$, which is negative in the stable and positive in the unstable case.

Figure~\ref{fig:jac_mean_mae_noiseless} shows that \model{}s trained on small graphs, $n^\mathrm{train}=64$, and deployed on larger graphs, $64\leq n^\mathrm{test}\leq4096$, with the same properties as the training graph, $(\gamma^\mathrm{test}, \, \beta^\mathrm{test})=(\gamma^\mathrm{train}, \beta^\mathrm{train})=(\gamma, \beta)$, approach locally stable fixed points but that these fixed points can differ from those of the ground truth dynamical system.

In the case of very degree heterogeneous graphs with weak clustering, $(\gamma, \, \beta)=(2.1, \, 0.1)$, the largest eigenvalue of the Jacobian (Fig.~\ref{fig:jac_mean_mae_noiseless}\textbf{(a)}) is negative for all dynamical systems, but increases on average with the size of the system. Mean MAE is low at small test graphs but increases with test graph size (Fig.~\ref{fig:jac_mean_mae_noiseless}\textbf{(b)}). The predictions remain most accurate for the \model~trained on the SIS model (blue circles), and worsen most for models trained on the ND (green diamonds) and BD (purple hexagons) models. This suggests that \model{}s approach a different but stable fixed point if the test graph is much larger than the training graph. We observe a very similar behavior of the largest eigenvalue (Fig.~\ref{fig:jac_mean_mae_noiseless}\textbf{(c)}) and MAE at the fixed point (Fig.~\ref{fig:jac_mean_mae_noiseless}\textbf{(d)}) in case of very degree heterogeneous graphs with strong clustering, $(\gamma, \, \beta)=(2.1, \, 4.1)$.

For graphs with moderate degree heterogeneity and clustering, $(\gamma, \, \beta)=(3.0, \, 1.1)$, we reach similar conclusions on stability (Fig.~\ref{fig:jac_mean_mae_noiseless}\textbf{(e)}), but find that the MAE at the fixed point (Fig.~\ref{fig:jac_mean_mae_noiseless}\textbf{(f)}) increases less strongly in all systems---especially for the ND and BD models.

On less degree heterogeneous graphs with weak clustering, $(\gamma, \, \beta)=(3.9, \, 0.1)$, or strong clustering, $(\gamma, \, \beta)=(3.9, \, 4.1)$, we find that the Jacobian's largest eigenvalue remains negative irrespective of the dynamical system the \model{} was trained on (Fig.~\ref{fig:jac_mean_mae_noiseless}\textbf{(g)}, \textbf{(i)}). In contrast to the more degree heterogeneous cases the MAE at the fixed point increases only slowly or stays constant with test graph size (Fig.~\ref{fig:jac_mean_mae_noiseless}\textbf{(h)}, \textbf{(j)}). This indicates that in the less degree heterogeneous case \model{}s are able to accurately identify fixed points even on graphs that are larger than the training graph.

In summary, our analysis of the local stability of \model{}s trained on different dynamical systems and graphs with different properties shows that fixed points are accurately captured in the less degree heterogeneous regime, and serves as a note of caution in the degree heterogeneous regime: Even though \model{}s reach stable fixed points, these might differ from the true fixed point of the system. We discuss how robust these findings are to noise in the training data in Appendix~\ref{appendix:jacobian}, where we also provide error quantiles.

\subsection{Robustness on partially observed graphs}\label{sec:results:missingness}

\begin{figure*}[tb]
    \centering
    \includegraphics[width=0.965\textwidth]{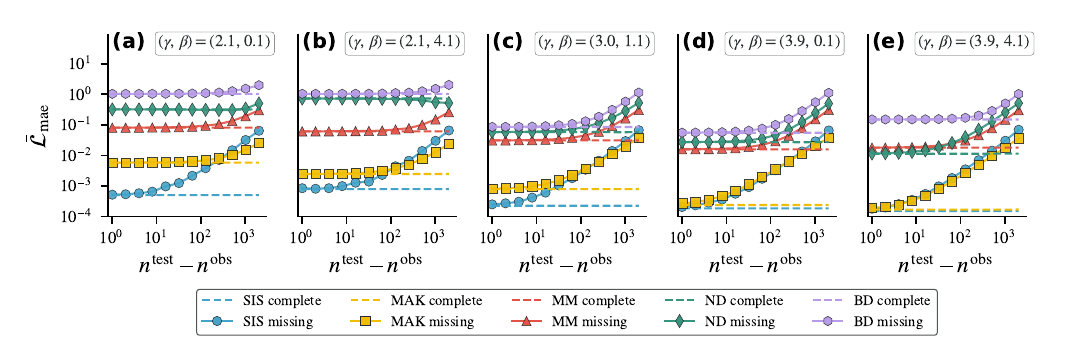}
    \caption{\textbf{Robustness of predictions to unobserved nodes.} The robustness of \model{}s trained on small graphs,  $n^\mathrm{train}=64$, to unobserved nodes in larger test graphs, $n^\mathrm{test}=8192$, with the same parameters as the training graph, $(\gamma^\mathrm{test}, \, \beta^\mathrm{test}) =(\gamma^\mathrm{train},\, \beta^\mathrm{train}) =(\gamma, \, \beta)$, depends on the number of observed nodes $n^\mathrm{obs}$, graph properties (\textbf{(a)} $(\gamma, \, \beta)=(2.1, \, 0.1)$, \textbf{(b)} $(\gamma, \, \beta)=(2.1, \, 4.1)$, \textbf{(c)} $(\gamma, \, \beta)=(3.0, \, 1.1)$, \textbf{(d)} $(\gamma, \, \beta)=(3.9, \, 0.1)$, \textbf{(e)} $(\gamma, \, \beta)=(3.9, \, 4.1)$), and the dynamical system (SIS (blue circles), MAK (yellow squares), MM (red triangles), ND (green diamonds), BD (purple hexagons)). Overall, the mean node-wise MAE, $\bar{\mathcal{L}}_\mathrm{mae}$, if $n^\mathrm{test}-n^\mathrm{obs}$ nodes remain unobserved (solid line) stays equivalent to the baseline value for a completely observed graph (dashed line) longer on very degree heterogeneous than on less degree heterogeneous graphs with clustering playing a minor role. However, the MAE tends to be higher in the very degree heterogeneous than the less degree heterogeneous setting. This means robustness to unobserved nodes depends on a complex interplay of the type of dynamical system and graph properties.} 
    \label{fig:missing_mean_mae_noiseless}
\end{figure*}

In large complex systems observing the entirety of the system's state and structure can be challenging. Here, we investigate what problems arise when \model{}s trained on small, $n^\mathrm{train}=64$, but completely observed systems, are employed on larger test graphs,  $n^\mathrm{test}=8192$, with the same properties, $(\gamma^\mathrm{train}, \, \beta^\mathrm{train})=(\gamma^\mathrm{test}, \, \beta^\mathrm{test})=(\gamma, \, \beta)$, in which only $2048\leq n^\mathrm{obs}\leq n^\mathrm{test}$ nodes, chosen uniformly at random, are observed.

Figure~\ref{fig:missing_mean_mae_noiseless} shows that mean MAE at observed nodes increases with the number of unobserved nodes (solid lines) relative to the performance on the completely observed test graph (dashed lines) but that the onset of this increase depends on the dynamical system and graph parameters. This increase is to be understood relative to the performance on the fully-observed test graphs, as the axes are logarithmically scaled.
%
For very degree heterogeneous graphs with weak clustering ($(\gamma, \, \beta)=(2.1, \, 0.1)$, Fig.~\ref{fig:missing_mean_mae_noiseless}\textbf{(a)}) and strong clustering ($(\gamma, \, \beta)=(2.1, \, 4.1)$, Fig.~\ref{fig:missing_mean_mae_noiseless}\textbf{(b)}), the MAE of the \model{}s trained on the SIS model (blue circles) starts to increase when as few as $10$ nodes are unobserved, whereas this increase starts at $100$ unobserved nodes for \model{}s trained on the MAK model (yellow squares). \model{}s trained on the MM (red triangles), ND (green diamonds), or BD (purple hexagons) model show high MAE even on fully observed graphs (in line with Section~\ref{sec:results:size_generalization_global}), but little further increase in MAE.
%
If degree heterogeneity and clustering are moderate ($(\gamma, \, \beta)=(3.0, \, 1.1)$, Fig.~\ref{fig:missing_mean_mae_noiseless}\textbf{(c)}) MAE on the fully observed graph is lower across dynamical systems, but MAE increases at fewer unobserved nodes for the \model{} trained on the MAK, MM, ND, or BD models.
%
In the case of less degree heterogeneity and weak clustering ($(\gamma, \, \beta)=(3.9, \, 0.1)$, Fig.~\ref{fig:missing_mean_mae_noiseless}\textbf{(d)}) \model{}s show lower MAE across dynamical systems than for other graph parameters if all nodes are observed. However, for \model{}s trained on the SIS or MAK model, MAE increases if as few as $10$ nodes are unobserved. In the case of \model{}s trained on the MM, ND, or BD model this increase starts at $100$ unobserved nodes.  
%
On graphs with less degree heterogeneity and strong clustering ($(\gamma, \, \beta)=(3.9, \, 4.1)$, Fig.~\ref{fig:missing_mean_mae_noiseless}\textbf{(e)}) \model{}s trained on the SIS and MAK models behave similar to the case of weak clustering, but for \model{}s on the MM and ND models MAE increases earlier and for the one trained on the BD model it is higher even on the fully observed graph.

These experiments paint a complex picture of the interplay between the type of dynamical systems, the properties of the test graphs, and the number of unobserved nodes: Overall, \model{}s can tolerate more unobserved nodes on more degree heterogeneous graphs, potentially because activity concentrates on few hubs that are unlikely to be affected by random removal of nodes. On less degree heterogeneous graphs relative performance starts to decrease at a smaller number of unobserved nodes. However, as shown in Section~\ref{sec:results:size_generalization_global}, overall performance on fully observed graphs is worse in the very degree heterogeneous than in the less degree heterogeneous case.


\section{Discussion}

Extracting governing equations directly from time series data using \model{}s is a powerful modeling paradigm for understanding complex systems with graph-structured interactions across domains, from epidemic spreading to population dynamics. However, a detailed understanding of their capabilities and failure modes is vital to incorporating \model{}s into scientific workflows.
%
Here, we have employed four different evaluation strategies to understand the behavior of \model{}s, tailored to training data from dynamical systems in BB form (Eq.~\eqref{eq:bb-dynamics}), across five dynamical systems and graphs spanning different sizes as well as different levels of degree heterogeneity and average clustering.

First, we study how well \model{}s trained on small graphs can generalize to much larger graphs and reveal the central role of degree heterogeneity. While \model{}s generally show good size generalization when graphs are degree homogeneous, size generalization is limited to \model{}s trained on few dynamical systems in the degree heterogeneous case. We conjecture that the increase in activity with node degree pushes the model into regions of state space that were not observed during training when making predictions at hubs, thereby impairing predictive performance. While training on small graphs can dramatically lower training time and cost, our findings serve as a reminder to examine graph properties when attempting to exploit these efficiency gains. They also exemplify how insights from dynamical systems theory can help us understand the failure modes of machine learning models.

In addition, we examine the ability of \model{}s to generalize to graphs that differ in their degree heterogeneity and average clustering from those encountered during training. Across dynamical systems, we find that generalization towards graphs with less degree heterogeneity and clustering is possible but performance becomes fragile as degree heterogeneity or average clustering increase---a potentially worrying finding, since high degree heterogeneity and clustering are hallmarks of complex networks. This second evaluation strategy also showcases the utility of random graph models, which allow for precise control over graph properties, as a tool for evaluating machine learning models.

Moreover, by examining the fixed points of \model{}s and their stability we reveal that \model{}s reach stable fixed points for all five dynamical systems but also find that these fixed points might differ from the ones of the ground truth dynamical system depending on the graph's degree heterogeneity. These findings highlight the need to move beyond predictive performance when testing \model{}s, when they are employed on diverse tasks.

Finally, we show that, across dynamical systems and graph parameters, \model{}s are sensitive to the presence of unobserved nodes. Even if \model{}s perform well on the fully observed graph, performance deteriorates quickly as the fraction of observed nodes decreases. As mapping large complex systems in their entirety can be infeasible, these results call for the development of new training procedures and architectural changes to enable trustworthy deployment of \model{}s in messy real-world applications. 
%

Our study has several limitations that directly point to future work. First, we only focus on dynamical systems in BB form (Eq.~\eqref{eq:bb-dynamics}), a class that is broad enough to encompass many commonly studied dynamical systems but structured enough to explicitly inform the design of our \model{} architecture. Dynamical systems outside of this class---e.g., the Kuramoto model or Laplacian diffusion, for which the interaction term depends on node state differences---may be more adequately studied with architectures that parameterize the interaction term as non-separable, thereby adjusting the neural ODE's inductive bias.
Second, while the $\mathbb{S}^1$-model allows us to generate graphs with key properties of realistic complex networks (i.e., sparsity, degree heterogeneity, strong clustering, and small average shortest path length), empirical graphs often possess additional structure, such as communities or degree correlations, which can impact the behavior of dynamical systems~\cite{lambiotte2021_modularitydynamicscomplex}.
Using real-world networks may indeed aid in assessing the impact of mesoscopic graph structure on \model{} performance, but adequate null models, i.e., random graph models, offer an alternative and more principled approach~\cite{karrer2011_stochasticblockmodelscommunity, peixoto2019_bayesianstochastic, mahadevan2006_systematictopology, orsini2015_quantifying}.

%
Moreover, we focus on a specific neural ODE architecture and training hyperparameters. We argue that by modeling the architecture after the BB form (Eq.~\eqref{eq:bb-dynamics}), we create a strong inductive bias for the learning of these types of systems and thus examine a best-case scenario. However, as the design space of neural ODEs and graph machine learning models is vast~\cite{kidger2022_neuraldifferentialequations, you2020_designspacegraph}, exploring how some of the observed weaknesses can be mitigated through advanced architectures or training procedures is an important next step.

Taken together, our results offer a nuanced perspective on \model{}s as tools for complex systems research, emphasizing their potential to generalize across network sizes and properties but also their limitations on degree heterogeneous and partially-observed graphs. More broadly, by extending model evaluation beyond predictive performance alone and leveraging powerful, synthetic data generation, our evaluation framework can serve as a template for future assessments of data-driven methods for dynamical systems.

\section{Methods}\label{sec:methods}

In the following, we introduce the specific form of dynamical system from which we generate training and test data in greater detail. We proceed by introducing our neural ODE architecture, \model{}, and the training procedure. Then, we provide essential background on the $\mathbb{S}^1$-model and describe how we sample training and test graphs. Finally, we explain our four evaluation strategies. A guide to the relevant mathematical notation can be found in Appendix~\ref{appendix:notation}. We make all code available under~\url{https://github.com/moritz-laber/neural-ode-generalization}.

\subsection{Dynamical systems in Barab\'asi-Barzel form}\label{sec:methods:network_dynamics}

Prior work~\cite{barzel2013_universalitynetworkdynamics, hens2019_spatiotemporalsignalpropagation, meena2023_emergentstabilitycomplex} has highlighted the role of ODEs of the form of Eq.~\eqref{eq:bb-dynamics}, sometimes called the Barab\'asi-Barzel (BB) form, for understanding complex systems whose structure can be abstractly represented as a simple, undirected graph $G=(\mathcal{V}, \mathcal{E})$ with node set $\mathcal{V}$ and edge set $\mathcal{E}$. We denote the number of nodes as $n=|\mathcal{V}|$, the number of edges as $m=|\mathcal{E}|$, and the graph's adjacency matrix as $A\in\{0, 1\}^{n\times n}$.

Each node is characterized by a time-dependent state $x_i(t)$ taking values in the real numbers or a subset thereof. This means at time $t$ the state of the entire system is described by a vector $\mathbf{x}(t)=(x_1(t), \dots, x_n(t))^\mathsf{T}\in\mathbb{R}^n$. For example, in a simple model of population dynamics each node $i\in \mathcal{V}$ could represent a habitat and $x_i(t)$ the abundance of a species at time $t$.

We assume that in a time interval $[t_0, t_1]$, the time evolution of these node states is described by, 
\begin{widetext}
\begin{equation}\label{eq:bb-dynamics-methods}
    \frac{\mathrm{d}x_i(t)}{\mathrm{d}t}= F_i(\mathbf{x})=f(x_i(t))+\sum_{j=1}^{n}h^\mathrm{ego}(x_i(t))A_{ij}h^\mathrm{alt}(x_j(t))~~~~\forall\, i\in\mathcal{V}, 
\end{equation}
\end{widetext}
starting from an initial condition
\begin{equation}\label{eq:initial-condition}
    \mathbf{x}(t_0)=\mathbf{\tilde{x}}, 
\end{equation}
and with the functions $f, \, h^\mathrm{ego}, \, h^\mathrm{alt}:\mathbb{R}\to\mathbb{R}$ jointly characterizing the vector field. We call $f$ the self-dynamics and refer to $h^\mathrm{ego}, \, h^\mathrm{alt}$ as the contributions of the ego node $i$ and alter nodes $\mathcal{N}_i=\{j\in\mathcal{V}:A_{ij}=1\}$ to the interaction term.

In choosing this functional form, we make several assumptions. First, there is no explicit time dependence of the interaction, i.e., the system of differential equations is autonomous. Second, all nodes obey the same dynamics. This means the form of $f, \, h^\mathrm{ego}, \, h^\mathrm{alt}$ does not depend on the node $i$ under consideration. Third, the change of node $i$'s state at time $t$ depends only on node $i$'s state at time $t$ and the state of node $i$'s neighbors $j\in\mathcal{N}_i$ at time $t$. This assumption is often referred to as locality, and intuitively means that interactions are mediated by the edges of the graph. Finally, we assume that the interaction term factorizes in two functions $h^\mathrm{ego}$ and $h^\mathrm{alt}$ that only depend on the ego node $i$ and its neighbors respectively. This factorization is of course not unique. While many dynamical systems from various domains are well described by this form, it excludes some prominent examples such as Laplacian diffusion and the Kuramoto model, in which the interaction of nodes $i$ and $j$ depends on the difference of $x_i(t)$ and $x_j(t)$.

Here, we consider five dynamical systems from epidemiology, chemistry, gene-regulation, population dynamics, and neuroscience~\cite{hens2019_spatiotemporalsignalpropagation, vasiliauskaite2024_generalizationneuralnetwork}. We summarize their exact functional form in Tab.~\ref{tab:dynamics}.

\begin{table*}[tb]
    \centering
    \begin{tabular}{|c | c | c | c | c | c | c | c | c | c| c| c |}
         \hline
         \textbf{Name} & $f(x)$               & $h^\mathrm{ego}(x)$       & $h^\mathrm{alt}(x)$                       & $\nu$ & $\rho_1$ & $\rho_2$ & $\alpha_1$ & $\alpha_2$ & $\tilde{x}_\mathrm{min}$ & $\tilde{x}_\mathrm{max}$ & $\xi$\\ \hline
         SIS  & $-\rho_1 x$             & $1-x$                       & $\rho_2 x$                              & -     & $1.0$ & $1.2$     & -      & -     & $0.0$ & $0.5$ & $0.005$ \\ 
         \hline
         MAK  & $\nu - \rho_1 x$        & $-x$                      & $\rho_2 x$                                 & $2.0$ & $2.0$ & $1.0$     & -      & -     & $0.0$ & $1.0$  & $0.005$\\
         \hline 
         MM   & $-\rho_1 x^{\alpha_1}$  & $1$                       & $\rho_2 x^{\alpha_2} / (1+x^{\alpha_2})$   & -     & $1.0$  & $1.0$    & $2.0$  & $0.5$ & $0.0$ & $5.0$ &  $0.025$ \\
         \hline
         BD   & $-\rho_1 x^{\alpha_1}$  & $1$                       & $\rho_2 x^{\alpha_2}$                      & -     & $0.5$  & $1.0$    & $2.0$  & $0.5$ & $0.0$ & $10.0$ & $0.050$ \\
         \hline
         ND   & $ \nu\tanh(x)-\rho_1 x$ & $1$                       & $\rho_2 \tanh(x)$                          & $0.5$ & $4.0$  & $1.0$    & -      & -     & $0.0$ & $5.0$  & $0.025$\\
         \hline
    \end{tabular}
    \caption{\textbf{Overview of studied dynamical systems}. The functional form of five dynamical systems, Susceptible-Infected-Susceptible (SIS) model, Mass-Action Kinetics (MAK), Michaelis-Menten (MM) model, Birth-Death (BD) process, and Neuronal Dynamics (ND) in terms of the functions $f, \, h^\mathrm{ego}, \, h^\mathrm{alt}$, and the values of scalar parameters $\nu, \, \rho_1, \, \rho_2, \, \alpha_1, \, \alpha_2$, kept fixed throughout the manuscript. Initial conditions are sampled uniformly at random from the interval $(\tilde{x}_\mathrm{min}, \tilde{x}_\mathrm{max})$. To generate noisy training data we add zero-mean Gaussian noise with standard deviation $\xi$ to solutions of the dynamical system.}
    \label{tab:dynamics}
\end{table*}

To generate training or test data on a fixed graph with adjacency matrix $A$, we sample the initial condition for each node $i$ uniformly at random in the interval $[\tilde{x}_\mathrm{min}, \tilde{x}_\mathrm{max}]$, and solve the resulting initial value problem numerically. We use a Dormand-Prince 5/4 solver based on a 5th-order Runge-Kutta method with 4th-order sub-routine for step size adjustment, as implemented in the \texttt{diffrax} package~\cite{kidger2022_neuraldifferentialequations} with relative tolerance $\epsilon_\mathrm{rel}=10^{-4}$, absolute tolerance $\epsilon_\mathrm{abs}=10^{-4}$ and initial step size $\Delta t_\mathrm{init}=10^{-5}$. We store the result at $n_\tau^\mathrm{train}=256$ regularly spaced time points $\tau_r\in[0, \, 1]$.

To generate training data, we sample $n_G^\mathrm{train}=1$ graph at a given setting of the parameters $n^\mathrm{train}, \, \bar{k}^\mathrm{train}, \, \gamma^\mathrm{train}, \, \beta^\mathrm{train}$ of the $\mathbb{S}^1$-model and $n_\mathrm{init}^\mathrm{train}=256$ independent initial conditions. Section~\ref{sec:methods:s1-model} describes how we generate these graphs. We gather the node-wise time series resulting from solving Eq.~\eqref{eq:bb-dynamics-methods} in tensors, $X^\mathrm{train}\in\mathbb{R}^{n_\mathrm{init}^\mathrm{train}\times n_\tau^\mathrm{train} \times n^\mathrm{train}}$, where $n^\mathrm{train}$ is the number of nodes in each training graph. For training, we use small graphs $n^\mathrm{train}=64$.

To test the robustness of our results to noise, we also create a noisy version of the training data by adding independent Gaussian noise with zero mean and a standard deviation $\xi$, depending on the dynamical system as reported in Tab.~\ref{tab:dynamics}, to each entry of $X^\mathrm{train}$. We denote the noisy version of the training data as $X^{\prime \, \mathrm{train}}\in\mathbb{R}^{n^\mathrm{train}_\mathrm{init}\times n_\tau^\mathrm{train}\times n^\mathrm{train}}$.

\subsection{Neural ODE architecture}\label{sec:methods:graph_neural_odes}

In creating our neural ODE architecture, \model{}, we mirror Eq.~\eqref{eq:bb-dynamics-methods}, i.e., we assume a vector field
\begin{widetext}
    \begin{equation}\label{eq:nODE}
    \frac{\mathrm{d}x_i(t)}{\mathrm{d}t}=F_{\omega, i}(\mathbf{x}(t), A)=f_\omega(x_i(t))+\sum_{j=1}^n h_\omega^{\mathrm{ego}}(x_i(t))A_{ij}h_\omega^{\mathrm{alt}}(x_j(t)), 
\end{equation}
\end{widetext}
parameterized by three multilayer perceptrons (MLPs) $f_\omega, \, h_\omega^\mathrm{ego}, \, h_\omega^\mathrm{alt}:\mathbb{R}\to\mathbb{R}$. We use $\omega$ to denote the parameters of the entire architecture $F_\omega$, and emphasize that $f_\omega, \, h_\omega^\mathrm{ego}, \, h_\omega^\mathrm{alt}$ have different weight matrices and biases. Each of the MLPs has $3$ layers of width $64$, leading to a total of $3(2\cdot 64+2\cdot 64^2+3\cdot 64+2)=25,542$ parameters per \model{}. We implement the entire model in the differentiable programming language \texttt{jax}~\cite{bradbury2024_JAX} using the \texttt{equinox} library~\cite{kidger2022_neuraldifferentialequations}. 

By parameterizing the vector field in the form of Eq.~\eqref{eq:nODE}, we make similar assumptions to the ones discussed in the previous section (Section~\ref{sec:methods:network_dynamics}). However, we can give them additional meaning from a machine learning perspective. First, we assume that the vector field has no explicit time dependence. This allows us to use the same set of parameters $\omega$ to make predictions at each time point $t$, i.e., employ weight sharing, and also means we do not need to provide time as an input either directly or in form of a temporal encoding. Second, we assume that the functional form of the vector field is the same at all nodes. This assumption means that parameters $\omega$ are shared across nodes and is standard practice in graph machine learning. Third, the vector field at a given $i$ node only depends on neighbors $j$ in its neighborhood $\mathcal{N}_i$. This modeling assumption resembles the design of message-passing neural networks (MPNNs). However, MPNNs do not usually assume that the interaction term, called the message function in the graph machine learning literature, factorizes. This form of $F_\omega$ allows us to use graphs of different size and structure during testing and training. This setting, called inductive learning, is central to our endeavor of understanding generalization from one graph structure to another.

When making predictions with \model{}s, we use a Dormand-Prince 5/4 solver with adaptive step size from the \texttt{diffrax} package~\cite{kidger2022_neuraldifferentialequations} with relative tolerance $\epsilon_\mathrm{rel}=10^{-4}$, absolute tolerance $\epsilon_\mathrm{abs}=10^{-4}$, and an initial step size of $\Delta t_\mathrm{init}=10^{-5}$. The same numerical method is used to generate the training and test data (see Section~\ref{sec:methods:network_dynamics} and Section~\ref{sec:methods:evaluation}).

\subsection{Training}\label{sec:methods:training}

Having described the architecture, we now turn to the question of how to find a good set of weights $\omega$, i.e., training. We start from the node-wise time series gathered in the tensor $X^\mathrm{train}\in\mathbb{R}^{n^\mathrm{train}_\mathrm{init}\times n_\tau^\mathrm{train}\times n^\mathrm{train}}$, where $n_\mathrm{init}^\mathrm{train}=256$ is the number of independent initial conditions, $n_\tau^\mathrm{train}$ the number of discrete time points, and $n^\mathrm{train}=64$ is the number of nodes in the training graph. We denote the value of the state of the $i$th node at the $r$th time point $\tau_r$ when solving from the $a$th initial condition as $x_i^{(a)}(\tau_r)$.

For each dynamical system and set of $\mathbb{S}^1$-model parameters, we train a different \model{}. We draw the initial weights independently at random from a normal distribution with zero mean and standard deviation $\sigma_\mathrm{init}=\sqrt{\frac{0.01}{d_\mathrm{in}}}$, where the fan-in factor $d_\mathrm{in}$ is the number of neurons in the preceding layer. We initialize all biases to zero.

The training process proceeds iteratively: 
First, we choose a random permutation of $\{1,\dots, n_\mathrm{init}^\mathrm{train}\}$ and shuffle the first, i.e., initial condition, dimension of $X^\mathrm{train}$ according to that permutation. We continue to process the data in batches of $n_b=32$ different initial conditions, and re-shuffle the batch dimension when all initial conditions have been used. We do not subset the second, i.e., time, or third, i.e., node, dimension of $X^\mathrm{train}$. We denote a batch of the training data as $X^{\mathrm{train}, b}\in \mathbb{R}^{n_b\times n_\tau^\mathrm{train} \times n^\mathrm{train}}$.
We then numerically solve the ODE defined by Eq.~\eqref{eq:nODE} from the initial conditions,
\begin{equation}
    \mathbf{\tilde{x}}^\mathrm{(a')} = (X^{\mathrm{train}, b}_{a', 0, 1},\dots, X^{\mathrm{train}, b}_{a', 0, n^\mathrm{train}})^\mathsf{T},
\end{equation}
for $a'\in\{1,\dots,n_b\}$, and save the numerical approximation to the solution at all times $\{\tau_r\}_{r=1}^{n_\tau^\mathrm{train}}$. We denote the tensor formed by these model predictions as $\hat{X}^\mathrm{train}\in\mathbb{R}^{n_b\times n_\tau^\mathrm{train} \times n^\mathrm{train}}$. 
We compare the prediction $\hat{X}^\mathrm{train}$ to the batch $X^{\mathrm{train}, b}$ using the root mean squared error
\begin{widetext}
    \begin{equation}\label{eq:lossfunction}
    \mathcal{L}(\hat{X}^\mathrm{train}, X^{\mathrm{train}, b})
    =
    \left(
    \frac{1}{n_b n_\tau^\mathrm{train} n^\mathrm{train}}
    \sum_{a = 1}^{n_b}
    \sum_{i = 1}^{n^\mathrm{train}}
    \sum_{r = 1}^{n_\tau^\mathrm{train}}
    (\hat{X}^\mathrm{train}_{a, r, i} - X^{\mathrm{train}, b}_{a, r, i})^2
    \right)^{1/2}, 
\end{equation}
\end{widetext}
as a loss function and update the values of the \model{} parameters $\omega$ using a gradient-based optimization algorithm \texttt{AdaGrad} as implemented in the \texttt{optax} library~\cite{deepmind2020_optax} that adaptively tunes the learning rate for each dimension of the weight space. We set the global scaling factor $\eta$ of the learning rate to $\eta=10^{-3}$, and otherwise use default parameters: an initial accumulator value of $g_0=0.01$ and a constant of $\varepsilon=10^{-7}$ that prevents division by zero. A single gradient update constitutes one step of the training algorithm. We use a total of $S=10^5$ such steps.
%
As we are using a differentiable numerical solver implemented in the \texttt{diffrax}~\cite{kidger2022_neuraldifferentialequations} we can compute the gradients $\nabla_\omega \mathcal{L}(\hat{X}^\mathrm{train}, X^{\mathrm{train}, b})$ of the loss function efficiently using automatic differentiation.

To provide intuition for the trained \model{}s, we visualize examples of test time series and predictions in Appendix~\ref{appendix:timeseries} and visually compare the learned vectorfield to the analytical one in Appendix~\ref{appendix:vectorfield}.

\subsection{Generating graphs}\label{sec:methods:s1-model}
The properties of dynamical systems on graphs as well as graph machine learning models can strongly vary with the properties of the underlying graph. This makes it desirable to generate graphs from random graph models to systematically vary their properties. The $\mathbb{S}^1$-model stands out as a model with tunable average degree, degree heterogeneity, and clustering coefficient that can generate graphs of varying size~\cite{serrano2008_selfsimilaritycomplexnetworks, krioukov2010_hyperbolicgeometrycomplex, vanderhoorn2018_sparsemaximumentropyrandom, boguna2020_smallworldsclustering}. Degree heterogeneity and clustering are among the most important properties of real-world networks~\cite{albert2002_statisticalmechanicscomplex, newman2010_networksintroduction, dorogovtsev2022_naturecomplexnetworks, boguna2021_networkgeometry}. The $\mathbb{S}^1$-model has the added benefit of being fully characterized by only four numbers that serve as the input to the model: The number of nodes $n$, the average degree $\bar{k}$, the power law exponent $\gamma$ of the degree distribution, and the inverse temperature $\beta$.

We generate graphs from the hypercanonical $\mathbb{S}^1$-model as follows: We assign each node $i$ two hidden variables $\kappa_i$ and $\theta_i$. The former, $\kappa_i$, is sampled independently for each node from the Pareto distribution $\kappa_i\sim\mathrm{Pareto}[\kappa_\mathrm{min}, \gamma-1]$ where $\kappa_\mathrm{min}$ is related to the average degree $\bar{k}$ as $\kappa_\mathrm{min}=\frac{\gamma - 2}{\gamma-1}\bar{k}$ and $\gamma$ is the exponent of the Pareto distribution's probability density function. The hidden variable $\kappa_i$ controls a node's expected degree under the model. The parameter $\bar{k}$ thus controls how sparse the sampled graphs are, while $\gamma$ controls their degree heterogeneity, i.e., the extent to which the graph contains hubs that possess many more connections than an average node in the graph.

The latter hidden variable, $\theta_i$, is chosen independently for each node and uniformly at random, $\theta_i\sim \mathrm{U}[[0, 2\pi)]$. One can think of $\theta$s as angular coordinates on a circle, and define the distance $\theta_{ij}$ between nodes $i$ and $j$ as
\begin{equation}
    \theta_{ij}=\pi-|\pi - |\theta_i-\theta_j||.
\end{equation}
In the $\mathbb{S}^1$-model, a connection between two distinct nodes $i$ and $j$ with hidden variables ($\kappa_i$, $\theta_i$) and $(\kappa_j, \theta_j)$ respectively, exists with probability $p_{ij}$ depending on the distance $\theta_{ij}$ between their angular coordinates and their expected degrees. The exact functional form of $p_{ij}$ depends on the value of the inverse temperature $\beta$.

In the cold regime, $\beta\in (1, \infty)$, the connection probability is given by 
\begin{equation}\label{eq:pij_cold}
    p_{ij}= \frac{1}{1+(\frac{n\theta_{ij}}{\mu\kappa_i\kappa_j})^{\beta}}, 
\end{equation}
where $\mu$ is the logarithm of the chemical potential and is chosen such that $\kappa_i$ equals a node's expected degree by solving 
\begin{equation}\label{eq:chemical_potential}
    \bar{k} = \frac{2}{n}\sum_{i=1}^n\sum_{j=i+1}^n p_{ij}(\mu), 
\end{equation}
where we made the dependence of $p_{ij}$ on $\mu$ explicit. While this equation can be solved analytically in the thermodynamic limit $n\to\infty$, this is a poor approximation for small graphs~\cite{vanderkolk2022_anomaloustopologicalphase, boguna2020_smallworldsclustering}. We thus solve Eq.~\eqref{eq:chemical_potential} numerically using Newton's methods as implemented in \texttt{scipy.optimize}~\cite{virtanen2020_scipy10fundamental}. We use the value of $\mu$ in the thermodynamic limit as an initial guess. 

In the hot regime, $\beta\in(0, 1)$, the connection probability is given by
\begin{equation}\label{eq:pij_hot}
    p_{ij}= \frac{1}{1+\frac{(n\theta_{ij})^{\beta}}{\mu\kappa_i\kappa_j}}, 
\end{equation}
where again $\mu$ is the logarithm of the chemical potential and again found numerically by solving Eq.~\eqref{eq:chemical_potential} substituting Eq.~\eqref{eq:pij_hot} for $p_{ij}(\mu)$.

Graphs sampled from the $\mathbb{S}^1$-model are simple, undirected graphs. Depending on the values of $\gamma$ and $\beta$, these graphs exhibit very different properties. For values $\gamma\in(2, 3]$ degrees are very heterogeneous, and the second moment of the degree distribution $\langle k ^2\rangle$ diverges. As $\gamma>3$ the second moment is finite and the graphs approach degree homogeneity as $\gamma\to\infty$. In the hot regime ($\beta<1$), these graphs have vanishing average clustering coefficient $\bar{c}$ in the thermodynamic limit $n\to\infty$, thus this regime is sometimes called non-geometric. Non-vanishing clustering emerges in a phase transition at $\beta=1$, and for $\beta\in(1, \infty)$ graphs have non-vanishing clustering in the thermodynamic limit. We summarize these properties in Tab.~\ref{tab:S1_properties}.

Our five types of training graphs cover these different behaviors. All training graphs have $n^\mathrm{train}=64$ nodes and average degree $\bar{k}^\mathrm{train}=10$, but differ in their degree heterogeneity and their clustering. We use graphs with high degree heterogeneity and weak clustering, $(\gamma^\mathrm{train}, \, \beta^\mathrm{train})=(2.1, \, 0.1)$, high degree heterogeneity and strong clustering, $(\gamma^\mathrm{train}, \, \beta^\mathrm{train})=(2.1, \, 4.1)$, less degree heterogeneity and weak clustering, $(\gamma^\mathrm{train}, \, \beta^\mathrm{train})=(3.9, \, 0.1)$, less degree heterogeneity and strong clustering, $(\gamma^\mathrm{train}, \, \beta^\mathrm{train})=(3.9, \, 4.1)$, and an intermediate setting, $(\gamma^\mathrm{train}, \, \beta^\mathrm{train})=(3.0, \, 1.1)$. For each of these settings we sample $n_G^\mathrm{train}=1$ training graph.

Our test graphs explore many further parameter values. We vary their size $n^\mathrm{test}\in\{64, 128, \dots, 8192\}$, degree heterogeneity $\gamma^\mathrm{test}\in\{2.1, 2.4, \dots, 3.9\}$, and inverse temperature $\beta^\mathrm{test}\in\{0.1, 0.6, \dots, 4.1\}$. We fix the average degree to $\bar{k}^\mathrm{test}=10$. We sample $n_G^\mathrm{test}=100$ test graphs for each parameter setting.

\begin{table}[tb]
    \centering
    \begin{tabular}{|c|c|c|}
    \hline
     & $\beta \in (0, 1)$ & $\beta \in (1, \infty)$ \\
    \hline
    $\gamma \in (2, 3]$      & $\langle k^2\rangle\to\infty$, $\bar{c}\to 0$ & $\langle k^2\rangle\to\infty$, $\bar{c}> 0$ \\
    \hline
    $\gamma \in (3, \infty)$ & $\langle k^2\rangle<\infty$, $\bar{c}\to 0$ & $\langle k^2\rangle<\infty$, $\bar{c}>0$ \\
    \hline
    \end{tabular}
    \caption{\textbf{Properties of graphs from the $\mathbb{S}^1$-model}. The degree heterogeneity, measured by the second moment $\langle k^2\rangle$ of the degree distribution, and the average clustering coefficient $\bar{c}$ of graphs sampled from the $\mathbb{S}^1$-model differ substantially depending on the inverse temperature $\beta$ and the exponent $\gamma$ of the degree distribution. We report their behavior in the thermodynamic limit, $n\to\infty$.}
    \label{tab:S1_properties}
\end{table}

\subsection{Evaluation strategies}\label{sec:methods:evaluation}

To evaluate the performance of \model{}s, we rely on four different strategies, including generalization across graph sizes, graph parameters, local stability analysis, and robustness to unobserved nodes at test time.

First, we study the ability of a \model~trained on a small graph with $n^\mathrm{train}=64$ nodes to generalize to larger graphs with $n^\mathrm{test}\in\{64, 128, \dots, 8192\}$ nodes with the same parameters as the training graph, i.e., that have the same values of average degree $\bar{k}^\mathrm{test}=\bar{k}^\mathrm{train}=10$, degree heterogeneity $\gamma^\mathrm{test}=\gamma^\mathrm{train}$, and inverse temperature $\beta^\mathrm{test}=\beta^\mathrm{train}$. We use $n_G^\mathrm{test}=100$ test graphs for a given setting of the $\mathbb{S}^1$-model parameters, and an independently sampled initial condition on each of them, $n_\mathrm{init}^\mathrm{test}=1$. We solve the resulting initial value problem with the same solver used to generate the training data (see Section~\ref{sec:methods:network_dynamics}). We store the solution at $n_\tau^\mathrm{test}=256$ discrete time points. We gather these time series in a tensor $X^\mathrm{test}\in\mathbb{R}^{n_G^\mathrm{test}\times n_\tau^\mathrm{test} \times n^\mathrm{test}}$. We make predictions using a \model{} trained on the same type of dynamical system used to generate the test data and from the same initial conditions used to generate the test data. We gather the resulting time series in a tensor, $\hat{X}^\mathrm{test}\in\mathbb{R}^{n_G^\mathrm{test}\times n_\tau^\mathrm{test}\times n^\mathrm{test}}$.

We use different notions of error to evaluate the predictions, each of which is defined as a summary of a node-wise error. The mean absolute error (MAE) at node $i$ in the $g$th test graph is, 
\begin{equation}\label{eq:mae}
    \mathcal{L}^{\mathrm{mae}}_{g, i}
    =
    \frac{1}{n^\mathrm{test}_\tau}
    \sum_{r = 1}^{n_\tau^\mathrm{test}}
    |X^\mathrm{test}_{g, r, i} - \hat{X}^\mathrm{test}_{g, r, i}|.
\end{equation}
Similarly, the mean relative error (MRE) at node $i$ in the $g$th test graph is, 
\begin{equation}\label{eq:mre}
    \mathcal{L}^{\mathrm{mre}}_{g, i}
    =
        \frac{1}{n_\tau^\mathrm{test}}
    \sum_{r = 1}^{n_\tau^\mathrm{test}}
    |X^\mathrm{test}_{g, r, i} - \hat{X}^\mathrm{test}_{g, r, i}|/|X^\mathrm{test}_{g, r, i}|, 
\end{equation}
where we restrict summation indices such that division by zero does not occur.
Finally, the root mean squared error (RMSE) at node $i$ in the $g$th graph is given by
\begin{equation}\label{eq:rmse}
    \mathcal{L}^{\mathrm{rmse}}_{g, i}
    =
    \left(
    \frac{1}{n_\tau^\mathrm{test}}
    \sum_{r = 1}^{n_\tau^\mathrm{test}}
    (X^\mathrm{test}_{g, r, i} - \hat{X}^\mathrm{test}_{g, r, i})^2
    \right)^{1/2}.
\end{equation}
We summarize each of these quantities $q\in\{\mathrm{mae}, \, \mathrm{mre}, \, \mathrm{rmse}\}$ either through their mean over test graphs and nodes
\begin{equation}\label{eq:mean-error}
    \bar{\mathcal{L}_q}
    =
    \frac{1}{n_G^\mathrm{test}n^\mathrm{test}}
    \sum_{g = 1}^{n_G^\mathrm{test}}
    \sum_{i = 1}^{n^\mathrm{test}}
    \mathcal{L}_{g, i}^{q}
\end{equation}
or similarly through quantiles, in particular their median $\tilde{\mathcal{L}}_q$ and interquartile range.

Our second evaluation strategy focuses on the ability of \model{}s trained on small graphs with $n^\mathrm{train}=64$ nodes to generalize to graphs with different degree heterogeneity, controlled by $\gamma^\mathrm{test}$, and average clustering, controlled by $\beta^\mathrm{test}$, that are either of the same size as the training graph, $n^\mathrm{test}=n^\mathrm{train}=64$, or substantially larger, $n^\mathrm{test}=8192>n^\mathrm{train}$. We keep the average degree fixed at $\bar{k}^\mathrm{test}=\bar{k}^\mathrm{train}=10$. We vary $\beta^\mathrm{test}\in\{0.1, 0.6, \dots, 4.1\}$ and $\gamma^\mathrm{test}\in\{2.1, 2.4, \dots, 3.9\}$. For each setting of the $\mathbb{S}^1$-model parameters, we compare \model{} predictions $\hat{X}^\mathrm{test}$ with the test data $X^\mathrm{test}$ on $n_G^\mathrm{test}=100$ test graphs each with $n_\mathrm{init}^\mathrm{test}=1$ independent initial condition using $n_\tau^\mathrm{test}=256$ time points. To this end, we use the mean of the MAE (Eq.~\eqref{eq:mae}) or median of the MRE (Eq.~\eqref{eq:mre}) normalized to their value on test graphs with the same parameters, $(\gamma^\mathrm{test}, \, \beta^\mathrm{test})=(\gamma^\mathrm{train}, \, \beta^\mathrm{train})$, and size, $n^\mathrm{test}=n^\mathrm{train}$, as the training graph. We denote these quantities as $\bar{\mathcal{L}}_\mathrm{mae}^\prime$ and $\tilde{\mathcal{L}}_\mathrm{mre}^\prime$. The former is defined as 
\begin{equation}
    \bar{\mathcal{L}}^\prime_q
    =
    \frac{\bar{\mathcal{L}}_q(n^\mathrm{test}, \bar{k}^\mathrm{test}, \gamma^\mathrm{test}, \beta^\mathrm{test})}
    {\bar{\mathcal{L}}_q(n^\mathrm{train}, \bar{k}^\mathrm{train}, \gamma^\mathrm{train}, \beta^\mathrm{train})}, 
\end{equation}
where we made the dependence of the error on the parameters of the graph explicit and emphasize that the denominator is the test error on graphs with the same parameters as the training graph, not a training error. The latter, $\tilde{\mathcal{L}}_\mathrm{mre}^\prime$, is defined analogously.

Third, we investigate the degree to which \model{}s trained to approximate dynamical systems on small graphs, $n^\mathrm{train}=64$, can approximate fixed points, $\mathbf{x}^\star$, and local stability properties of dynamical systems of the same type on equally sized or larger graphs, $n^\mathrm{test}\in\{64, 4096\}$, that have the same parameters as the training graph, $\bar{k}^\mathrm{test}=\bar{k}^\mathrm{train}=10$, $(\gamma^\mathrm{test}, \, \beta^\mathrm{test})=(\gamma^\mathrm{train}, \, \beta^\mathrm{train})$. As fixed-points, $\mathbf{x}^\star$, are defined by $F(\mathbf{x}^\star)=0$ for the ground truth dynamics and $F_\omega(\mathbf{\hat{x}}^\star)=0$ for the \model{}s, we employ the following strategy to identify them on all $n_G^\mathrm{test}=100$ test graphs. On the $g$th graph, we solve the ground truth system and the one defined by the \model{} from $n_\mathrm{init}^\mathrm{test}=1$ initial condition up to $t_1=10$ to obtain $\mathbf{x}_{g}(t=t_1)$ and $\hat{\mathbf{x}}_{g}(t=t_1)$ respectively. We then use these values to initialize a custom implementation of Newton's method with Armijo line search to find a root $\mathbf{x}_{g}^\star$ of $F(\mathbf{x})$ and $\hat{\mathbf{x}}_{g}^\star$ of $F_\omega(\mathbf{x})$. These roots are numerical approximations of the respective fixed points.

We compare these fixed points using the mean MAE, 
\begin{equation}\label{eq:mae_fixedpoint}
    \bar{\mathcal{L}}_\mathrm{mae}^\star
    =
    \frac{1}{n_G^\mathrm{test}n^\mathrm{test}}
    \sum_{g = 1}^{n_G^\mathrm{test}}
    \sum_{i = 1}^{n^\mathrm{test}}
    |x^{\star}_{g, i} - \hat{x}^\star_{g, i}|,
\end{equation}
and the analogously defined median MAE, $\tilde{\mathcal{L}}^\star_\mathrm{mae}$.

An important property of fixed points is their stability. A fixed point is stable if the largest eigenvalue $\lambda_1$ of the Jacobian $J$, with entries
\begin{equation}\label{eq:jacobian_entries}
    J_{ij}(\mathbf{x})
    =
    \frac{\partial F_i}{\partial x_j}(\mathbf{x})
\end{equation}
evaluated at the fixed point $\mathbf{x}^\star$, is negative. For dynamical systems of the BB form (Eq.~\eqref{eq:bb-dynamics-methods}) the entries of $J$ can be computed in closed form, 
\begin{widetext}
\begin{equation}\label{eq:jacobian}
    J_{ij}(\mathbf{x})
    =
    \frac{\partial F_i(\mathbf{x})}{\partial x_j}
    =
    \delta_{ij}
    \left(
    \frac{\partial f(x_i)}{\partial x_i}
    +
    \frac{\partial h^\mathrm{ego}(x_i)}{\partial x_i}
    \left(
    \sum_{l = 1}^{n}A_{il}
    h^\mathrm{alt}(x_l)
    \right)
    \right)
    +
    h^\mathrm{ego} (x_i) A_{ij}\frac{\partial h^\mathrm{alt}(x_j)}{\partial x_j}, 
\end{equation}
\end{widetext}
where $\delta_{ij}$ is the Kronecker-$\delta$. As our \model{}s also follow the BB form (Eq.~\eqref{eq:nODE}) their Jacobian $\hat{J}(\mathbf{x})$ has the same form as Eq.~\eqref{eq:jacobian}, but $f, \, h^\mathrm{ego}, \, h^\mathrm{alt}$ are replaced by neural networks $f_\omega, \, h^\mathrm{ego}_\omega, \, h^\mathrm{alt}_\omega$. To form the \model{}'s Jacobian we thus need to evaluate the partial derivatives of these neural networks with respect to their inputs. We use forward-mode automatic differentiation in \texttt{JAX} for this purpose. Finally, we compute the spectrum of $\hat{J}(\mathbf{x})$ at the fixed point $\hat{\mathbf{x}}^\star$ using $QR$-decomposition as implemented in \texttt{jax.numpy.linalg.eig} to obtain the largest eigenvalue $\hat{\lambda}_1$.

Our final evaluation strategy examines the robustness of \model{}s to missing data at the time of deployment. To this end, we use the test graphs of size $n^\mathrm{test}=8192$ that otherwise have the same parameters as the training graph $\bar{k}^\mathrm{test}=\bar{k}^\mathrm{train}$, $(\gamma^\mathrm{test}, \, \beta^\mathrm{test})=(\gamma^\mathrm{train}, \, \beta^\mathrm{train})$. However, when making predictions with \model{}s, we select a subset $\mathcal{V}^\mathrm{obs} \subset \mathcal{V}$ of ``observed nodes'' uniformly at random from all nodes $\mathcal{V}$ and present the \model{} only with the initial condition at these nodes and restrict the graph to the corresponding node-induced subgraph $G^\mathrm{obs}=(\mathcal{V}^\mathrm{obs}, \mathcal{E}\cap (\mathcal{V}^\mathrm{obs}\times \mathcal{V}^\mathrm{obs}))$ to make predictions $\hat{X}^\mathrm{test}$. We rely on the mean and median node-wise MAE (Eq.~\eqref{eq:mae}, Eq.~\eqref{eq:mean-error}) to compare the \model{}'s predictions at the observed nodes $\mathcal{V}^\mathrm{obs}$ to the numerical solution of the respective dynamical system $X^\mathrm{test}$, obtained using the entire graph. This means the sum over nodes in Eq.~\eqref{eq:mean-error} is restricted to the observed nodes $\mathcal{V}^\mathrm{obs}$. We vary the number of observed nodes $n^\mathrm{obs}=|\mathcal{V}^\mathrm{obs}|$ from $n^\mathrm{obs}=n^\mathrm{test}$ to $n^\mathrm{test}/4=2048$. For each value of $n^\mathrm{obs}$ on each of the $n_G^\mathrm{test}=100$ test graphs, we sample an independent initial condition $n_\mathrm{init}^\mathrm{test}=1$ and a new set of observed nodes $\mathcal{V}^\mathrm{obs}$.

\section{Acknowledgments}

We thank Melanie Weber for her feedback on an early version of this work. M.L.~and T.E.R.~are supported by the Inaugural Joseph E.~Aoun Endowment.

\bibliography{bib}

\clearpage
\appendix
\onecolumngrid
\setcounter{figure}{0}
\setcounter{table}{0}
\renewcommand{\thefigure}{A\arabic{figure}}
\renewcommand{\thetable}{A\arabic{table}}

\onecolumngrid
\phantomsection

\begin{center}
{\large\bfseries Supplementary Information for}\\[0.35em]
{\large\bfseries When do neural ordinary differential equations generalize on complex networks?}\\[0.75em]

{\normalsize
Moritz Laber,
Tina Eliassi-Rad,
Brennan Klein
}\\[0.35em]

{\small \texttt{\url{laber.m@northeastern.edu}}}
\end{center}

\section{Notation}\label{appendix:notation}

We summarize the mathematical notation in Tab.~\ref{tab:notation}.

\begin{table*}[b]
    \centering
    \begin{adjustbox}{max totalheight=0.75\textheight, keepaspectratio}
    \begin{tabular}{c|l}
    \textbf{Symbol}                                    &   \textbf{Definition} \\ \hline
     $G=(\mathcal{V}, \mathcal{E})$                    & a graph with node set $\mathcal{V}$ and edge set $\mathcal{E}$. \\
     $m=|\mathcal{E}|$                                 & the number of edges in a graph. \\
     $\kappa_i\in[0, \infty)$                           & the expected degree/hidden variable assigned to node $i\in\mathcal{V}$. \\
     $\theta_i\in[0, 2\pi]$                             & the angular coordinate assigned to node $i\in\mathcal{V}$. \\
     $\theta_{ij}$                                     & the distance of nodes $i$ and $j$ on the unit circle, $\mathbb{S}^1$. \\
     $p_{ij}$                                          & the probability that an edge exists between nodes $i$ and $j$. \\
     $\mu$                                             &  the $\log$-chemical potential of a graph. \\ 
     $n, \, n^\mathrm{train}, \, n^\mathrm{test}$          & the number of nodes in a graph in general, in the training set, in the test set. \\
     $\bar{k}, \, \bar{k}^\mathrm{train}, \, \bar{k}^\mathrm{test}$ & the average degree of a graph in general, in the training set, in the test set. \\
     $\gamma, \, \gamma^\mathrm{train}, \, \gamma^\mathrm{test}$    & the exponent of the degree distribution in general, in the training set, in the test set. \\
     $\beta$, $\beta^\mathrm{train}$, $\beta^\mathrm{test}$     & the inverse temperature of a graph in general, in the training set, in the test set. \\
     $A\in\mathbb{R}^{n\times n}$                               & a graph's adjacency matrix. \\
     $\mathcal{N}_i = \{j\in\mathcal{V}:A_{ij}=1\}$             & the set of neighbors of node $i$. \\
     $k_i$                                                      & the degree of node $i$. \\
     $\langle k^2\rangle$                                       & the second moment of the degree distribution. \\ 
     $c_i$                                                      & the clustering coefficient of node $i$. \\
     $\bar{c}$                                                  & the average local clustering coefficient of a graph. \\
     $\nu, \, \rho_1, \, \rho_2, \, \alpha_1, \, \alpha_2$      & scalar parameters appearing in the definition of different dynamical systems. \\
     $t$                                                        & continuous time parameter. \\
     $\tau_r$                                                   & single selected time points $\tau_r$ at which the time series are observed. \\
     $n_\tau, \, n_\tau^\mathrm{train}, \, n_\tau^\mathrm{test}$    & the number of discrete time points in general, for training, for testing. \\
     $n_G^\mathrm{train}\, , n_G^\mathrm{test}$                   & the number of graphs in the training and test set. \\
     $n_\mathrm{init}^\mathrm{train}, \, n_\mathrm{init}^\mathrm{test}$                     & the number of initial conditions in the training and test set. \\
     $\mathbf{x}(t)=(x_1(t), \dots, x_n(t))^\mathsf{T}\in\mathbb{R}^{n}$                    & the ground truth vector of node states. \\
     $\mathbf{\hat{x}}(t)=(\hat{x}_1(t), \dots, \hat{x}_n(t))^\mathsf{T}\in\mathbb{R}^{n}$  & the vector of node states predicted by the \model{}. \\
     $X^\mathrm{train}\in\mathbb{R}^{n_\mathrm{init}^\mathrm{train}\times n_\tau^\mathrm{train}\times n^\mathrm{train}}$    & the training data. \\
     $X^\mathrm{test}\in\mathbb{R}^{n_G^\mathrm{test}\times n_\tau^\mathrm{test}\times n^\mathrm{test}}$                    & the test data. \\
     $\hat{X}^\mathrm{train}, \, \hat{X}^\mathrm{test}$                                     & the predictions on the training and test data. \\
     $\mathbf{x}^\star = (x_1^\star, \dots, x_n^\star)^\mathsf{T}$                          & the fixed point of the ground truth dynamical system. \\
     $\mathbf{\hat{x}}^\star=(\hat{x}_1^\star, \dots, \hat{x}_n^\star)^\mathsf{T}$          & the fixed point of the \model{}. \\
     $\omega$                                                         & the parameters of the \model{} architecture. \\
     $F, \, F_\omega$                                                   & the vector field of the ground truth dynamical system and the \model{}. \\
     $f, \, f_\omega$                         & the self-dynamics of the ground truth dynamical system and the \model{}. \\
     $h^\mathrm{ego}, \, h^\mathrm{ego}_\omega$                         & the ego node contribution to the ground truth dynamical system and the \model{}. \\
     $h^\mathrm{alt}, \, h^\mathrm{alt}_\omega$                         & the alter node contribution to the ground truth dynamical system and the \model{}. \\
     $\mathcal{V}^\mathrm{obs}, \, n^\mathrm{obs}=|\mathcal{V}^\mathrm{obs}|$     & the observed nodes and their number in the fourth evaluation strategy. \\
     $\mathcal{L}^\mathrm{mae}_{g, i}, \, \mathcal{L}^\mathrm{mre}_{g, i}, \, \mathcal{L}^\mathrm{rmse}_{g, i}$ & the node-wise MAE, MRE, RMSE of node $i$ on the $g$th graph. \\
     $\bar{\mathcal{L}}_\mathrm{mae}, \, \bar{\mathcal{L}}_\mathrm{mre}, \, \bar{\mathcal{L}}_\mathrm{rmse}$ & the mean node-wise MAE, MRE, RMSE. \\
     $\tilde{\mathcal{L}}_\mathrm{mae}, \, \tilde{\mathcal{L}}_\mathrm{mre}, \, \tilde{\mathcal{L}}_\mathrm{rmse}$ & the median node-wise MAE, MRE, RMSE. \\
     $\bar{\mathcal{L}}_\mathrm{mae}^\prime, \, \bar{\mathcal{L}}_\mathrm{mre}^\prime, \, \bar{\mathcal{L}}_\mathrm{rmse}^\prime$ & the normalized mean node-wise MAE, MRE, RMSE. \\
     $\tilde{\mathcal{L}}_\mathrm{mae}^\prime, \, \tilde{\mathcal{L}}_\mathrm{mre}^\prime, \, \tilde{\mathcal{L}}_\mathrm{rmse}^\prime$ & the normalized median node-wise MAE, MRE, RMSE. \\
     $\bar{\mathcal{L}}_\mathrm{mae}^\star$                            & the MAE of the fixed point approximation. \\
     $\mathcal{L}$                                                     & the loss function during training. \\
     $J, \, \hat{J}$                                                       & the Jacobian of the ground truth dynamical system and the \model{}. \\
     $\lambda_1, \, \hat{\lambda}_1$                 & the largest eigenvalue of the ground truth Jacobian and the \model{}'s Jacobian. \\
     $\langle \dots \rangle_{\mathcal{N}_i}$                               & average over the neighborhood of node $i$. \\
     $\langle \dots \rangle_\mathcal{N}$                                   & graph wide average of $\langle \dots\rangle_{\mathcal{N}_i}$. \\
     $\epsilon_\mathrm{abs}$, $\epsilon_\mathrm{rel}$, $\Delta t_\mathrm{init}$         & ODE solver hyperparameters: absolute tolerance, relative tolerance, initial step size. \\
     $n_b$, $\eta$, $g_0$, $\varepsilon$, $S$                              & optimizer hyperparameters: step size, learning rate scale, initial accumulator, division offset, num. steps. \\
     $\sigma_\mathrm{init}$, $d_\mathrm{in}$                               & weight initialization hyperparameters: variance of the initialization, fan-in of a neuron.
    \end{tabular}
    \end{adjustbox}
    \caption{\textbf{Notation.} A summary of the mathematical notation we use.}
    \label{tab:notation}
\end{table*}

\section{Scaling of fixed point activity with node degree}\label{appendix:scaling}

Here, we derive the scaling of node $i$'s fixed point activity $x_i^\star$ with its degree $k_i=\sum_{j=1}^n A_{ij}$, where $A$ is the adjacency matrix. We use the notation $\langle x\rangle _{\mathcal{N}_i} = \frac{1}{k_i}\sum_{j =1}^n A_{ij}x_j$ to denote the average over neighbors. We denote $x_\mathcal{N}$ the average of this quantity over the graph, $\langle x\rangle_\mathcal{N}=\frac{1}{n}\sum_{i} \langle x\rangle_{\mathcal{N}_i}$. We make a mean-field-like approximation, popularized by Meena et al.~\cite{meena2023_emergentstabilitycomplex}, 
\begin{equation}\label{eq:mean-field-approx}
    \langle x \rangle_{\mathcal{N}_i}\approx x_\mathcal{N}, 
\end{equation}
for all nodes $i$, and with $x_\mathcal{N}$ independent of $i$. Below, we use that $F(\mathbf{x})=0$ implies that $F_i(\mathbf{x})=0$ for all $i\in\{1, \dots, n\}$. All systems under consideration permit a trivial fixed point $\mathbf{x}^\star=0$. We will not consider this fixed point in our analyses, as its scaling with degree is trivial.

First, consider the SIS model. In this case, using the fixed point conditions and Eq.~\eqref{eq:mean-field-approx} yields

\begin{equation}
\begin{split}
    0 &= -\rho_1 x_i^\star +\rho_2 (1-x_i^\star)\sum_{j=1}^{n}A_{ij}x_j^\star \\
    0 &= -\rho_1 x_i^\star + \rho_2(1-x_i^\star) k_i \langle x^\star\rangle_{\mathcal{N}}\\
    x_i^\star &= \left(1 + \frac{\rho_1}{\rho_2 k_i \langle x^\star\rangle_\mathcal{N}}\right)^{-1}. 
\end{split}
\end{equation}

Note, that as $k_i\to\infty$, we have $x_i^\star\to 1$, i.e., the fixed point activity remains bounded.

The analogous calculation for the MAK model yields, 

\begin{equation}
\begin{split}
    0 &=\nu - \rho_1 x_i^\star -\rho_2 x_i^\star \sum_{j=1}^{n}A_{ij}x_j^\star\\
    0 &=\nu - (\rho_1 +\rho_2 k_i\langle x^\star\rangle_\mathcal{N})x_i^\star\\
    x_i^\star &=\frac{\nu}{\rho_1 + \rho_2 k_i\langle x^\star\rangle_\mathcal{N}}.
\end{split}
\end{equation}

In the case of the MAK model, large degree nodes hold a smaller share of the overall activity, and in the limit $k_i\to\infty$ one has $x_i^\star\to 0$.

In the case of the MM model, we obtain

\begin{equation}
\begin{split}
    0 &= -\rho_1 x_i^{\star\alpha_1} + \rho_2 \sum_{j=1}^{n} \frac{x_j^{\star \alpha_2}}{ 1+x_j^{\star\alpha_2}}\\
    x_i^\star &= \left(\frac{\rho_2}{\rho_1}k_i\left\langle \frac{x^{\star\alpha_2}}{1+x^{\star \alpha_2}}\right\rangle_\mathcal{N}\right)^{1/\alpha_1}. 
\end{split}
\end{equation}

This yields a scaling of the fixed point activity with degree as $k_i^{1/\alpha_1}$, which scales like $\sqrt{k_i}$ for the setting in Tab.~\ref{tab:dynamics}.

For the ND model, we have

\begin{equation}
\begin{split}
    0 & = \nu\tanh(x_i^\star)-\rho_1 x_i^\star+\rho_2\sum_{j=1}^{n}A_{ij}\tanh(x_j^\star)\\
    x_i^\star &= \frac{\nu}{\rho_1}\tanh(x_i^\star) + \frac{\rho_2}{\rho_1}k_i\langle\tanh(x)\rangle_\mathcal{N}.
\end{split}
\end{equation}

Noting that $\tanh(x)\to1$ as $x_i\to\infty$, the leading order contribution to the activity is linear $k_i$.

Finally, turning our attention to the BD model, we find

\begin{equation}
\begin{split}
    0 &= -\rho_1  x_i^{\star\alpha_1}  + \rho_2 \sum_{j=1}^{n} A_{ij} x_{j}^{\star\alpha_2}\\
    x_i^{\star}&= \left(\frac{\rho_2}{\rho_1}k_i\langle x^{\star \alpha_2}\rangle_\mathcal{N}\right)^{1/\alpha_1}.
\end{split}
\end{equation}

In the BD model, a node's fixed point activity scales like $k_i^{1/\alpha_1}$ and thus with the settings of Tab.~\ref{tab:dynamics} like $\sqrt{k_i}$.


\FloatBarrier
\section{Additional graph-level results on size generalization}\label{appendix:size_generalization}

\subsection{Results on noise-free training data}

Here, we show additional results on the ability of \model{}s to generalize across graph sizes. All \model{}s were trained on small graphs, $n^\mathrm{train}=64$, and are evaluated on larger graphs of size $n^\mathrm{test}$ generated with the same parameters as the training graph, $(\gamma^\mathrm{test}, \, \beta^\mathrm{test})=(\gamma^\mathrm{train}, \, \beta^\mathrm{train})=(\gamma, \, \beta)$. Results are based on $n_G^\mathrm{test}=100$ test graphs.

\begin{figure*}[tb]
    \centering
    \includegraphics[width=\textwidth]{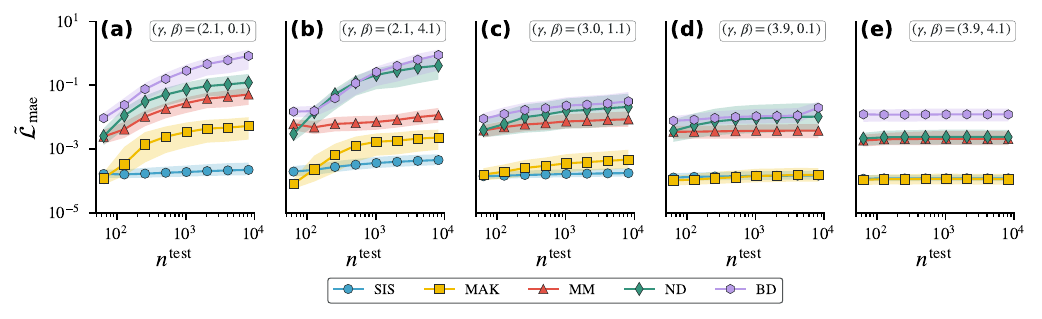}
    \caption{\textbf{Graph-level perspective on size generalization through the median MAE.} The ability of \model{}s trained to approximate different dynamical systems (SIS (blue circles), MAK (yellow squares), MM (red triangles), ND (green diamonds), BD (purple hexagons)) on small graphs, $n^\mathrm{train}=64$, to make accurate predictions on larger graphs, $n^\mathrm{test}$, with the same properties as the training graph, $(\gamma^\mathrm{test},\,\beta^\mathrm{test})=(\gamma^\mathrm{train}, \, \beta^\mathrm{train})=(\gamma, \, \beta)$, differs between \textbf{(a)} very degree heterogeneous graphs with weak clustering, $(\gamma, \, \beta)=(2.1, \, 0.1)$, \textbf{(b)} very degree heterogeneous graphs with strong clustering, $(\gamma, \, \beta)=(2.1, \, 4.1)$, \textbf{(c)} graphs with moderate degree heterogeneity and clustering, $(\gamma, \, \beta)=(3.0, \, 1.1)$, \textbf{(d)} less degree heterogeneous graphs with weak clustering $(\gamma, \, \beta)=(3.9, \, 0.1)$, and \textbf{(e)} less degree heterogeneous graphs with strong clustering, $(\gamma, \, \beta)=(3.9, \, 4.1)$. The median node-wise MAE, $\tilde{\mathcal{L}}_\mathrm{mae}$, over $n_G^\mathrm{test}=100$ test graphs, stays constant or increases slowly on less degree heterogeneous graphs independent of clustering but increases noticeably on more degree heterogeneous graphs for most \model{}s. Those trained on the SIS model show the smallest increase in median MAE. This means degree heterogeneity is a limiting factor for size generalization of \model{}s predicting dynamical systems on graphs.}
\label{fig:size_generalization_median_mae_noiseless}
\end{figure*}

Figure~\ref{fig:size_generalization_median_mae_noiseless} shows the median and interquartile range for the mean node-wise MAE. The results for the median mirror those for the mean presented in the main text Section~\ref{sec:results:size_generalization_global} with a strong increase in MAE for the ND (green diamonds) and BD (purple hexagons) model on very degree heterogeneous graphs (Fig.~\ref{fig:size_generalization_median_mae_noiseless}\textbf{(a)}, \textbf{(b)}), lower MAE in the intermediate case (Fig.~\ref{fig:size_generalization_median_mae_noiseless}\textbf{(c)}), and low and barely increasing MAE in the less degree heterogeneous case (Fig.~\ref{fig:size_generalization_median_mae_noiseless}\textbf{(d)}, \textbf{(e)}). The uncertainty tends to decrease with degree heterogeneity.

\begin{figure*}[tb]
    \centering
    \includegraphics[width=\textwidth]{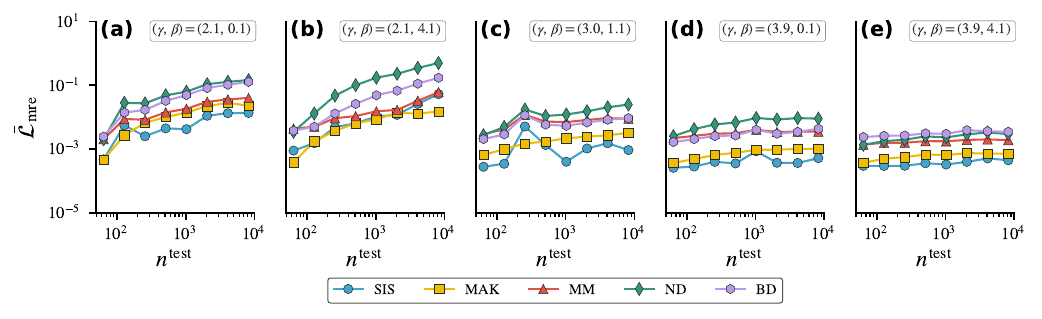}
    \caption{\textbf{Graph-level perspective on size generalization through the mean MRE.} The ability of \model{}s trained to approximate different dynamical systems (SIS (blue circles), MAK (yellow squares), MM (red triangles), ND (green diamonds), BD (purple hexagons)) on small graphs, $n^\mathrm{train}=64$, to make accurate predictions on larger graphs, $n^\mathrm{test}$, with the same properties as the training graph, $(\gamma^\mathrm{test},\,\beta^\mathrm{test})=(\gamma^\mathrm{train}, \, \beta^\mathrm{train})=(\gamma, \, \beta)$, differs between \textbf{(a)} very degree heterogeneous graphs with weak clustering, $(\gamma, \, \beta)=(2.1, \, 0.1)$, \textbf{(b)} very degree heterogeneous graphs with strong clustering, $(\gamma, \, \beta)=(2.1, \, 4.1)$, \textbf{(c)} graphs with moderate degree heterogeneity and clustering, $(\gamma, \, \beta)=(3.0, \, 1.1)$, \textbf{(d)} less degree heterogeneous graphs with weak clustering $(\gamma, \, \beta)=(3.9, \, 0.1)$, and \textbf{(e)} less degree heterogeneous graphs with strong clustering, $(\gamma, \, \beta)=(3.9, \, 4.1)$. The mean node-wise MRE, $\bar{\mathcal{L}}_\mathrm{mre}$, over $n_G^\mathrm{test}=100$ test graphs, stays constant or increases slowly on less degree heterogeneous graphs independent of clustering but increases noticeably on more degree heterogeneous graphs for most \model{}s. Those trained on the SIS model show the smallest increase in mean MRE. This means degree heterogeneity is a limiting factor for size generalization of \model{}s predicting dynamical systems on graphs.}
    \label{fig:size_generalization_mean_mre_noiseless}
\end{figure*}

Figure~\ref{fig:size_generalization_mean_mre_noiseless}, which shows the mean of the node-wise MRE, exhibits overall similar trends as the node-wise mean MAE. However, on very degree heterogeneous graphs (Fig.~\ref{fig:size_generalization_mean_mre_noiseless}\textbf{(a)}, \textbf{(b)}) the mean MRE also shows an increase in the error of \model{} trained on the SIS model. However, their error is still small overall. The results in the other cases---intermediate (Fig.~\ref{fig:size_generalization_mean_mre_noiseless}\textbf{(c)}) and low (Fig.~\ref{fig:size_generalization_mean_mre_noiseless}\textbf{(d)}, \textbf{(e)}) degree heterogeneity---mirror those of the MAE more closely. Overall, the mean MRE is a more noisy measure than the MAE, probably due to fluctuations of the small node states in the denominator.

\begin{figure*}[tb]
    \centering
    \includegraphics[width=\textwidth]{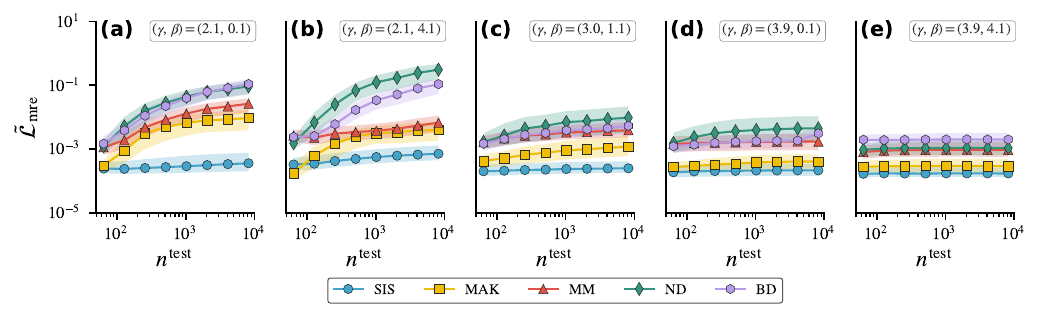}
    \caption{\textbf{Graph-level perspective on size generalization through the median MRE.} The ability of \model{}s trained to approximate different dynamical systems (SIS (blue circles), MAK (yellow squares), MM (red triangles), ND (green diamonds), BD (purple hexagons)) on small graphs, $n^\mathrm{train}=64$, to make accurate predictions on larger graphs, $n^\mathrm{test}$, with the same properties as the training graph, $(\gamma^\mathrm{test},\,\beta^\mathrm{test})=(\gamma^\mathrm{train}, \, \beta^\mathrm{train})=(\gamma, \, \beta)$, differs between \textbf{(a)} very degree heterogeneous graphs with weak clustering, $(\gamma, \, \beta)=(2.1, \, 0.1)$, \textbf{(b)} very degree heterogeneous graphs with strong clustering, $(\gamma, \, \beta)=(2.1, \, 4.1)$, \textbf{(c)} graphs with moderate degree heterogeneity and clustering, $(\gamma, \, \beta)=(3.0, \, 1.1)$, \textbf{(d)} less degree heterogeneous graphs with weak clustering $(\gamma, \, \beta)=(3.9, \, 0.1)$, and \textbf{(e)} less degree heterogeneous graphs with strong clustering, $(\gamma, \, \beta)=(3.9, \, 4.1)$. The median node-wise MRE, $\tilde{\mathcal{L}}_\mathrm{mre}$, over $n_G^\mathrm{test}=100$ test graphs, stays constant or increases slowly on less degree heterogeneous graphs independent of clustering but increases noticeably on more degree heterogeneous graphs for most \model{}s. Those trained on the SIS model show the smallest increase in median MRE. This means degree heterogeneity is a limiting factor for size generalization of \model{}s predicting dynamical systems on graphs.}
    \label{fig:size_generalization_median_mre_noiseless}
\end{figure*}

The median of the MRE (Fig.~\ref{fig:size_generalization_median_mre_noiseless}) is more stable. Moreover, the increase of the error for the SIS model in the degree heterogeneous case (Fig.~\ref{fig:size_generalization_mean_mre_noiseless}\textbf{(a)}, \textbf{(b)}) is again lower, suggesting that the increase in the mean MRE is driven by outliers. In the intermediate (Fig.~\ref{fig:size_generalization_mean_mre_noiseless}\textbf{(c)}) and degree homogeneous case (Fig.~\ref{fig:size_generalization_mean_mre_noiseless}\textbf{(d)}, \textbf{(e)}) the increase in error is small or absent, as for the MAE.

\begin{figure*}[tb]
    \centering
    \includegraphics[width=\textwidth]{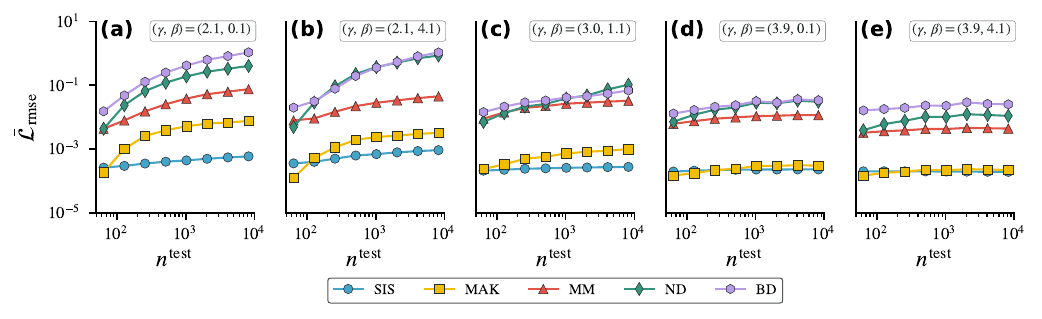}
    \caption{\textbf{Graph-level perspective on size generalization through the mean RMSE.} The ability of \model{}s trained to approximate different dynamical systems (SIS (blue circles), MAK (yellow squares), MM (red triangles), ND (green diamonds), BD (purple hexagons)) on small graphs, $n^\mathrm{train}=64$, to make accurate predictions on larger graphs, $n^\mathrm{test}$, with the same properties as the training graph, $(\gamma^\mathrm{test},\,\beta^\mathrm{test})=(\gamma^\mathrm{train}, \, \beta^\mathrm{train})=(\gamma, \, \beta)$, differs between \textbf{(a)} very degree heterogeneous graphs with weak clustering, $(\gamma, \, \beta)=(2.1, \, 0.1)$, \textbf{(b)} very degree heterogeneous graphs with strong clustering, $(\gamma, \, \beta)=(2.1, \, 4.1)$, \textbf{(c)} graphs with moderate degree heterogeneity and clustering, $(\gamma, \, \beta)=(3.0, \, 1.1)$, \textbf{(d)} less degree heterogeneous graphs with weak clustering $(\gamma, \, \beta)=(3.9, \, 0.1)$, and \textbf{(e)} less degree heterogeneous graphs with strong clustering, $(\gamma, \, \beta)=(3.9, \, 4.1)$. The mean node-wise RMSE, $\bar{\mathcal{L}}_\mathrm{rmse}$, over $n_G^\mathrm{test}=100$ test graphs, stays constant or increases slowly on less degree heterogeneous graphs independent of clustering but increases noticeably on more degree heterogeneous graphs for most \model{}s. Those trained on the SIS model show the smallest increase in mean RMSE. This means degree heterogeneity is a limiting factor for size generalization of \model{}s predicting dynamical systems on graphs.}
    \label{fig:size_generalization_mean_rmse_noiseless}
\end{figure*}

Figure~\ref{fig:size_generalization_mean_rmse_noiseless} shows the mean node-wise RMSE. It increases substantially in the very degree heterogeneous case for \model{}s trained on most of the dynamical systems (Fig.~\ref{fig:size_generalization_mean_rmse_noiseless}\textbf{(a)}, \textbf{(b)}), less so in the intermediate case (Fig.~\ref{fig:size_generalization_mean_rmse_noiseless}\textbf{(c)}), and barely in the less degree heterogeneous case (Fig.~\ref{fig:size_generalization_mean_rmse_noiseless}\textbf{(d)}, \textbf{(e)}).

\begin{figure*}[tb]
    \centering
    \includegraphics[width=\textwidth]{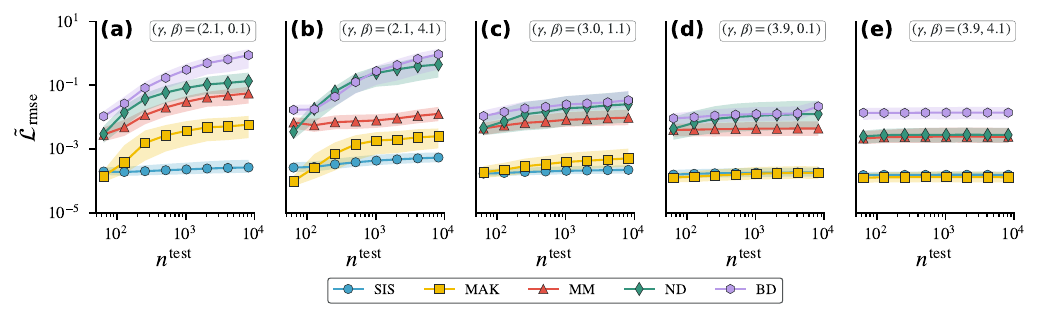}
    \caption{\textbf{Graph-level perspective on size generalization through the median RMSE.} The ability of \model{}s trained to approximate different dynamical systems (SIS (blue circles), MAK (yellow squares), MM (red triangles), ND (green diamonds), BD (purple hexagons)) on small graphs, $n^\mathrm{train}=64$, to make accurate predictions on larger graphs, $n^\mathrm{test}$, with the same properties as the training graph, $(\gamma^\mathrm{test},\,\beta^\mathrm{test})=(\gamma^\mathrm{train}, \, \beta^\mathrm{train})=(\gamma, \, \beta)$, differs between \textbf{(a)} very degree heterogeneous graphs with weak clustering, $(\gamma, \, \beta)=(2.1, \, 0.1)$, \textbf{(b)} very degree heterogeneous graphs with strong clustering, $(\gamma, \, \beta)=(2.1, \, 4.1)$, \textbf{(c)} graphs with moderate degree heterogeneity and clustering, $(\gamma, \, \beta)=(3.0, \, 1.1)$, \textbf{(d)} less degree heterogeneous graphs with weak clustering $(\gamma, \, \beta)=(3.9, \, 0.1)$, and \textbf{(e)} less degree heterogeneous graphs with strong clustering, $(\gamma, \, \beta)=(3.9, \, 4.1)$. The median node-wise RMSE, $\tilde{\mathcal{L}}_\mathrm{rmse}$, over $n_G^\mathrm{test}=100$ test graphs, stays constant or increases slowly on less degree heterogeneous graphs independent of clustering but increases noticeably on more degree heterogeneous graphs for most \model{}s. Those trained on the SIS model show the smallest increase in median RMSE. This means degree heterogeneity is a limiting factor for size generalization of \model{}s predicting dynamical systems on graphs.}
    \label{fig:size_generalization_median_rmse_noiseless}
\end{figure*}

The median and interquartile range of the node-wise RMSE shown in Fig.~\ref{fig:size_generalization_median_rmse_noiseless} paint a similar picture across graphs with high degree heterogeneity (Fig.~\ref{fig:size_generalization_median_rmse_noiseless}\textbf{(a)}), \textbf{(b)}), intermediate degree heterogeneity (Fig.~\ref{fig:size_generalization_median_rmse_noiseless}\textbf{(c)}), and less degree heterogeneity (Fig.~\ref{fig:size_generalization_median_rmse_noiseless}\textbf{(d)}, \textbf{(e)}).
%
Together these findings corroborate that degree heterogeneity is central for driving performance degradation when generalizing across graphs of different size. They also show that the results presented in the main text are not sensitive to which error metric is used, nor how we aggregate across nodes and graphs. 


\clearpage
\FloatBarrier
\subsection{Results on noisy training data}\label{appendix:subsec:size_generalization_noiseless}

Here, we demonstrate that our findings on size generalization are robust to noise in the training data. We consider \model{}s trained on small graphs, $n^\mathrm{train}=64$, and with independent Gaussian noise of zero mean and standard deviation listed in Tab.~\ref{tab:dynamics} added to the training data. We evaluate them on larger graphs generated with the same parameters as the training graph, $(\gamma^\mathrm{test}, \, \beta^\mathrm{test})=(\gamma^\mathrm{train}, \, \beta^\mathrm{train})=(\gamma, \, \beta)$. Results are based on $n_G^\mathrm{test}=100$ test graphs.

\begin{figure*}[tb]
    \centering
    \includegraphics[width=\textwidth]{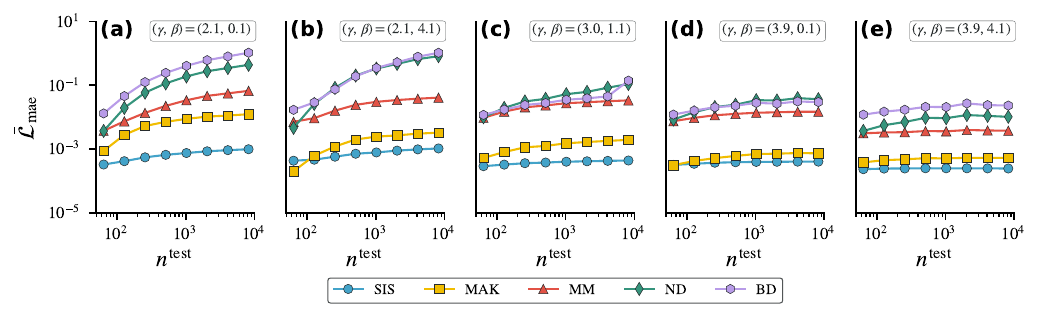}
    \caption{\textbf{Robustness of size generalization to noise through the lens of mean MAE.} The ability of \model{}s trained to approximate different dynamical systems (SIS (blue circles), MAK (yellow squares), MM (red triangles), ND (green diamonds), BD (purple hexagons)) on small graphs, $n^\mathrm{train}=64$, to make accurate predictions on larger graphs, $n^\mathrm{test}$, with the same properties as the training graph, $(\gamma^\mathrm{test},\,\beta^\mathrm{test})=(\gamma^\mathrm{train}, \, \beta^\mathrm{train})=(\gamma, \, \beta)$, differs between \textbf{(a)} very degree heterogeneous graphs with weak clustering, $(\gamma, \, \beta)=(2.1, \, 0.1)$, \textbf{(b)} very degree heterogeneous graphs with strong clustering, $(\gamma, \, \beta)=(2.1, \, 4.1)$, \textbf{(c)} graphs with moderate degree heterogeneity and clustering, $(\gamma, \, \beta)=(3.0, \, 1.1)$, \textbf{(d)} less degree heterogeneous graphs with weak clustering $(\gamma, \, \beta)=(3.9, \, 0.1)$, and \textbf{(e)} less degree heterogeneous graphs with strong clustering, $(\gamma, \, \beta)=(3.9, \, 4.1)$. The mean node-wise MAE, $\bar{\mathcal{L}}_\mathrm{mae}$, over $n_G^\mathrm{test}=100$ test graphs, stays constant or increases slowly on less degree heterogeneous graphs independent of clustering but increases noticeably on more degree heterogeneous graphs for most \model{}s. Those trained on the SIS model show the smallest increase in mean MAE. This means degree heterogeneity is a limiting factor for size generalization of \model{}s predicting dynamical systems on graphs.}
    \label{fig:size_generalization_mean_mae_noisy}
\end{figure*}

\begin{figure*}[tb]
    \centering
    \includegraphics[width=\textwidth]{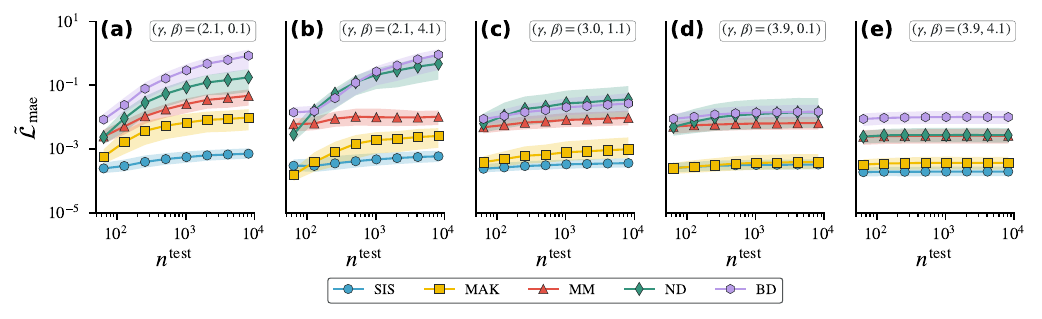}
    \caption{\textbf{Robustness of size generalization to noise through the lens of median MAE.} The ability of \model{}s trained to approximate different dynamical systems (SIS (blue circles), MAK (yellow squares), MM (red triangles), ND (green diamonds), BD (purple hexagons)) on small graphs, $n^\mathrm{train}=64$, to make accurate predictions on larger graphs, $n^\mathrm{test}$, with the same properties as the training graph, $(\gamma^\mathrm{test},\,\beta^\mathrm{test})=(\gamma^\mathrm{train}, \, \beta^\mathrm{train})=(\gamma, \, \beta)$, differs between \textbf{(a)} very degree heterogeneous graphs with weak clustering, $(\gamma, \, \beta)=(2.1, \, 0.1)$, \textbf{(b)} very degree heterogeneous graphs with strong clustering, $(\gamma, \, \beta)=(2.1, \, 4.1)$, \textbf{(c)} graphs with moderate degree heterogeneity and clustering, $(\gamma, \, \beta)=(3.0, \, 1.1)$, \textbf{(d)} less degree heterogeneous graphs with weak clustering $(\gamma, \, \beta)=(3.9, \, 0.1)$, and \textbf{(e)} less degree heterogeneous graphs with strong clustering, $(\gamma, \, \beta)=(3.9, \, 4.1)$. The median node-wise MAE, $\tilde{\mathcal{L}}_\mathrm{mae}$, over $n_G^\mathrm{test}=100$ test graphs, stays constant or increases slowly on less degree heterogeneous graphs independent of clustering but increases noticeably on more degree heterogeneous graphs for most \model{}s. Those trained on the SIS model show the smallest increase in median MAE. This means degree heterogeneity is a limiting factor for size generalization of \model{}s predicting dynamical systems on graphs.}
    \label{fig:size_generalization_median_mae_noisy}
\end{figure*}

The mean of the node-wise MAE (Fig.~\ref{fig:size_generalization_mean_mae_noisy}) as well as the median and interquartile range (Fig.~\ref{fig:size_generalization_median_mae_noisy}) follow similar patterns as in the noiseless case. Only in the very degree heterogeneous case (Fig.~\ref{fig:size_generalization_mean_mae_noisy}\textbf{(a)}, (Fig.~\ref{fig:size_generalization_median_mae_noisy}\textbf{(a)}) do we observe a slightly stronger increase in MAE for \model{}s trained on the SIS model. For all other cases (Fig.~\ref{fig:size_generalization_mean_mae_noisy}\textbf{(b)}, \textbf{(c)}, \textbf{(d)}, \textbf{(e)}, Fig.~\ref{fig:size_generalization_median_mae_noisy}\textbf{(b)}, \textbf{(c)}, \textbf{(d)}, \textbf{(e)}), we observe no major differences from the noiseless case.

\begin{figure*}[tb]
    \centering
    \includegraphics[width=\textwidth]{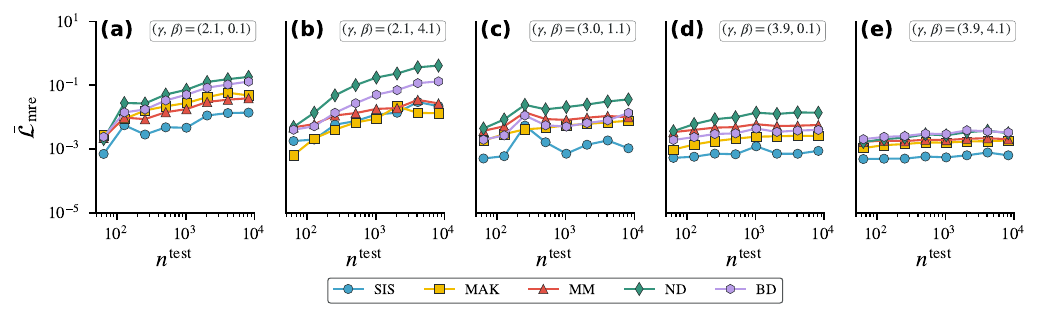}
    \caption{\textbf{Robustness of size generalization to noise through the lens of mean MRE.} The ability of \model{}s trained to approximate different dynamical systems (SIS (blue circles), MAK (yellow squares), MM (red triangles), ND (green diamonds), BD (purple hexagons)) on small graphs, $n^\mathrm{train}=64$, to make accurate predictions on larger graphs, $n^\mathrm{test}$, with the same properties as the training graph, $(\gamma^\mathrm{test},\,\beta^\mathrm{test})=(\gamma^\mathrm{train}, \, \beta^\mathrm{train})=(\gamma, \, \beta)$, differs between \textbf{(a)} very degree heterogeneous graphs with weak clustering, $(\gamma, \, \beta)=(2.1, \, 0.1)$, \textbf{(b)} very degree heterogeneous graphs with strong clustering, $(\gamma, \, \beta)=(2.1, \, 4.1)$, \textbf{(c)} graphs with moderate degree heterogeneity and clustering, $(\gamma, \, \beta)=(3.0, \, 1.1)$, \textbf{(d)} less degree heterogeneous graphs with weak clustering $(\gamma, \, \beta)=(3.9, \, 0.1)$, and \textbf{(e)} less degree heterogeneous graphs with strong clustering, $(\gamma, \, \beta)=(3.9, \, 4.1)$. The mean node-wise MRE, $\bar{\mathcal{L}}_\mathrm{mre}$, over $n_G^\mathrm{test}=100$ test graphs, stays constant or increases slowly on less degree heterogeneous graphs independent of clustering but increases noticeably on more degree heterogeneous graphs for most \model{}s. Those trained on the SIS model show the smallest increase in mean MRE. This means degree heterogeneity is a limiting factor for size generalization of \model{}s predicting dynamical systems on graphs.}
    \label{fig:size_generalization_mean_mre_noisy}
\end{figure*}

\begin{figure*}[tb]
    \centering
    \includegraphics[width=\textwidth]{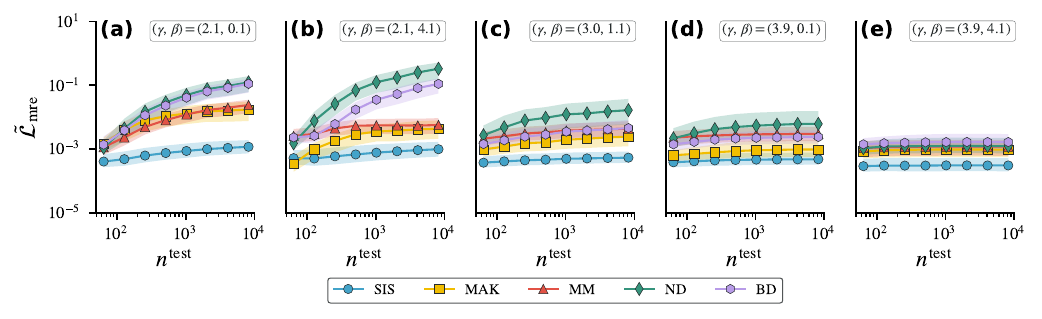}
    \caption{\textbf{Robustness of size generalization to noise through the lens of median MRE.}  The ability of \model{}s trained to approximate different dynamical systems (SIS (blue circles), MAK (yellow squares), MM (red triangles), ND (green diamonds), BD (purple hexagons)) on small graphs, $n^\mathrm{train}=64$, to make accurate predictions on larger graphs, $n^\mathrm{test}$, with the same properties as the training graph, $(\gamma^\mathrm{test},\,\beta^\mathrm{test})=(\gamma^\mathrm{train}, \, \beta^\mathrm{train})=(\gamma, \, \beta)$, differs between \textbf{(a)} very degree heterogeneous graphs with weak clustering, $(\gamma, \, \beta)=(2.1, \, 0.1)$, \textbf{(b)} very degree heterogeneous graphs with strong clustering, $(\gamma, \, \beta)=(2.1, \, 4.1)$, \textbf{(c)} graphs with moderate degree heterogeneity and clustering, $(\gamma, \, \beta)=(3.0, \, 1.1)$, \textbf{(d)} less degree heterogeneous graphs with weak clustering $(\gamma, \, \beta)=(3.9, \, 0.1)$, and \textbf{(e)} less degree heterogeneous graphs with strong clustering, $(\gamma, \, \beta)=(3.9, \, 4.1)$. The median node-wise MRE, $\tilde{\mathcal{L}}_\mathrm{mre}$, over $n_G^\mathrm{test}=100$ test graphs, stays constant or increases slowly on less degree heterogeneous graphs independent of clustering but increases noticeably on more degree heterogeneous graphs for most \model{}s. Those trained on the SIS model show the smallest increase in median MRE. This means degree heterogeneity is a limiting factor for size generalization of \model{}s predicting dynamical systems on graphs.}
    \label{fig:size_generalization_median_mre_noisy}
\end{figure*}

Comparing the mean (Fig.~\ref{fig:size_generalization_mean_mre_noisy}) as well as the median (Fig.~\ref{fig:size_generalization_median_mre_noisy}) and interquartile range of the mean node-wise MRE to their noiseless counterpart (Fig.~\ref{fig:size_generalization_mean_mre_noiseless}, Fig.~\ref{fig:size_generalization_median_mre_noiseless}) we do not observe any major differences across the very degree heterogeneous (Fig.~\ref{fig:size_generalization_mean_mre_noisy}\textbf{(a)}, \textbf{(b)}, Fig.~\ref{fig:size_generalization_median_mre_noisy}\textbf{(a)}, \textbf{(b)}), intermediate (Fig.~\ref{fig:size_generalization_mean_mre_noisy}\textbf{(c)}, Fig.~\ref{fig:size_generalization_median_mre_noisy}\textbf{(c)}), and less degree heterogeneous graphs (Fig.~\ref{fig:size_generalization_mean_mre_noisy}\textbf{(d)}, \textbf{(e)}, Fig.~\ref{fig:size_generalization_median_mre_noisy}\textbf{(d)}, \textbf{(e)}).

\begin{figure*}[tb]
    \centering
    \includegraphics[width=\textwidth]{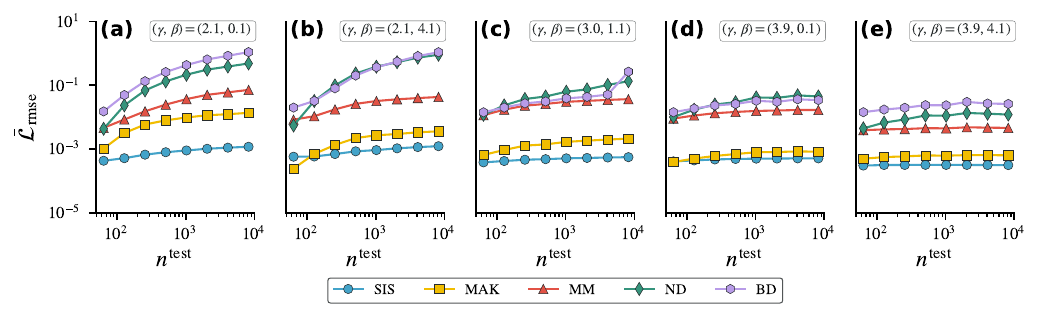}
    \caption{\textbf{Robustness of size generalization to noise through the lens of mean RMSE.} The ability of \model{}s trained to approximate different dynamical systems (SIS (blue circles), MAK (yellow squares), MM (red triangles), ND (green diamonds), BD (purple hexagons)) on small graphs, $n^\mathrm{train}=64$, to make accurate predictions on larger graphs, $n^\mathrm{test}$, with the same properties as the training graph, $(\gamma^\mathrm{test},\,\beta^\mathrm{test})=(\gamma^\mathrm{train}, \, \beta^\mathrm{train})=(\gamma, \, \beta)$, differs between \textbf{(a)} very degree heterogeneous graphs with weak clustering, $(\gamma, \, \beta)=(2.1, \, 0.1)$, \textbf{(b)} very degree heterogeneous graphs with strong clustering, $(\gamma, \, \beta)=(2.1, \, 4.1)$, \textbf{(c)} graphs with moderate degree heterogeneity and clustering, $(\gamma, \, \beta)=(3.0, \, 1.1)$, \textbf{(d)} less degree heterogeneous graphs with weak clustering $(\gamma, \, \beta)=(3.9, \, 0.1)$, and \textbf{(e)} less degree heterogeneous graphs with strong clustering, $(\gamma, \, \beta)=(3.9, \, 4.1)$. The mean node-wise RMSE, $\bar{\mathcal{L}}_\mathrm{rmse}$, over $n_G^\mathrm{test}=100$ test graphs, stays constant or increases slowly on less degree heterogeneous graphs independent of clustering but increases noticeably on more degree heterogeneous graphs for most \model{}s. Those trained on the SIS model show the smallest increase in mean RMSE. This means degree heterogeneity is a limiting factor for size generalization of \model{}s predicting dynamical systems on graphs.}
    \label{fig:size_generalization_mean_rmse_noisy}
\end{figure*}

\begin{figure*}[tb]
    \centering
    \includegraphics[width=\textwidth]{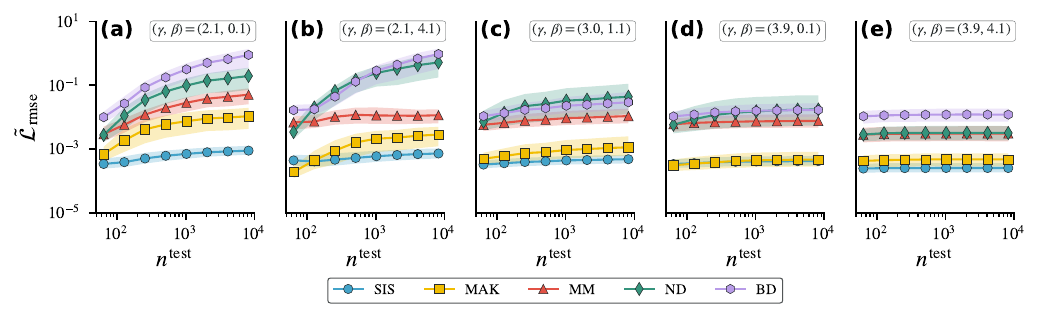}
    \caption{\textbf{Robustness of size generalization to noise through the lens of median RMSE.} The ability of \model{}s trained to approximate different dynamical systems (SIS (blue circles), MAK (yellow squares), MM (red triangles), ND (green diamonds), BD (purple hexagons)) on small graphs, $n^\mathrm{train}=64$, to make accurate predictions on larger graphs, $n^\mathrm{test}$, with the same properties as the training graph, $(\gamma^\mathrm{test},\,\beta^\mathrm{test})=(\gamma^\mathrm{train}, \, \beta^\mathrm{train})=(\gamma, \, \beta)$, differs between \textbf{(a)} very degree heterogeneous graphs with weak clustering, $(\gamma, \, \beta)=(2.1, \, 0.1)$, \textbf{(b)} very degree heterogeneous graphs with strong clustering, $(\gamma, \, \beta)=(2.1, \, 4.1)$, \textbf{(c)} graphs with moderate degree heterogeneity and clustering, $(\gamma, \, \beta)=(3.0, \, 1.1)$, \textbf{(d)} less degree heterogeneous graphs with weak clustering $(\gamma, \, \beta)=(3.9, \, 0.1)$, and \textbf{(e)} less degree heterogeneous graphs with strong clustering, $(\gamma, \, \beta)=(3.9, \, 4.1)$. The median node-wise RMSE, $\tilde{\mathcal{L}}_\mathrm{rmse}$, over $n_G^\mathrm{test}=100$ test graphs, stays constant or increases slowly on less degree heterogeneous graphs independent of clustering but increases noticeably on more degree heterogeneous graphs for most \model{}s. Those trained on the SIS model show the smallest increase in median RMSE. This means degree heterogeneity is a limiting factor for size generalization of \model{}s predicting dynamical systems on graphs.}
    \label{fig:size_generalization_median_rmse_noisy}
\end{figure*}

Turning finally to the mean (Fig.~\ref{fig:size_generalization_mean_rmse_noisy}) and median (Fig.~\ref{fig:size_generalization_median_rmse_noisy}) of the node-wise RMSE, we find that they exhibit very similar behavior across values of the $\mathbb{S}^1$-model parameters, i.e., on degree heterogeneous graphs with weak clustering (Fig.~\ref{fig:size_generalization_mean_rmse_noisy}\textbf{(a)}, Fig.~\ref{fig:size_generalization_median_rmse_noisy}\textbf{(a)}), degree heterogeneous graphs with strong clustering
(Fig.~\ref{fig:size_generalization_mean_rmse_noisy}\textbf{(b)}, Fig.~\ref{fig:size_generalization_median_rmse_noisy}\textbf{(b)}), graphs with intermediate degree heterogeneity and clustering (Fig.~\ref{fig:size_generalization_mean_rmse_noisy}\textbf{(c)}, Fig.~\ref{fig:size_generalization_median_rmse_noisy}\textbf{(c)}), less degree heterogeneous graphs with weak clustering (Fig.~\ref{fig:size_generalization_mean_rmse_noisy}\textbf{(d)}, Fig.~\ref{fig:size_generalization_median_rmse_noisy}\textbf{(d)}), and less degree heterogeneous graphs with strong clustering (Fig.~\ref{fig:size_generalization_mean_rmse_noisy}\textbf{(e)}, Fig.~\ref{fig:size_generalization_median_rmse_noisy}\textbf{(e)}), as in the case of noise-free training data. Only, in the case of the mean node-wise RMSE on graphs with intermediate degree heterogeneity and clustering (Fig.~\ref{fig:size_generalization_mean_rmse_noisy}\textbf{(c)}) do we notice an uptick in error on the largest graphs $n^\mathrm{test}=8192$.

Overall, these results confirm that our findings are robust to noisy training data---an important consideration when employing \model{}s on real-world data.

\clearpage
\FloatBarrier
\section{Node-level results on size generalization}\label{appendix:size_generalization_local}

In this appendix, we provide a more fine-grained perspective on size generalization to understand which nodes contribute to the degradation of performance with graph size. We describe our approach in Appendix~\ref{appendix:subsec:size_generalization_local_methods}, discuss results for noiseless training data in Appendix~\ref{appendix:subsec:size_generalization_local_noiseless}, and robustness to noise in Appendix~\ref{appendix:subsec:size_generalization_local_noisy}.

\subsection{Methods}\label{appendix:subsec:size_generalization_local_methods}
To supplement global summary statistics with a more fine-grained perspective, we compute averages for nodes within a certain range of degree $k$ or clustering coefficient $c$. Let $k_{g, i}$ be the degree of node $i$ in the $g$th test graph, and $(k^{(1)}, \dots, k^{(B+1)})$ a sequence of potentially non-uniformly spaced bin edges. We define the average error $\bar{\mathcal{L}}_q(k)$ for $q\in\{\mathrm{mae}, \, \mathrm{mre}, \, \mathrm{rmse}\}$ for nodes of degree $k\in[k_b, k_{b+1})$ in the $b$th bin as
\begin{equation}
    \bar{\mathcal{L}}_q(k)
    =
    \frac{1}{C}
    \sum_{g = 1}^{n_G^\mathrm{test}}
    \sum_{i = 1}^{n^\mathrm{test}}
    \mathcal{L}_{g, i}^{q}\, \mathbf{1}_{\{k_{b}\leq k_{g, i}<k_{b+1}\}}, 
\end{equation}
where $\mathbf{1}_{\{\cdot\}}$ is an indicator function and the normalization $C$ is given by the number of nodes in the bin
\begin{equation}
    C
    =
    \sum_{g = 1}^{n_G^\mathrm{test}}
    \sum_{i = 1}^{n^\mathrm{test}}
    \mathbf{1}_{\{k_{b} \leq k_{g, i} < k_{b+1}\}}.
\end{equation}
The average error for nodes of a given clustering coefficient, $\bar{\mathcal{L}}_q(c)$ can be defined mutatis mutandis.

\subsection{Results on noise-free training data}\label{appendix:subsec:size_generalization_local_noiseless}

\begin{figure*}[tb]
    \centering
    \includegraphics[width=\textwidth]{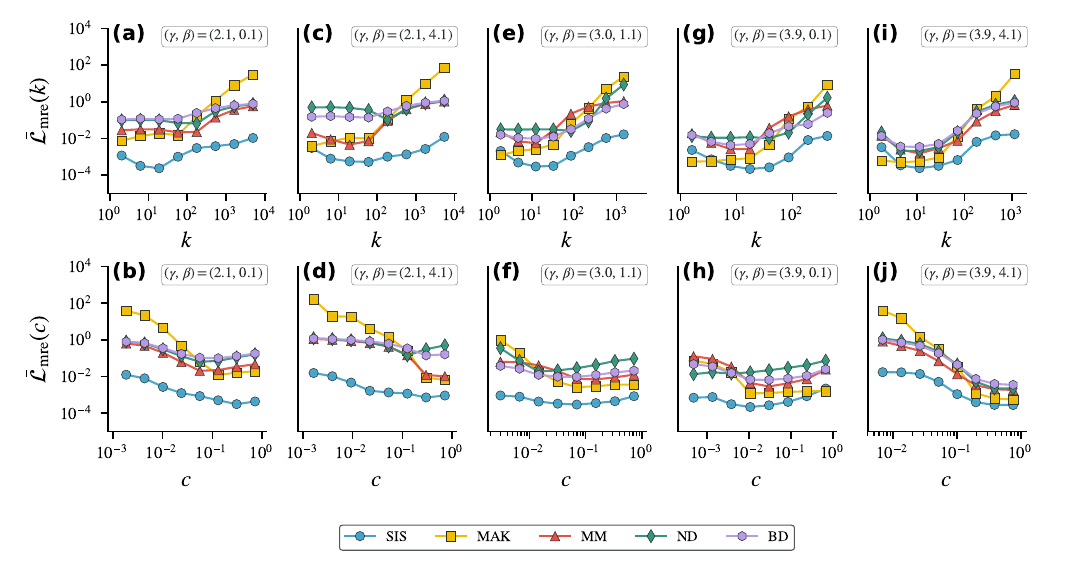}
    \caption{\textbf{Node-level perspective on size generalization through MRE.} The mean node-wise MRE $\bar{\mathcal{L}}_\mathrm{mre}(\cdot)$ on $n_G^\mathrm{test}=100$ large test graphs, $n^\mathrm{test}=8192$, can vary widely with node degree $k$ (upper row) or clustering $c$ (lower row) for \model{}s trained to approximate different dynamical systems (SIS (blue circles), MAK (yellow squares), MM (red triangles), ND (green diamonds), BD (purple hexagons)) on small graphs, $n^\mathrm{train}=64$, with the same parameters as the training graph, $(\gamma^\mathrm{train},\,\beta^\mathrm{train})=(\gamma^\mathrm{test},\,\beta^\mathrm{test})=(\gamma,\,\beta)$. The exact form of this dependence changes between \textbf{(a)}, \textbf{(b)} very degree heterogeneous graphs with weak clustering ($(\gamma, \, \beta)=(2.1, \, 0.1)$), \textbf{(c)}, \textbf{(d)} very degree heterogeneous graphs with strong clustering ($(\gamma, \, \beta)=(2.1, \, 4.1)$), \textbf{(e)}, \textbf{(f)} graphs with moderate degree heterogeneity and clustering ($(\gamma, \, \beta)=(3.0, \, 1.1)$), \textbf{(g)}, \textbf{(h)} less degree heterogeneous graphs with weak clustering ($(\gamma, \, \beta)=(3.9, \, 0.1)$), and \textbf{(i)}, \textbf{(j)} less degree heterogeneous graphs with strong clustering ($(\gamma, \, \beta)=(3.9, \, 4.1)$). However, overall the mean of the node-wise MRE increases as a function of the degree $k$, and either stays constant or decreases as a function of the clustering coefficient $c$. Hubs, nodes of high degree and low clustering, are thus of special importance for shaping the performance of \model{}s.}
    \label{fig:node_wise_mean_mre_noiseless}
\end{figure*}

\begin{figure*}[tb]
    \centering    \includegraphics[width=\textwidth]{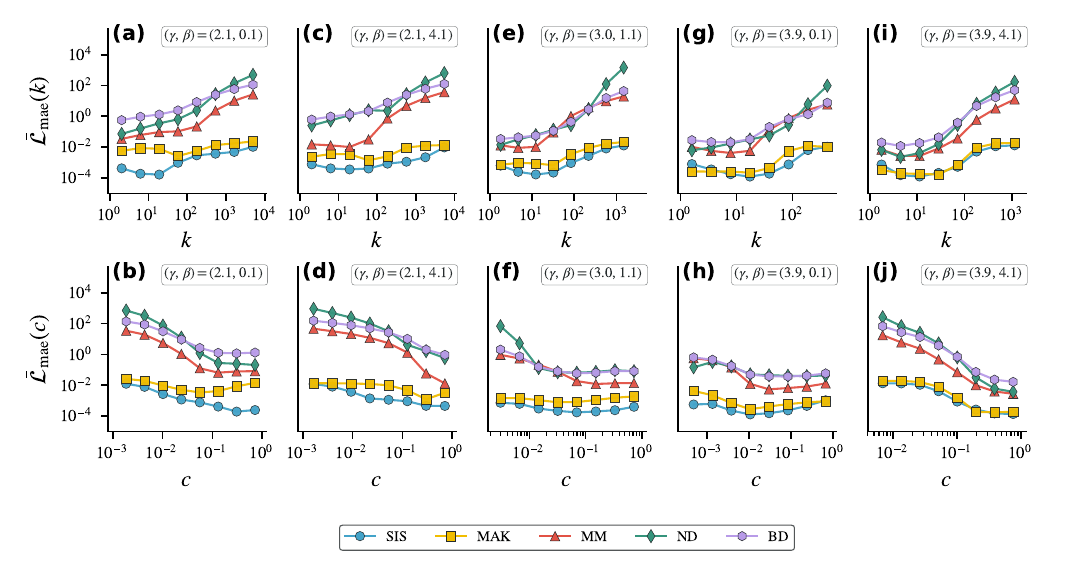}
    \caption{\textbf{Node-level perspective on size generalization through MAE.} The mean node-wise MAE $\bar{\mathcal{L}}_\mathrm{mae}(\cdot)$ on $n_G^\mathrm{test}=100$ large test graphs, $n^\mathrm{test}=8192$, can vary widely with node degree $k$ (upper row) or clustering $c$ (lower row) for \model{}s trained to approximate different dynamical systems (SIS (blue circles), MAK (yellow squares), MM (red triangles), ND (green diamonds), BD (purple hexagons)) on small graphs, $n^\mathrm{train}=64$, with the same parameters as the training graph, $(\gamma^\mathrm{train},\,\beta^\mathrm{train})=(\gamma^\mathrm{test},\,\beta^\mathrm{test})=(\gamma,\,\beta)$. The exact form of this dependence changes between \textbf{(a)}, \textbf{(b)} very degree heterogeneous graphs with weak clustering ($(\gamma, \, \beta)=(2.1, \, 0.1)$), \textbf{(c)}, \textbf{(d)} very degree heterogeneous graphs with strong clustering ($(\gamma, \, \beta)=(2.1, \, 4.1)$), \textbf{(e)}, \textbf{(f)} graphs with moderate degree heterogeneity and clustering ($(\gamma, \, \beta)=(3.0, \, 1.1)$), \textbf{(g)}, \textbf{(h)} less degree heterogeneous graphs with weak clustering ($(\gamma, \, \beta)=(3.9, \, 0.1)$), and \textbf{(i)}, \textbf{(j)} less degree heterogeneous graphs with strong clustering ($(\gamma, \, \beta)=(3.9, \, 4.1)$) but overall the mean of the node-wise MAE increases as a function of the degree $k$, and either stays constant or decreases as a function of the clustering coefficient $c$. Hubs, nodes of high degree and low clustering, are thus of special importance for shaping the performance of \model{}s.}
    \label{fig:node_wise_mean_mae_noiseless}
\end{figure*}

\begin{figure*}[tb]
    \centering
    \includegraphics[width=\textwidth]{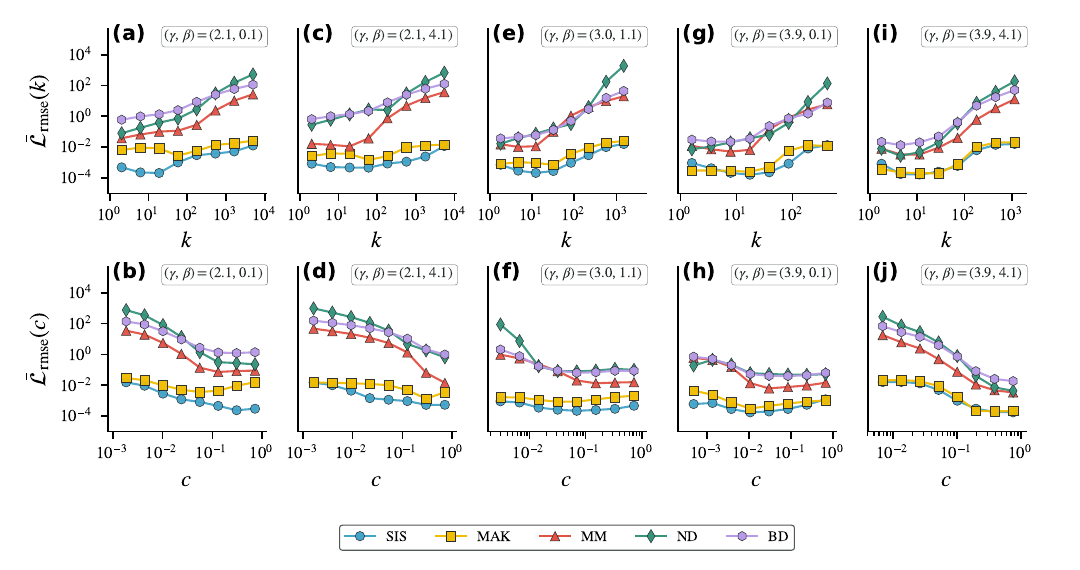}
    \caption{\textbf{Node-level perspective on size generalization through RMSE.} The mean node-wise RMSE $\bar{\mathcal{L}}_\mathrm{rmse}(\cdot)$ on $n_G^\mathrm{test}=100$ large test graphs, $n^\mathrm{test}=8192$, can vary widely with node degree $k$ (upper row) or clustering $c$ (lower row) for \model{}s trained to approximate different dynamical systems (SIS (blue circles), MAK (yellow squares), MM (red triangles), ND (green diamonds), BD (purple hexagons)) on small graphs, $n^\mathrm{train}=64$, with the same parameters as the training graph, $(\gamma^\mathrm{train},\,\beta^\mathrm{train})=(\gamma^\mathrm{test},\,\beta^\mathrm{test})=(\gamma,\,\beta)$. The exact form of this dependence changes between \textbf{(a)}, \textbf{(b)} very degree heterogeneous graphs with weak clustering ($(\gamma, \, \beta)=(2.1, \, 0.1)$), \textbf{(c)}, \textbf{(d)} very degree heterogeneous graphs with strong clustering ($(\gamma, \, \beta)=(2.1, \, 4.1)$), \textbf{(e)}, \textbf{(f)} graphs with moderate degree heterogeneity and clustering ($(\gamma, \, \beta)=(3.0, \, 1.1)$), \textbf{(g)}, \textbf{(h)} less degree heterogeneous graphs with weak clustering ($(\gamma, \, \beta)=(3.9, \, 0.1)$), and \textbf{(i)}, \textbf{(j)} less degree heterogeneous graphs with strong clustering ($(\gamma, \, \beta)=(3.9, \, 4.1)$) but overall the mean of the node-wise RMSE increases as a function of the degree $k$, and either stays constant or decreases as a function of the clustering coefficient $c$. Hubs, nodes of high degree and low clustering, are thus of special importance for shaping the performance of \model{}s.}
    \label{fig:node_wise_mean_rmse_noiseless}
\end{figure*}

Results were obtained on test graphs with $n^\mathrm{test}=8192$ nodes, much larger than the training graphs, $n^\mathrm{train}=64$, but with the same parameters $(\gamma^\mathrm{test}, \, \beta^\mathrm{test})=(\gamma^\mathrm{train}, \, \beta^\mathrm{train})=(\gamma, \, \beta)$. In Fig.~\ref{fig:node_wise_mean_mre_noiseless}, we display the node-wise mean relative error (MRE) among nodes of degree $k$ or among nodes of clustering coefficient $c$. We focus on MRE as the magnitude of node states can vary widely with degree (see Appendix~\ref{appendix:scaling}). We will comment on mean absolute error (MAE) and root mean squared error (RMSE) in Fig.~\ref{fig:node_wise_mean_mae_noiseless} and Fig.~\ref{fig:node_wise_mean_rmse_noiseless} respectively.

On very degree heterogeneous graphs with weak clustering ($(\gamma, \, \beta)=(2.1, \, 0.1)$, Fig.~\ref{fig:node_wise_mean_mre_noiseless}\textbf{(a)}) \model{}s trained on different dynamical systems behave differently as degree varies. The \model{} trained to approximate the SIS model (blue circles) shows overall lowest mean MRE, and exhibits a notable dip in MRE at nodes of degree approximately equal to the average degree of the training and test graphs, $k=\bar{k}^\mathrm{train}=\bar{k}^\mathrm{test}=10$. In the case of the MAK model (yellow squares) the MRE stays almost constant up to $k\approx10^2$ and then increases monotonically with $k$ and upon reaching the highest degree values exhibits the largest mean MRE of all \model{}s. The \model{} trained on the MM (red triangles), ND (green diamonds), and BD (purple hexagons) behave similarly to the one trained on the MM model, but their mean MRE at low degree nodes is higher while their MRE at high degree nodes is lower. The ranking in terms of MRE among the \model{}s for different dynamical systems is similar when considering clustering (Fig.~\ref{fig:node_wise_mean_mre_noiseless}\textbf{(b)}). However, MRE first decreases with increasing clustering and then stays constant or exhibits a mild increase.

For degree heterogeneous graphs with strong clustering ($(\gamma, \, \beta)=(2.1, \, 4.1)$, Fig~\ref{fig:node_wise_mean_mre_noiseless}\textbf{(c)}), \model{}s trained on the SIS, MAK, MM, and BD models behave similarly to the case of weak clustering. The ND model, on the other hand, shows a noticeable increase in MRE at low degree nodes. The dependence of MRE on the clustering (Fig~\ref{fig:node_wise_mean_mre_noiseless}\textbf{(d)}) is overall similar to the weakly clustered case.

In the case of intermediate degree heterogeneity and clustering ($(\gamma, \, \beta)=(3.0, \, 1.1)$, Fig~\ref{fig:node_wise_mean_mre_noiseless}\textbf{(e)}) the dependence of MRE on degree bears similarity to the case of very degree heterogeneous graphs with weak clustering for all \model{}s, but in general, MRE at high degree nodes is lower. Across dynamical changes with clustering (Fig~\ref{fig:node_wise_mean_mre_noiseless}\textbf{(f)}) are overall less pronounced than the changes with degree.

For less degree heterogeneous graphs with weak clustering ($(\gamma, \, \beta)=(3.9, \, 0.1)$, Fig~\ref{fig:node_wise_mean_mre_noiseless}\textbf{(g)}) all \model{}s show lower MRE at low degree nodes than in the more degree heterogeneous cases. MRE is higher at high degree nodes than at low degree nodes but passes through a minimum for \model{}s trained on the SIS, MM, or BD model. The performance of \model{}s trained on the SIS and ND model vary least with clustering (Fig~\ref{fig:node_wise_mean_mre_noiseless}\textbf{(h)}) with the latter exhibiting a slight increase towards higher values of clustering. The \model{}s trained on the MAK, MM, and BD models show an increase towards nodes of lower clustering.

Considering the dependence of the MRE on degree in less degree heterogeneous graphs with strong clustering ($(\gamma, \, \beta)=(3.9, \, 4.1)$, Fig~\ref{fig:node_wise_mean_mre_noiseless}\textbf{(i)}) , we observe a noticeable minimum around the average degree $\bar{k}=10$ for all models but the one trained on the MAK model. We observe a monotonic decrease in MRE with increasing clustering (Fig~\ref{fig:node_wise_mean_mre_noiseless}\textbf{(j)}) across dynamical systems.

MAE (Fig.~\ref{fig:node_wise_mean_mae_noiseless}) and RMSE (Fig.~\ref{fig:node_wise_mean_rmse_noiseless}) behave similar to each other, but show some differences to MRE (Fig.~\ref{fig:node_wise_mean_mre_noiseless}). Nevertheless, the global pattern of increasing error with degree and decrease or minor variation with clustering is preserved across error metrics.
%
On very degree heterogeneous graphs with weak clustering (Fig.~\ref{fig:node_wise_mean_mae_noiseless}\textbf{(a)}, Fig.~\ref{fig:node_wise_mean_rmse_noiseless}\textbf{(a)}) the error increases with degree $k$ for \model{}s trained on the ND, BD, and MM model, but shows a smaller increase for \model{}s trained on the MAK and SIS model. We observe a decrease of error with clustering for all \model{}s but the one trained on the MAK model, whose error varies least with clustering (Fig.~\ref{fig:node_wise_mean_mae_noiseless}\textbf{(b)}, Fig.~\ref{fig:node_wise_mean_rmse_noiseless}\textbf{(b)}).

Considering very degree heterogeneous graphs with strong clustering (Fig.~\ref{fig:node_wise_mean_mae_noiseless}\textbf{(c)}, \textbf{(d)}, Fig.~\ref{fig:node_wise_mean_rmse_noiseless}\textbf{(c)}, \textbf{(d)}), we make similar observations. 

When degree heterogeneity and clustering are moderate still increases with degree
(Fig.~\ref{fig:node_wise_mean_mae_noiseless}\textbf{(e)}, Fig.~\ref{fig:node_wise_mean_rmse_noiseless}\textbf{(e)}), however the \model{}s trained on the ND and BD model show lower error on low degree nodes than in the very degree heterogeneous case. The dependence on clustering is weaker across dynamical systems (Fig.~\ref{fig:node_wise_mean_mae_noiseless}\textbf{(f)}, Fig.~\ref{fig:node_wise_mean_rmse_noiseless}\textbf{(f)}).

In the less degree heterogeneous case with weak clustering, we observe a plateau of only weakly changing error for low to moderate degree nodes across dynamical systems followed by an increase in error for the \model{}s trained on the MM, ND, and BD models (Fig.~\ref{fig:node_wise_mean_mae_noiseless}\textbf{(g)}, Fig.~\ref{fig:node_wise_mean_rmse_noiseless}\textbf{(g)}). There is less variation in error with clustering than with degree (Fig.~\ref{fig:node_wise_mean_mae_noiseless}\textbf{(h)}, Fig.~\ref{fig:node_wise_mean_rmse_noiseless}\textbf{(h)}).

Finally, on less degree heterogeneous graphs with strong clustering the dependence of error on degree (Fig.~\ref{fig:node_wise_mean_mae_noiseless}\textbf{(i)}, Fig.~\ref{fig:node_wise_mean_rmse_noiseless}\textbf{(i)}) is similar to the case of weak clustering. However, there is a more pronounced decrease of error with clustering across dynamical systems (Fig.~\ref{fig:node_wise_mean_mae_noiseless}\textbf{(j)}, Fig.~\ref{fig:node_wise_mean_rmse_noiseless}\textbf{(j)}). 

Examining size generalization on the node level paints a more nuanced picture than the graph-level analysis. Even if one observes low test error in graph-level metrics on larger graphs, the predictive accuracy for individual nodes can vary widely. Compatible with the findings in Section~\ref{sec:results:size_generalization_global}, the states of hubs, nodes of high degree and low clustering, are hardest to predict. This could be the case because their node state $x_i$ grows beyond the parts of phase space observed during training, as hypothesized at the end of Section~\ref{sec:results:size_generalization_global}. Different error metrics highlight different aspects of this phenomenon. The MAE and RMSE highlight contributions to the bulk of the systems activity, while MRE emphasizes that hubs are hard to predict even when accounting for overall activity.

\FloatBarrier
\subsection{Results on noisy training data}\label{appendix:subsec:size_generalization_local_noisy}

To gauge the robustness of our findings in Appendix~\ref{appendix:subsec:size_generalization_local_noiseless} to noise in the training data, we consider again models trained on test graphs of small size, $n^\mathrm{train}=64$, and deployed on much larger graphs, $n^\mathrm{test}=8192$, with the same parameters, $(\gamma^\mathrm{test}, \, \beta^\mathrm{test})=(\gamma^\mathrm{train}, \, \beta^\mathrm{train})=(\gamma, \, \beta)$, as the training graphs. However, for these models, the training data was corrupted by independent Gaussian noise with standard deviation reported in Tab.~\ref{tab:dynamics}.

\begin{figure*}[tb]
    \centering
    \includegraphics[width=\textwidth]{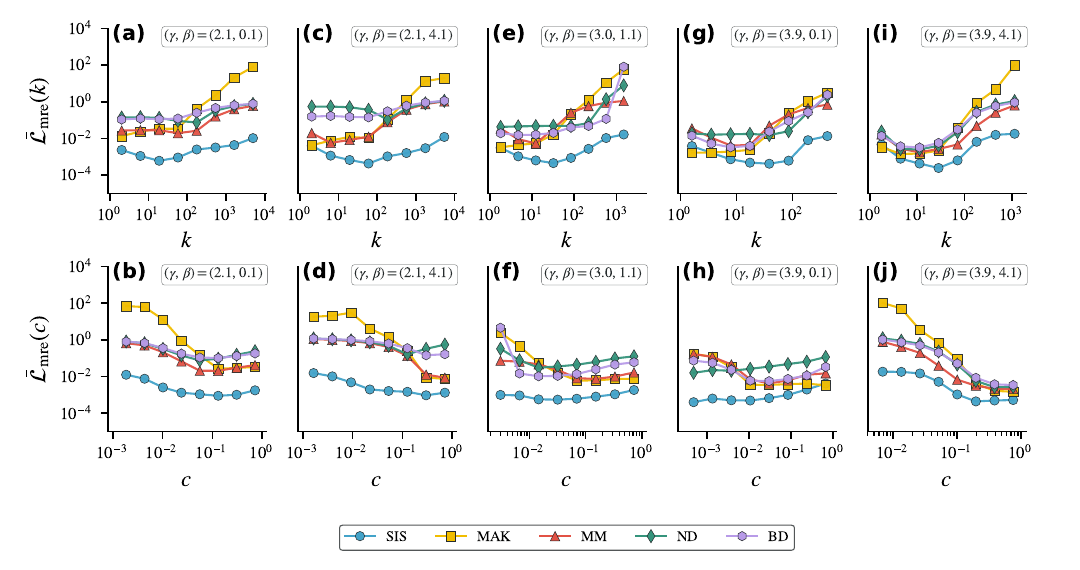}
    \caption{\textbf{Robustness of the node-level perspective on size generalization through MRE to noise in the training data}. The mean node-wise MAE $\bar{\mathcal{L}}_\mathrm{mre}(\cdot)$ on $n_G^\mathrm{test}=100$ large test graphs, $n^\mathrm{test}=8192$, can vary widely with node degree $k$ (upper row) or clustering $c$ (lower row) for \model{}s trained to approximate different dynamical systems (SIS (blue circles), MAK (yellow squares), MM (red triangles), ND (green diamonds), BD (purple hexagons)) on small graphs, $n^\mathrm{train}=64$, with the same parameters as the training graph, $(\gamma^\mathrm{train},\,\beta^\mathrm{train})=(\gamma^\mathrm{test},\,\beta^\mathrm{test})=(\gamma,\,\beta)$. The exact form of this dependence changes between \textbf{(a)}, \textbf{(b)} very degree heterogeneous graphs with weak clustering ($(\gamma, \, \beta)=(2.1, \, 0.1)$), \textbf{(c)}, \textbf{(d)} very degree heterogeneous graphs with strong clustering ($(\gamma, \, \beta)=(2.1, \, 4.1)$), \textbf{(e)}, \textbf{(f)} graphs with moderate degree heterogeneity and clustering ($(\gamma, \, \beta)=(3.0, \, 1.1)$), \textbf{(g)}, \textbf{(h)} less degree heterogeneous graphs with weak clustering ($(\gamma, \, \beta)=(3.9, \, 0.1)$), and \textbf{(i)}, \textbf{(j)} less degree heterogeneous graphs with strong clustering ($(\gamma, \, \beta)=(3.9, \, 4.1)$) but overall the mean of the node-wise MRE increases as a function of the degree $k$, and either stays constants or decreases as a function of the clustering coefficient $c$. Hubs, nodes of high degree and low clustering, are thus of special importance for shaping the performance of \model{}s.}
    \label{fig:node_wise_mean_mre_noisy}
\end{figure*}

Examining the MRE, we find results that are largely the same across dynamical systems and graphs with high degree heterogeneity and weak clustering (Fig.~\ref{fig:node_wise_mean_mre_noisy}\textbf{(a)}, \textbf{(b)}), graphs with high degree heterogeneity and strong clustering (Fig.~\ref{fig:node_wise_mean_mre_noisy}\textbf{(c)}, \textbf{(d)}), graphs with intermediate degree heterogeneity and clustering (Fig.~\ref{fig:node_wise_mean_mre_noisy}\textbf{(e)}, \textbf{(f)}), graphs with less degree heterogeneity and weak clustering (Fig.~\ref{fig:node_wise_mean_mre_noisy}\textbf{(g)}, \textbf{(h)}), and graphs with less degree heterogeneity and strong clustering (Fig.~\ref{fig:node_wise_mean_mre_noisy}\textbf{(i)}, \textbf{(j)}). Only in two cases do we observe differences: For \model{}s trained on the MAK model, the error increases for the nodes of largest degree and lowest clustering with respect to the noiseless case when graphs are very degree heterogeneous and have strong clustering (Fig.~\ref{fig:node_wise_mean_mre_noisy}\textbf{(c)}, \textbf{(d)}). The same is true for \model{}s trained on the BD model in the case of graphs with intermediate degree heterogeneity and clustering (Fig.~\ref{fig:node_wise_mean_mre_noisy}\textbf{(e)}, \textbf{(f)}).

\begin{figure*}[tb]
    \centering
    \includegraphics[width=\textwidth]{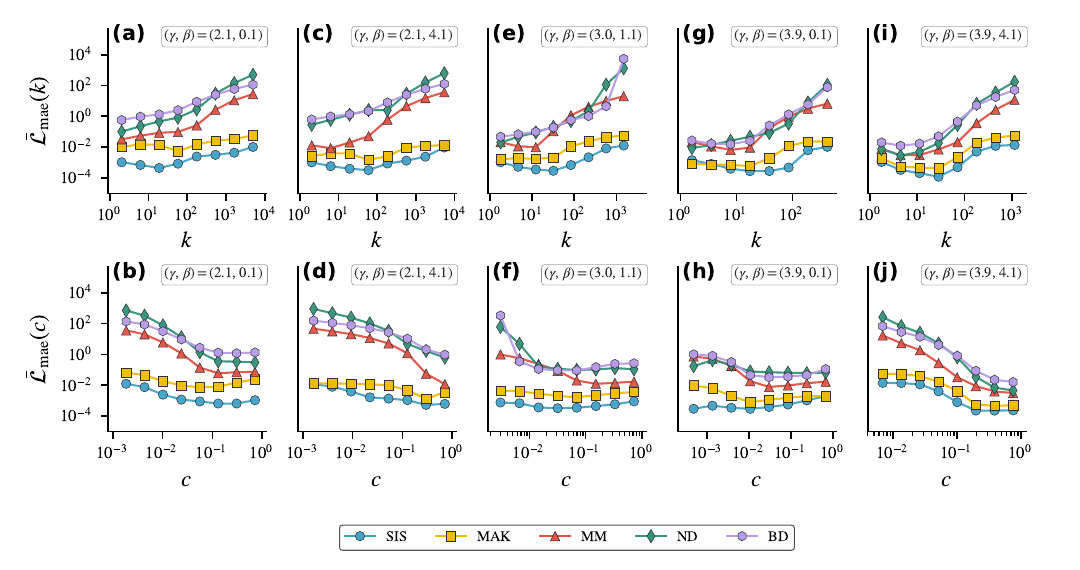}
    \caption{\textbf{Robustness of the node-level perspective on size generalization through MAE to noise in the training data}. The mean node-wise MAE $\bar{\mathcal{L}}_\mathrm{mae}(\cdot)$ on $n_G^\mathrm{test}=100$ large test graphs, $n^\mathrm{test}=8192$, can vary widely with node degree $k$ (upper row) or clustering $c$ (lower row) for \model{}s trained to approximate different dynamical systems (SIS (blue circles), MAK (yellow squares), MM (red triangles), ND (green diamonds), BD (purple hexagons)) on small graphs, $n^\mathrm{train}=64$, with the same parameters as the training graph, $(\gamma^\mathrm{train},\,\beta^\mathrm{train})=(\gamma^\mathrm{test},\,\beta^\mathrm{test})=(\gamma,\,\beta)$. The exact form of this dependence changes between \textbf{(a)}, \textbf{(b)} very degree heterogeneous graphs with weak clustering ($(\gamma, \, \beta)=(2.1, \, 0.1)$), \textbf{(c)}, \textbf{(d)} very degree heterogeneous graphs with strong clustering ($(\gamma, \, \beta)=(2.1, \, 4.1)$), \textbf{(e)}, \textbf{(f)} graphs with moderate degree heterogeneity and clustering ($(\gamma, \, \beta)=(3.0, \, 1.1)$), \textbf{(g)}, \textbf{(h)} less degree heterogeneous graphs with weak clustering ($(\gamma, \, \beta)=(3.9, \, 0.1)$), and \textbf{(i)}, \textbf{(j)} less degree heterogeneous graphs with strong clustering ($(\gamma, \, \beta)=(3.9, \, 4.1)$) but overall the mean of the node-wise MAE increases as a function of the degree $k$, and either stays constants or decreases as a function of the clustering coefficient $c$. Hubs, nodes of high degree and low clustering, are thus of special importance for shaping the performance of \model{}s.}
    \label{fig:node_wise_mean_mae_noisy}
\end{figure*}

\begin{figure*}[tb]
    \centering
    \includegraphics[width=\textwidth]{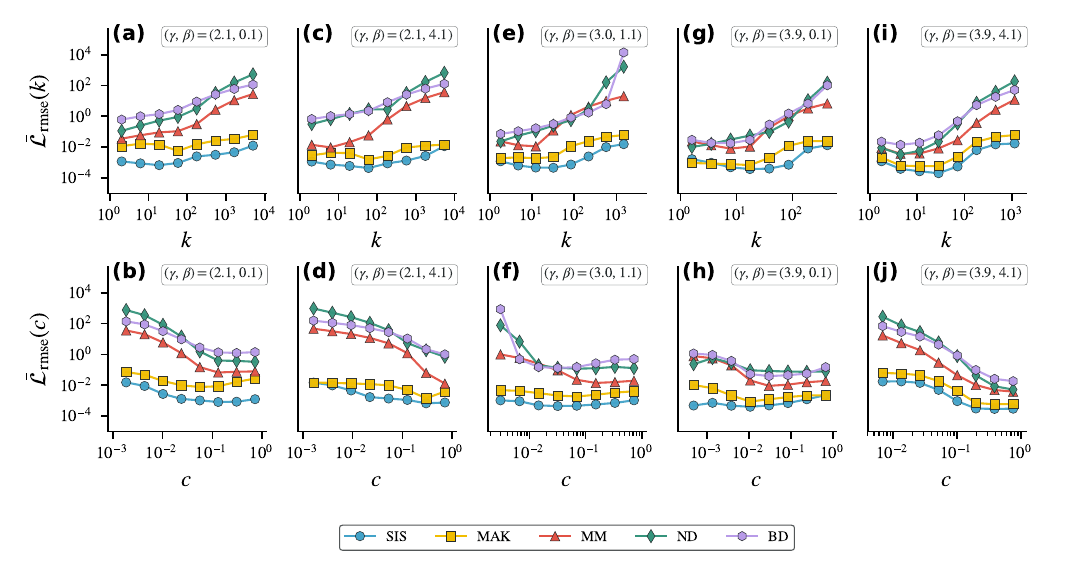}
    \caption{\textbf{Robustness of the node-level perspective on size generalization through RMSE to noise in the training data}. The mean node-wise RMSE $\bar{\mathcal{L}}_\mathrm{rmse}(\cdot)$ on $n_G^\mathrm{test}=100$ large test graphs, $n^\mathrm{test}=8192$, can vary widely with node degree $k$ (upper row) or clustering $c$ (lower row) for \model{}s trained to approximate different dynamical systems (SIS (blue circles), MAK (yellow squares), MM (red triangles), ND (green diamonds), BD (purple hexagons)) on small graphs, $n^\mathrm{train}=64$, with the same parameters as the training graph, $(\gamma^\mathrm{train},\,\beta^\mathrm{train})=(\gamma^\mathrm{test},\,\beta^\mathrm{test})=(\gamma,\,\beta)$. The exact form of this dependence changes between \textbf{(a)}, \textbf{(b)} very degree heterogeneous graphs with weak clustering ($(\gamma, \, \beta)=(2.1, \, 0.1)$), \textbf{(c)}, \textbf{(d)} very degree heterogeneous graphs with strong clustering ($(\gamma, \, \beta)=(2.1, \, 4.1)$), \textbf{(e)}, \textbf{(f)} graphs with moderate degree heterogeneity and clustering ($(\gamma, \, \beta)=(3.0, \, 1.1)$), \textbf{(g)}, \textbf{(h)} less degree heterogeneous graphs with weak clustering ($(\gamma, \, \beta)=(3.9, \, 0.1)$), and \textbf{(i)}, \textbf{(j)} less degree heterogeneous graphs with strong clustering ($(\gamma, \, \beta)=(3.9, \, 4.1)$) but overall the mean of the node-wise RMSE increases as a function of the degree $k$, and either stays constants or decreases as a function of the clustering coefficient $c$. Hubs, nodes of high degree and low clustering, are thus of special importance for shaping the performance of \model{}s.}
    \label{fig:node_wise_mean_rmse_noisy}
\end{figure*}

Similar to the noiseless case, MAE (Fig.~\ref{fig:node_wise_mean_mae_noisy}) and RMSE (Fig.~\ref{fig:node_wise_mean_rmse_noisy}) show the same behavior. In addition, their behavior is similar to the noiseless case across dynamical systems and graph properties (Fig.~\ref{fig:node_wise_mean_mae_noisy}\textbf{(a)}-\textbf{(j)}, Fig.~\ref{fig:node_wise_mean_rmse_noisy}\textbf{(a)}-\textbf{(j)}). The increase in error on very high degree or low clustering nodes for \model{}s trained on the BD model on graphs with intermediate degree heterogeneity and clustering (Fig.~\ref{fig:node_wise_mean_mae_noisy}\textbf{(e)}, \textbf{(f)}, Fig.~\ref{fig:node_wise_mean_rmse_noisy}\textbf{(e)}, \textbf{(f)}) is also observed in these error metrics. However, the analogous behavior in the case of \model{}s trained on the MAK model on very degree heterogeneous graphs with strong clustering (Fig.~\ref{fig:node_wise_mean_mae_noisy}\textbf{(c)}, \textbf{(d)}, Fig.~\ref{fig:node_wise_mean_rmse_noisy}\textbf{(c)}, \textbf{(d)}) is not visible when inspecting MAE or RMSE. This is because the activity decreases with degree in this model, meaning that deviations are less visible without normalization.

Together, these observations corroborate the results of Appendix~\ref{appendix:subsec:size_generalization_local_noiseless} and show their robustness to noise in the training data.

\FloatBarrier
\clearpage
\section{Additional results on generalization across graph properties}\label{appendix:property_generalization}

Here, we consider the ability of \model{}s trained on small graphs, $n^\mathrm{train}=64$, to generalize across graph parameters, $(\gamma^\mathrm{train}, \, \beta^\mathrm{train})\neq(\gamma^\mathrm{test}, \, \beta^\mathrm{test})$, on graphs of the same size, $n^\mathrm{test}=64$, and substantially larger size, $n^\mathrm{test}=8192$, than the training graph. All results for a given parameter setting are averaged over $n_G^\mathrm{test}=100$ test graph with independent initial conditions.

\begin{figure*}[b]
    \centering
    \includegraphics[width=\textwidth]{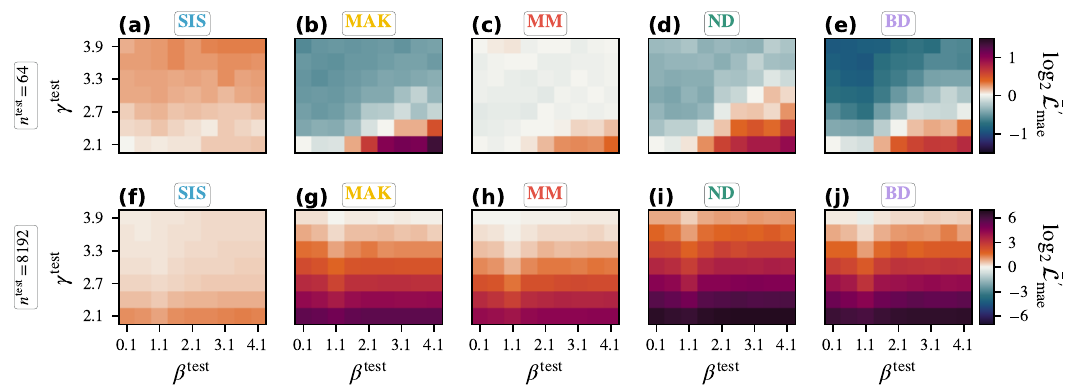}
    \caption{\textbf{Generalizing to graphs with different properties, $(\gamma^\mathrm{train}, \, \beta^\mathrm{train})=(2.1, \, 0.1)$}. The ability of \model{}s trained on small graphs, $n^\mathrm{train}=64$, with strong degree heterogeneity and weak clustering, $(\gamma^\mathrm{train}, \, \beta^\mathrm{train})=(2.1, \, 0.1)$, to generalize to graphs with different properties, $(\gamma^\mathrm{test}, \, \beta^\mathrm{test})$, on $n_G^\mathrm{test}=100$ test graphs of the same, $n^\mathrm{test}=64$, (upper row) or larger, $n^\mathrm{test}=8192$, (lower row) size is measured as the change in mean node-wise MAE (color coded) relative to its value on graphs with the same properties and size as the training graph $\bar{\mathcal{L}}^\prime_\mathrm{mae}$. On small graphs, results vary by dynamical system (\textbf{(a)} SIS, \textbf{(b)} MAK, \textbf{(c)} MM, \textbf{(d)} ND, \textbf{(e)} BD model) but generalization to very degree heterogeneous graphs with strong clustering appears most difficult. On larger graphs, normalized MAE decreases towards less degree heterogeneous graphs independent of the dynamical system (\textbf{(f)} SIS, \textbf{(g)} MAK, \textbf{(h)} MM, \textbf{(i)} ND, \textbf{(j)} BD). This means \model{}s are to some extent robust to changes in graph properties, especially if deployed on less degree heterogeneous graphs or graphs of the same size as the training graph.}
    \label{fig:parameters_mean_mae_2101}
\end{figure*}

\begin{figure*}[tb]
    \centering
    \includegraphics[width=\textwidth]{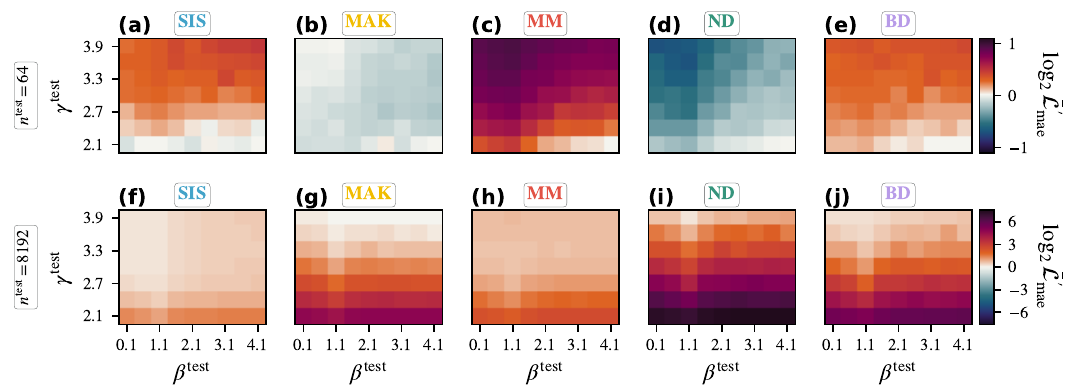}
    \caption{\textbf{Generalizing to graphs with different properties, $(\gamma^\mathrm{train}, \, \beta^\mathrm{train})=(2.1, \, 4.1)$}. The ability of \model{}s trained on small graphs, $n^\mathrm{train}=64$, with strong degree heterogeneity and strong clustering, $(\gamma^\mathrm{train}, \, \beta^\mathrm{train})=(2.1, \, 4.1)$, to generalize to graphs with different properties, $(\gamma^\mathrm{test}, \, \beta^\mathrm{test})$, on $n_G^\mathrm{test}=100$ test graphs of the same, $n^\mathrm{test}=64$, (upper row) or larger, $n^\mathrm{test}=8192$, (lower row) size is measured as the change in mean node-wise MAE (color coded) relative to its value on graphs with the same properties and size as the training graph $\bar{\mathcal{L}}^\prime_\mathrm{mae}$. On small graphs, results vary by dynamical system (\textbf{(a)} SIS, \textbf{(b)} MAK, \textbf{(c)} MM, \textbf{(d)} ND, \textbf{(e)} BD model). On larger graphs, normalized MAE decreases towards less degree heterogeneous graphs independent of the dynamical system (\textbf{(f)} SIS, \textbf{(g)} MAK, \textbf{(h)} MM, \textbf{(i)} ND, \textbf{(j)} BD). This means \model{}s are to some extent robust to changes in graph properties, especially if deployed on less degree heterogeneous or less clustered graphs of the same size as the training graph.}
    \label{fig:parameters_mean_mae_2141}
\end{figure*}

\begin{figure*}[tb]
    \centering
    \includegraphics[width=\textwidth]{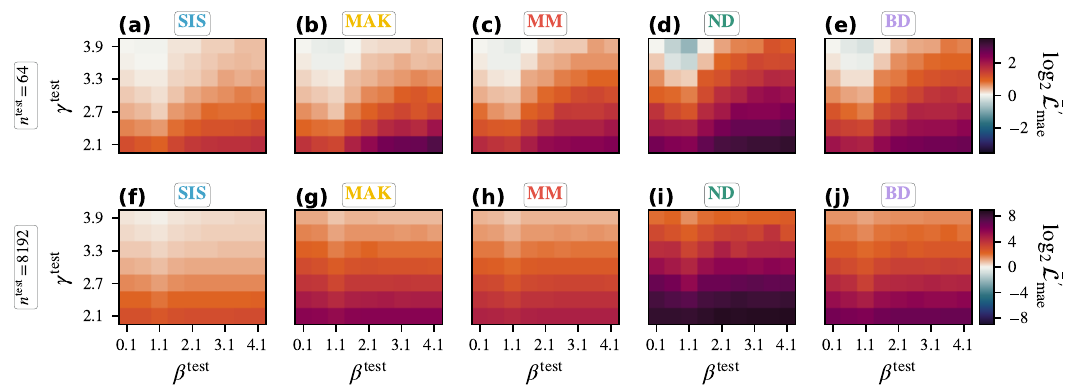}
    \caption{\textbf{Generalizing to graphs with different properties, $(\gamma^\mathrm{train}, \, \beta^\mathrm{train})=(3.9, \, 0.1)$}. The ability of \model{}s trained on small graphs, $n^\mathrm{train}=64$, with less degree heterogeneity and weak clustering, $(\gamma^\mathrm{train}, \, \beta^\mathrm{train})=(3.9, \, 0.1)$, to generalize to graphs with different properties, $(\gamma^\mathrm{test}, \, \beta^\mathrm{test})$, on $n_G^\mathrm{test}=100$ test graphs of the same, $n^\mathrm{test}=64$, (upper row) or larger, $n^\mathrm{test}=8192$, (lower row) size is measured as the change in mean node-wise MAE (color coded) relative to its value on graphs with the same properties and size as the training graph $\bar{\mathcal{L}}^\prime_\mathrm{mae}$. On small graphs, MAE increases as test graphs become more degree heterogeneous and have higher clustering, independent of the dynamical system (\textbf{(a)} SIS, \textbf{(b)} MAK, \textbf{(c)} MM, \textbf{(d)} ND, \textbf{(e)} BD model). On larger graphs, normalized MAE increases towards more degree heterogeneous graphs independent of the dynamical system (\textbf{(f)} SIS, \textbf{(g)} MAK, \textbf{(h)} MM, \textbf{(i)} ND, \textbf{(j)} BD). This means \model{}s are to some extent robust to changes in graph properties but increasing degree heterogeneity is a limiting factor for generalization across graph properties.}
    \label{fig:parameters_mean_mae_3901}
\end{figure*}

\begin{figure*}[tb]
    \centering
    \includegraphics[width=\textwidth]{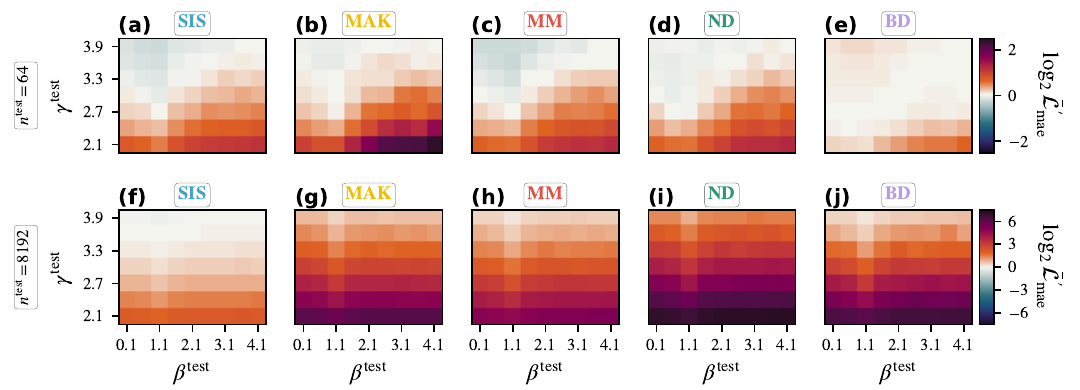}
    \caption{\textbf{Generalizing to graphs with different properties, $(\gamma^\mathrm{train}, \, \beta^\mathrm{train})=(3.9, \, 4.1)$}. The ability of \model{}s trained on small graphs, $n^\mathrm{train}=64$, with less degree heterogeneity and strong clustering, $(\gamma^\mathrm{train}, \, \beta^\mathrm{train})=(3.9, \, 4.1)$, to generalize to graphs with different properties, $(\gamma^\mathrm{test}, \, \beta^\mathrm{test})$, on $n_G^\mathrm{test}=100$ test graphs of the same, $n^\mathrm{test}=64$, (upper row) or larger, $n^\mathrm{test}=8192$, (lower row) size is measured as the change in mean node-wise MAE (color coded) relative to its value on graphs with the same properties and size as the training graph $\bar{\mathcal{L}}^\prime_\mathrm{mae}$. On small graphs, MAE increases as test graphs become more degree heterogeneous and have higher clustering, independent of the dynamical system (\textbf{(a)} SIS, \textbf{(b)} MAK, \textbf{(c)} MM, \textbf{(d)} ND, \textbf{(e)} BD model). On larger graphs, normalized MAE increases towards more degree heterogeneous graphs independent of the dynamical system (\textbf{(f)} SIS, \textbf{(g)} MAK, \textbf{(h)} MM, \textbf{(i)} ND, \textbf{(j)} BD) but only moderately for \model{}s trained on the SIS model. This means \model{}s are to some extent robust to changes in graph properties but increasing degree heterogeneity is a limiting factor for generalization across graph properties.}
    \label{fig:parameters_mean_mae_3941}
\end{figure*}

Having presented the results for training graphs with intermediate degree heterogeneity and clustering, $(\gamma^\mathrm{train}, \, \beta^\mathrm{train})=(3.0, \, 1.1)$, in Section~\ref{sec:results:property_generalization}, we now turn to training graphs with high degree heterogeneity and weak clustering ($(\gamma^\mathrm{train}, \, \beta^\mathrm{train})=(2.1, \, 0.1)$, Fig.~\ref{fig:parameters_mean_mae_2101}), high degree heterogeneity and strong clustering ($(\gamma^\mathrm{train}, \, \beta^\mathrm{train})=(2.1, \, 4.1)$, Fig.~\ref{fig:parameters_mean_mae_2141}), less degree heterogeneity and weak clustering ($(\gamma^\mathrm{train}, \, \beta^\mathrm{train})=(3.9, \, 0.1)$, Fig.~\ref{fig:parameters_mean_mae_3901}), and less degree heterogeneity and strong clustering ($(\gamma^\mathrm{train}, \, \beta^\mathrm{train})=(3.9, \, 4.1)$, Fig.~\ref{fig:parameters_mean_mae_3941}). 

We measure error in terms of the node-wise MAE, and display its mean normalized to the value on test graphs with the same size and properties as the training graph, denoted $\bar{\mathcal{L}}^\prime_\mathrm{mae}$.

If training graphs are very degree heterogeneous and have weak clustering, $(\gamma^\mathrm{train}, \, \beta^\mathrm{train})=(2.1, \, 0.1)$, the behavior of \model{}s on graphs with the same number of nodes as the training graph,($n^\mathrm{test}=64$, Fig.~\ref{fig:parameters_mean_mae_2101}\textbf{(a)}-\textbf{(e)}) depends on the dynamical system. For \model{}s trained on the SIS model the normalized MAE increases towards graphs less degree heterogeneous than the training graph, but only slightly. For \model{}s trained on the MM model it stays almost constant. In contrast,  for \model{}s trained on the MAK, ND, and BD model normalized MAE decreases slightly towards less degree heterogeneous graphs, but increases towards graphs with equal degree heterogeneity and higher clustering. This is in line with the findings presented in the main text (Section~\ref{sec:results:property_generalization}).

If graphs are much larger than the training graph ($n^\mathrm{test}=8192$, Fig.~\ref{fig:parameters_mean_mae_2101}\textbf{(f)}-\textbf{(j)}), we observe that \model{}s trained on the SIS model perform well throughout the entire space of $\mathbb{S}^1$-model parameters. For \model{}s trained on other dynamical systems, MAE is largest on very degree heterogeneous graphs but decreases towards less degree heterogeneous graphs. 

For \model{}s trained on very degree heterogeneous graphs with strong clustering, we again observe that performance changes only slightly as graph properties vary on graphs of the same size as the training graph  (Fig.~\ref{fig:parameters_mean_mae_2141}\textbf{(a)}-\textbf{(e)}). However, performance degrades for almost all dynamical systems away from the parameter values of the training graph. Only \model{}s trained on the MAK and ND model show better performance as the degree heterogeneity and clustering of test graphs decreases. 

On graphs much larger than the training graph  (Fig.~\ref{fig:parameters_mean_mae_2141}\textbf{(f)}-\textbf{(j)}), \model{}s behave similarly to the case of strong clustering. Only the performance of the \model{} trained on the MM model is better across $\mathbb{S}^1$-model parameters.

If the training graph is less degree heterogeneous and has weak clustering and \model{}s are evaluated on graphs of the same size as the training graph  (Fig.~\ref{fig:parameters_mean_mae_3901}\textbf{(a)}-\textbf{(e)}), the normalized MAE increases moderately towards graphs that are more degree heterogeneous and that have higher average clustering than the training graph. The increase in MAE is strongest for the ND model.

The behavior on larger test graphs (Fig.~\ref{fig:parameters_mean_mae_3901}\textbf{(f)}-\textbf{(j)}) shows once more an increase towards larger degree heterogeneity and only a weak dependence on clustering. 

In the case of training graphs with less degree heterogeneity and strong clustering, we find that when testing on graphs of the same size as the training graph (Fig.~\ref{fig:parameters_mean_mae_3941}\textbf{(a)}-\textbf{(e)}) MAE increases towards more degree heterogeneity and more clustering across dynamical systems. 

In the case of larger test graphs  (Fig.~\ref{fig:parameters_mean_mae_3941}\textbf{(f)}-\textbf{(j)}), degree heterogeneity is the main determining factor of MAE. Normalized MAE increases as degree heterogeneity increases. This increase is least severe for the \model{} trained on the SIS model and strongest for the \model{} trained on the ND model.

\begin{figure*}[b]
    \centering
    \includegraphics[width=\textwidth]{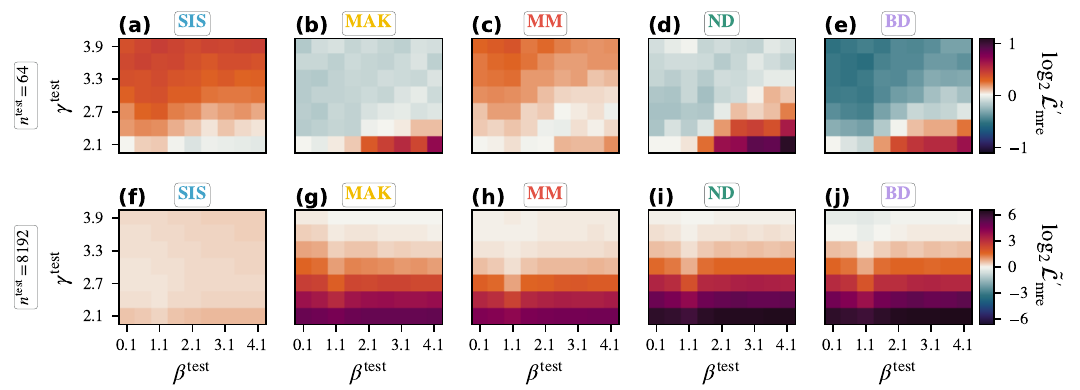}
    \caption{\textbf{Generalizing to graphs with different properties, $(\gamma^\mathrm{train}, \, \beta^\mathrm{train})=(2.1, \, 0.1)$}. The ability of \model{}s trained on small graphs, $n^\mathrm{train}=64$, with strong degree heterogeneity and weak clustering, $(\gamma^\mathrm{train}, \, \beta^\mathrm{train})=(2.1, \, 0.1)$, to generalize to graphs with different properties, $(\gamma^\mathrm{test}, \, \beta^\mathrm{test})$, on $n_G^\mathrm{test}=100$ test graphs of the same, $n^\mathrm{test}=64$, upper row) or larger, $n^\mathrm{test}=8192$, (lower row) size is measured as the change in median node-wise MRE (color coded) relative to its value on graphs with the same properties and size as the training graph $\tilde{\mathcal{L}}^\prime_\mathrm{mre}$. On small graphs, results vary by dynamical system (\textbf{(a)} SIS, \textbf{(b)} MAK, \textbf{(c)} MM, \textbf{(d)} ND, \textbf{(e)} BD model) but generalization to very degree heterogeneous graphs with strong clustering appears most difficult. On larger graphs, normalized MRE decreases towards less degree heterogeneous graphs independent of the dynamical system (\textbf{(f)} SIS, \textbf{(g)} MAK, \textbf{(h)} MM, \textbf{(i)} ND, \textbf{(j)} BD). This means \model{}s are to some extent robust to changes in graph properties, especially if deployed on less degree heterogeneous graphs or graphs of the same size as the training graph.}
    \label{fig:parameters_median_mre_2101}
\end{figure*}

\begin{figure*}[tb]
    \centering
    \includegraphics[width=\textwidth]{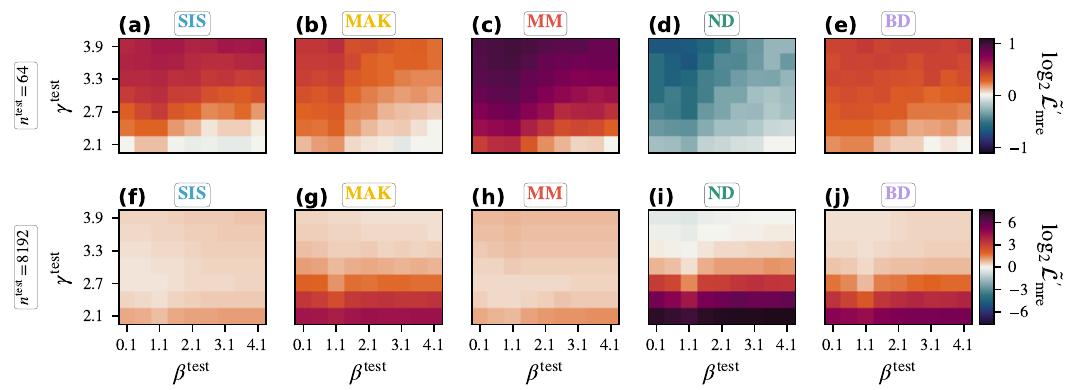}
    \caption{\textbf{Generalizing to graphs with different properties, $(\gamma^\mathrm{train}, \, \beta^\mathrm{train})=(2.1, \, 4.1)$}. The ability of \model{}s trained on small graphs, $n^\mathrm{train}=64$, with strong degree heterogeneity and strong clustering, $(\gamma^\mathrm{train}, \, \beta^\mathrm{train})=(2.1, \, 4.1)$, to generalize to graphs with different properties, $(\gamma^\mathrm{test}, \, \beta^\mathrm{test})$, on $n_G^\mathrm{test}=100$ test graphs of the same, $n^\mathrm{test}=64$, (upper row) or larger, $n^\mathrm{test}=8192$, (lower row) size is measured as the change in median node-wise MRE (color coded) relative to its value on graphs with the same properties and size as the training graph $\tilde{\mathcal{L}}^\prime_\mathrm{mre}$. On small graphs, all \model{}s but the one trained on the ND model show an increase in MRE towards less degree heterogeneous graphs (\textbf{(a)} SIS, \textbf{(b)} MAK, \textbf{(c)} MM, \textbf{(d)} ND, \textbf{(e)} BD model). This increase is however moderate. On larger graphs, normalized MRE decreases towards less degree heterogeneous graphs independent of the dynamical system (\textbf{(f)} SIS, \textbf{(g)} MAK, \textbf{(h)} MM, \textbf{(i)} ND, \textbf{(j)} BD). This means \model{}s are to some extent robust to changes in graph properties, especially if deployed on less degree heterogeneous and clustered graphs of the same size as the training graph.}
    \label{fig:parameters_median_mre_2141}
\end{figure*}

\begin{figure*}[tb]
    \centering
    \includegraphics[width=\textwidth]{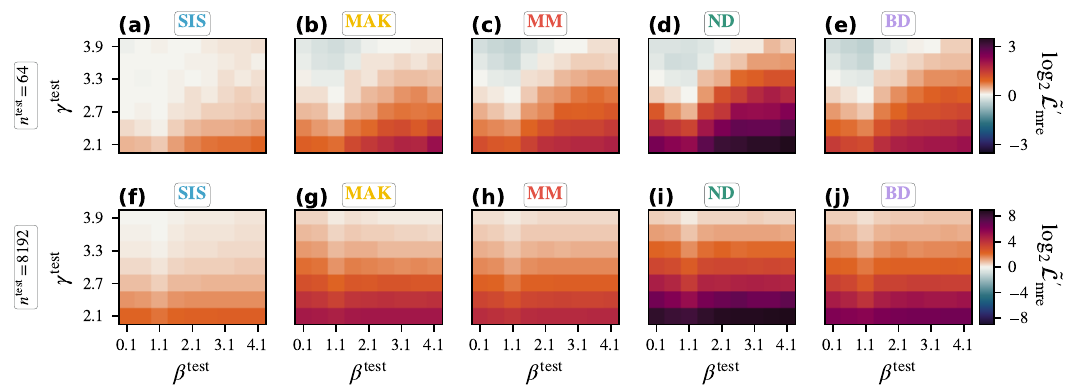}
    \caption{\textbf{Generalizing to graphs with different properties, $(\gamma^\mathrm{train}, \, \beta^\mathrm{train})=(3.0, \, 1.1)$.} The ability of \model{}s trained on small graphs, $n^\mathrm{train}=64$, with moderate degree heterogeneity and clustering, $(\gamma^\mathrm{train},\,\beta^\mathrm{train})=(3.0,\,1.1)$, to generalize to graphs with different properties, $(\gamma^\mathrm{test}, \, \beta^\mathrm{test})$, on $n_G^\mathrm{test}=100$ test graphs of the same, $n^\mathrm{test}=64$, (upper row) or larger, $n^\mathrm{test}=8192$, (lower row) size is measured as the change in median node-wise MRE (color coded) relative to its value on graphs with the same properties and size as the training graph $\tilde{\mathcal{L}}^\prime_\mathrm{mre}$. On small graphs, the performance improves towards lower degree heterogeneity and less clustered graphs, and decreases towards more degree heterogeneity and more clustering independent of the dynamical system (\textbf{(a)} SIS, \textbf{(b)} MAK, \textbf{(c)} MM, \textbf{(d)} ND, \textbf{(e)} BD model). On larger graphs, MRE increases throughout the $\mathbb{S}^1$-model parameter range for all dynamical systems (\textbf{(f)} SIS, \textbf{(g)} MAK, \textbf{(h)} MM, \textbf{(i)} ND, \textbf{(j)} BD), and most strongly towards higher degree heterogeneity. The increase is moderate for the \model{} trained on the SIS model but substantial for \model{}s trained on the ND model. This means \model{}s are to some extent robust to changes in graph properties, especially if deployed on less degree heterogeneous and clustered graphs of the same size as the training graph.}
    \label{fig:parameters_median_mre_3011}
\end{figure*}

\begin{figure*}[tb]
    \centering
    \includegraphics[width=\textwidth]{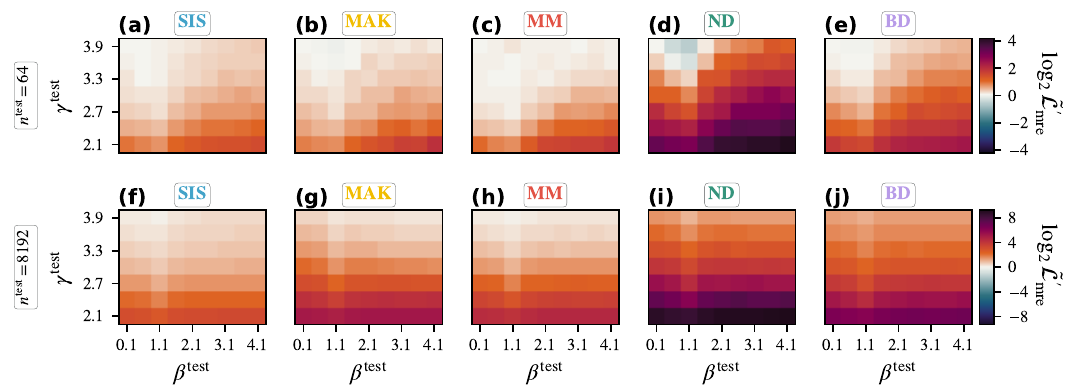}
    \caption{\textbf{Generalizing to graphs with different properties, $(\gamma^\mathrm{train}, \, \beta^\mathrm{train})=(3.9, \, 0.1)$}. The ability of \model{}s trained on small graphs, $n^\mathrm{train}=64$, with less degree heterogeneity and weak clustering, $(\gamma^\mathrm{train}, \, \beta^\mathrm{train})=(3.9, \, 0.1)$, to generalize to graphs with different properties, $(\gamma^\mathrm{test}, \, \beta^\mathrm{test})$, on $n_G^\mathrm{test}=100$ test graphs of the same, $n^\mathrm{test}=64$, (upper row) or larger, $n^\mathrm{test}=8192$, (lower row) size is measured as the change in median node-wise MRE (color coded) relative to its value on graphs with the same properties and size as the training graph $\tilde{\mathcal{L}}^\prime_\mathrm{mre}$. On small graphs, MRE increases as test graphs become more degree heterogeneous and have higher clustering, independent of the dynamical system (\textbf{(a)} SIS, \textbf{(b)} MAK, \textbf{(c)} MM, \textbf{(d)} ND, \textbf{(e)} BD model). On larger graphs, normalized MRE increases towards more degree heterogeneous graphs independent of the dynamical system (\textbf{(f)} SIS, \textbf{(g)} MAK, \textbf{(h)} MM, \textbf{(i)} ND, \textbf{(j)} BD). This means \model{}s are to some extent robust to changes in graph properties but increasing degree heterogeneity is a limiting factor.}
    \label{fig:parameters_median_mre_3901}
\end{figure*}

\begin{figure*}[tb]
    \centering
    \includegraphics[width=\textwidth]{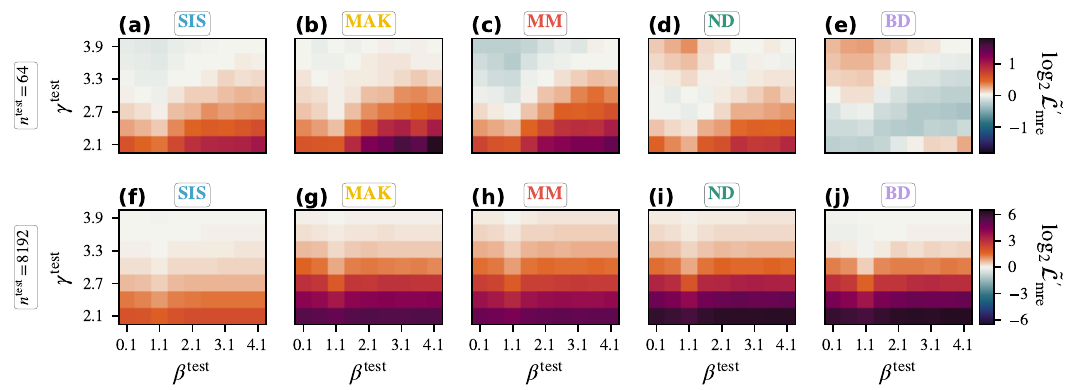}
    \caption{\textbf{Generalizing to graphs with different properties, $(\gamma^\mathrm{train}, \, \beta^\mathrm{train})=(3.9, \, 4.1)$}. The ability of \model{}s trained on small graphs, $n^\mathrm{train}=64$, with less degree heterogeneity and strong clustering, $(\gamma^\mathrm{train}, \, \beta^\mathrm{train})=(3.9, \, 4.1)$, to generalize to graphs with different properties, $(\gamma^\mathrm{test}, \, \beta^\mathrm{test})$, on $n_G^\mathrm{test}=100$ test graphs of the same, $n^\mathrm{test}=64$, (upper row) or larger, $n^\mathrm{test}=8192$, (lower row) size is measured as the change in median node-wise MRE (color coded) relative to its value on graphs with the same properties and size as the training graph $\tilde{\mathcal{L}}^\prime_\mathrm{mre}$. On small graphs, MRE increases as test graphs become more degree heterogeneous and have higher clustering for most dynamical systems (\textbf{(a)} SIS, \textbf{(b)} MAK, \textbf{(c)} MM, \textbf{(d)} ND, \textbf{(e)} BD model). On larger graphs, normalized MRE increases towards more degree heterogeneous graphs independent of the dynamical system (\textbf{(f)} SIS, \textbf{(g)} MAK, \textbf{(h)} MM, \textbf{(i)} ND, \textbf{(j)} BD) but only moderately for \model{}s trained on the SIS model. This means \model{}s are to some extent robust to changes in graph properties but increasing degree heterogeneity is a limiting factor for generalization across graph properties.}
    \label{fig:parameters_median_mre_3941}
\end{figure*}

To give an impression of the influence of using a relative compared to an absolute notion of error, we also present results for the normalized median MRE, $\tilde{\mathcal{L}}^\prime_\mathrm{mre}$. Overall, we find the main patterns of our analysis of the MAE hold also for the MRE on very degree heterogeneous graphs with weak clustering (Fig.~\ref{fig:parameters_median_mre_2101}), very degree heterogeneous graphs with strong clustering (Fig.~\ref{fig:parameters_median_mre_2141}), graphs with intermediate degree heterogeneity and clustering (Fig.~\ref{fig:parameters_median_mre_3011}), graphs with less degree heterogeneity and weak clustering (Fig.~\ref{fig:parameters_median_mre_3901}), and graphs with less degree heterogeneity and strong clustering (Fig.~\ref{fig:parameters_median_mre_3941}). We comment on the few differences that we observe.

If test graphs are very degree heterogeneous and have weak clustering (Fig.~\ref{fig:parameters_median_mre_2101}), we observe an increase in normalized MRE than MAE in the case of the \model{} trained on the MM model on test graphs of size $n^\mathrm{test}=64$. We make a similar observation for the \model{} trained on the MAK model in case of very degree heterogeneous graphs with strong clustering (Fig~\ref{fig:parameters_median_mre_2141}). In the case of graphs with less degree heterogeneity and strong clustering of the same size as the training graph (Fig~\ref{fig:parameters_median_mre_2141}\textbf{(a)}-\textbf{(e)}), we find that regions of constant or slightly increasing normalized mean MAE correspond to regions in which normalized median MRE slightly decreases. Moreover, we find that when considering relative instead of absolute error \model{}s trained on the BD models appear to recover their performance as $\gamma^\mathrm{test}\to 3.9$.

\clearpage
\FloatBarrier
\section{Additional results on fixed point approximation and stability}\label{appendix:jacobian}

In this appendix, we provide additional results on the stability of the \model{} fixed points, and their deviation from the true fixed point in terms of MAE. We evaluate \model{}s trained on small graphs, $n^\mathrm{train}=64$, on larger graphs with the same parameters as the training graph, $(\gamma^\mathrm{test}, \, \beta^\mathrm{test})=(\gamma^\mathrm{train}, \, \beta^\mathrm{train}) = (\gamma, \, \beta)$. All results are obtained on $n_G^\mathrm{test}=100$ test graphs with independent initial conditions.

\begin{figure*}[tb]
    \centering
    \includegraphics[width=0.875\textwidth]{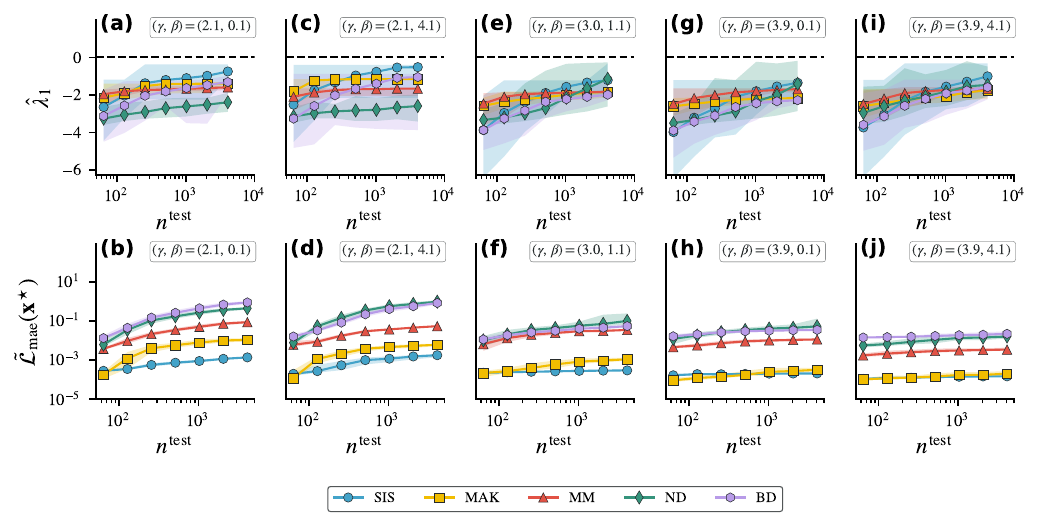}
    \caption{\textbf{Local Stability and fixed point approximation in terms of median MAE.} Across dynamical systems (SIS (blue circles), MAK (yellow squares), MM (red triangles), ND (green diamonds), BD (purple hexagons)) and graph parameters (\textbf{(a)}, \textbf{(b)} $(\gamma, \, \beta)=(2.1, \, 0.1)$, \textbf{(c)}, \textbf{(d)} $(\gamma, \, \beta)=(2.1, \, 4.1)$, \textbf{(e)}, \textbf{(f)} $(\gamma, \, \beta)=(3.0, \, 1.1)$, \textbf{(g)}, \textbf{(h)} $(\gamma, \, \beta)=(3.9, \, 0.1)$, \textbf{(i)}, \textbf{(j)} $(\gamma, \, \beta)=(3.9, \, 4.1)$) the largest eigenvalue $\hat{\lambda}_1$ of the Jacobian (upper row) remains negative (shaded band) for \model{}s trained on small graphs, $n^\mathrm{train}=64$, with the same parameters as the test graph, $(\gamma^\mathrm{train}, \,  \beta^\mathrm{train})=(\gamma^\mathrm{test}, \, \beta^\mathrm{test})=(\gamma, \, \beta)$, even though it increases on average (solid line) with the size of the test graph, $n^\mathrm{test}$. The median MAE (lower row) of the fixed point approximation, $\tilde{\mathcal{L}}_\mathrm{mae}^\star$, increases noticeably  across dynamical systems on very degree heterogeneous graphs independent of clustering but stays constant or increases only slightly in the less degree heterogeneous case. This means that even though \model{}s approach stable fixed points, these fixed points can differ from the dynamical system's true fixed point,  especially on very degree heterogeneous graphs.}
    \label{fig:jac_median_mae_noiseless}
\end{figure*}

In Fig.~\ref{fig:jac_median_mae_noiseless}\textbf{(a)}, \textbf{(c)}, \textbf{(e)}, \textbf{(g)}, \textbf{(i)}, we recapitulate the results on the size of the largest eigenvalue of the \model{}'s Jacobian $\hat{\lambda}_1$ from the main text (Section~\ref{sec:results:stability}). In Fig.~\ref{fig:jac_median_mae_noiseless}\textbf{(b)}, \textbf{(d)}, \textbf{(f)}, \textbf{(h)}, we show the median and interquartile range of the MAE on graphs with different parameters. The results for the median are very similar to those presented in the main text (Section~\ref{sec:results:stability}), and the interquartile range is small, indicating that the variation in MAE values is low.

\begin{figure*}[tb]
    \centering
    \includegraphics[width=0.875\textwidth]{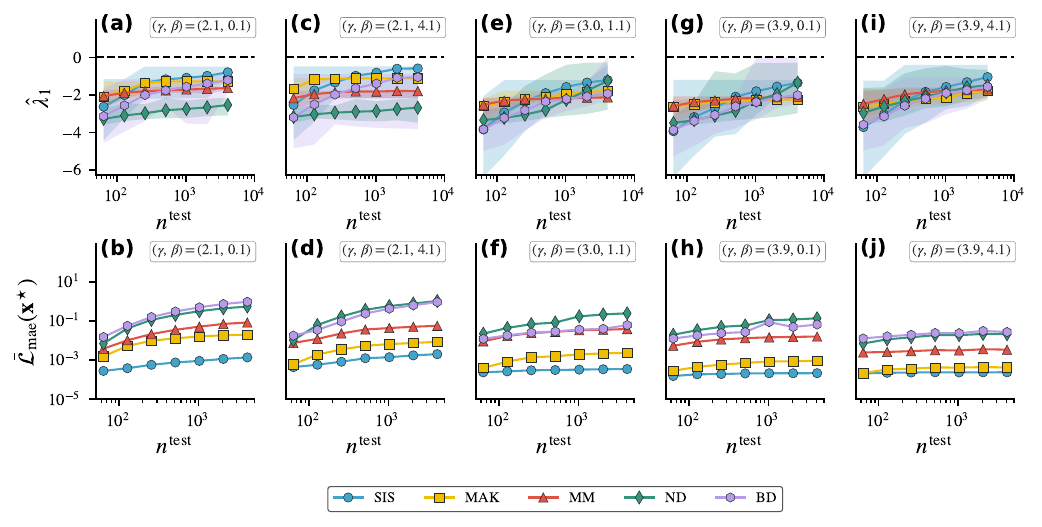}
    \caption{\textbf{Local Stability and fixed point approximation in terms of mean MAE for noisy training data.}   Across dynamical systems (SIS (blue circles), MAK (yellow squares), MM (red triangles), ND (green diamonds), BD (purple hexagons)) and graph parameters (\textbf{(a)}, \textbf{(b)} $(\gamma, \, \beta)=(2.1, \, 0.1)$, \textbf{(c)}, \textbf{(d)} $(\gamma, \, \beta)=(2.1, \, 4.1)$, \textbf{(e)}, \textbf{(f)} $(\gamma, \, \beta)=(3.0, \, 1.1)$, \textbf{(g)}, \textbf{(h)} $(\gamma, \, \beta)=(3.9, \, 0.1)$, \textbf{(i)}, \textbf{(j)} $(\gamma, \, \beta)=(3.9, \, 4.1)$) the largest eigenvalue $\hat{\lambda}_1$ of the Jacobian (upper row) stays negative most of the time (shaded band) for \model{}s trained on small graphs, $n^\mathrm{train}=64$, with the same parameters as the test graph, $(\gamma^\mathrm{train}, \,  \beta^\mathrm{train})=(\gamma^\mathrm{test}, \, \beta^\mathrm{test})=(\gamma, \, \beta)$, even though it increases on average (solid line) with the size of the test graph $n^\mathrm{test}$. The average MAE (lower row) of the fixed point approximation, $\bar{\mathcal{L}}_\mathrm{mae}^\star$, increases noticeably  across dynamical systems on very degree heterogeneous graphs independent of clustering but stays constant or increases only slightly in the less degree heterogeneous case. This means that even though \model{}s approach stable fixed points, these fixed points can differ from the dynamical system's true fixed point,  especially on very degree heterogeneous graphs.}
    \label{fig:jac_mean_mae_noisy}
\end{figure*}

\begin{figure*}[tb]
    \centering
    \includegraphics[width=0.875\textwidth]{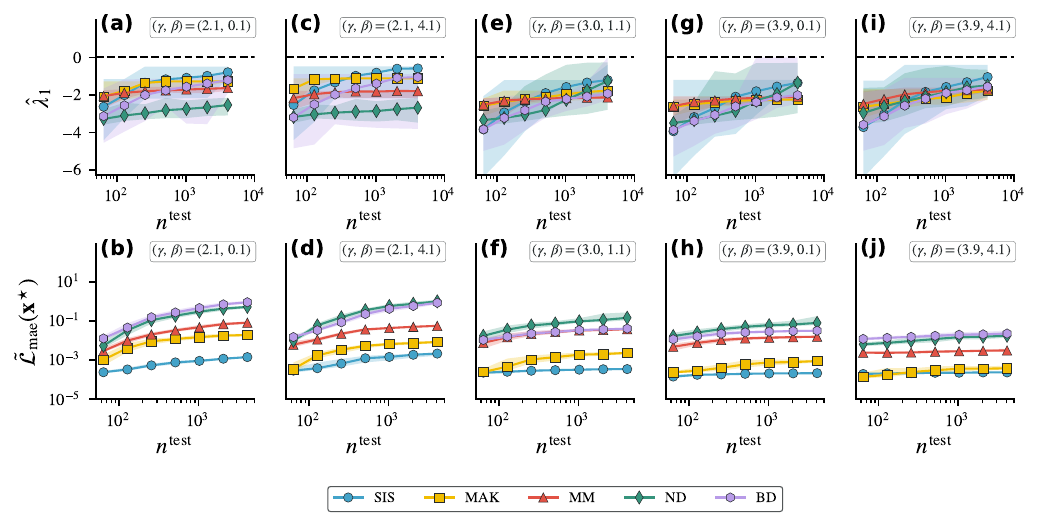}
    \caption{\textbf{Local Stability and fixed point approximation in terms of median MAE for noisy training data.} Across dynamical systems (SIS (blue circles), MAK (yellow squares), MM (red triangles), ND (green diamonds), BD (purple hexagons)) and graph parameters (\textbf{(a)}, \textbf{(b)} $(\gamma, \, \beta)=(2.1, \, 0.1)$, \textbf{(c)}, \textbf{(d)} $(\gamma, \, \beta)=(2.1, \, 4.1)$, \textbf{(e)}, \textbf{(f)} $(\gamma, \, \beta)=(3.0, \, 1.1)$, \textbf{(g)}, \textbf{(h)} $(\gamma, \, \beta)=(3.9, \, 0.1)$, \textbf{(i)}, \textbf{(j)} $(\gamma, \, \beta)=(3.9, \, 4.1)$) the largest eigenvalue $\hat{\lambda}_1$ of the Jacobian (upper row) stays mostly negative (shaded band) for \model{}s trained on small graphs, $n^\mathrm{train}=64$, with the same parameters as the test graph, $(\gamma^\mathrm{train}, \,  \beta^\mathrm{train})=(\gamma^\mathrm{test}, \, \beta^\mathrm{test})=(\gamma, \, \beta)$, even though it increases on average (solid line) with the size of the test graph, $n^\mathrm{test}$. The median MAE (lower row) of the fixed point approximation, $\tilde{\mathcal{L}}_\mathrm{mae}^\star$, increases noticeably  across dynamical systems on very degree heterogeneous graphs independent of clustering but stays constant or increases only slightly in the less degree heterogeneous case. This means that even though \model{}s approach stable fixed points, these fixed points can differ from the dynamical system's true fixed point,  especially on very degree heterogeneous graphs.}
    \label{fig:jac_median_mae_noisy}
\end{figure*}

In Fig.~\ref{fig:jac_mean_mae_noisy} and Fig.~\ref{fig:jac_median_mae_noisy}, we show the largest eigenvalue of the Jacobian for \model{}s trained on noisy training data, created by adding independent Gaussian noise with standard deviation reported in Tab.~\ref{tab:dynamics} to the ground truth dynamics. We only display cases in which the \model{} reached a stable fixed point. Overall, the displayed results for both the largest eigenvalue $\hat{\lambda}_1$ and MAE are similar on very degree heterogeneous graphs with weak clustering (Fig.~\ref{fig:jac_mean_mae_noisy}\textbf{(a)}, \textbf{(b)}, Fig.~\ref{fig:jac_median_mae_noisy}\textbf{(a)}, \textbf{(b)}), very degree heterogeneous graphs with strong clustering (Fig.~\ref{fig:jac_mean_mae_noisy}\textbf{(c)}, \textbf{(d)}, Fig.~\ref{fig:jac_median_mae_noisy}\textbf{(c)}, \textbf{(d)}), intermediate degree heterogeneity and clustering (Fig.~\ref{fig:jac_mean_mae_noisy}\textbf{(e)}, \textbf{(f)}, Fig.~\ref{fig:jac_median_mae_noisy}\textbf{(e)}, \textbf{(f)}), less degree heterogeneity and weak clustering (Fig.~\ref{fig:jac_mean_mae_noisy}\textbf{(g)}, \textbf{(h)}, Fig.~\ref{fig:jac_median_mae_noisy}\textbf{(g)}, \textbf{(h)}), and less degree heterogeneity and strong clustering (Fig.~\ref{fig:jac_mean_mae_noisy}\textbf{(i)}, \textbf{(j)}, Fig.~\ref{fig:jac_median_mae_noisy}\textbf{(i)}, \textbf{(j)}).

However, in a few cases, we found that \model{}s trained on the BD model did not reach stable fixed points on large graphs $n^\mathrm{test}=4096$. This was the case for the \model{} trained on a training graph with less degree heterogeneity and weak clustering, $(\gamma^\mathrm{train}, \, \beta^\mathrm{train})=(3.9, \, 0.1)$, for $2$ out of $100$ test graphs; this was also the case for a \model{} trained on a training graph with intermediate degree heterogeneity and clustering,  $(\gamma^\mathrm{train}, \, \beta^\mathrm{train})=(3.0, \, 1.1)$, for $4$ out of $100$ test graphs. This shows that for test graphs much larger than the training graph, noise in the training can negatively impact the ability of \model{}s to reproduce the stability properties of the dynamical system generating the training data.

\clearpage
\FloatBarrier
\section{Additional results on robustness to unobserved nodes}\label{appendix:missing_nodes}

We supplement our discussion of the increase in the mean of the node-wise MAE in the presence of unobserved nodes in Section~\ref{sec:results:missingness} with a discussion of its quantiles. We also show that our results are robust when we add noise to the training data.

\begin{figure*}[tb]
    \centering
    \includegraphics[width=\textwidth]{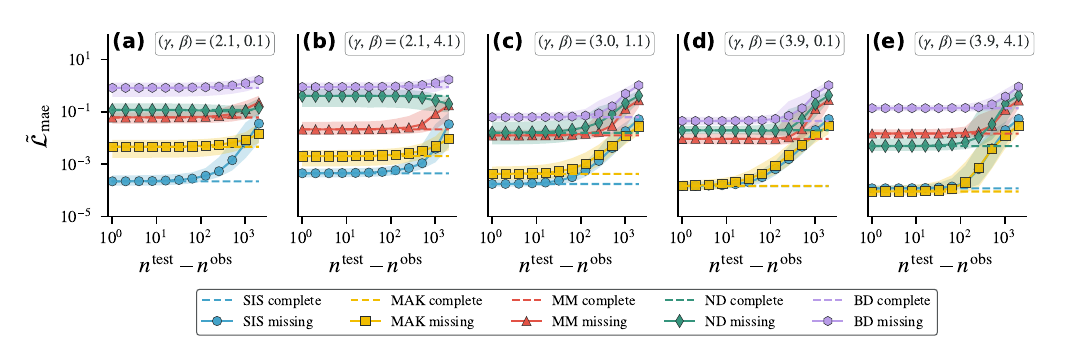}
    \caption{\textbf{Robustness of predictions to unobserved nodes in terms of median MAE.} The robustness of \model{}s trained on small graphs,  $n^\mathrm{train}=64$, to unobserved nodes in larger test graphs, $n^\mathrm{test}=8192$, with the same parameters as the training graph, $(\gamma^\mathrm{test}, \, \beta^\mathrm{test}) =(\gamma^\mathrm{train},\, \beta^\mathrm{train}) =(\gamma, \, \beta)$, depends on the number of observed nodes $n^\mathrm{obs}$, graph properties (\textbf{(a)} $(\gamma, \, \beta)=(2.1, \, 0.1)$, \textbf{(b)} $(\gamma, \, \beta)=(2.1, \, 4.1)$, \textbf{(c)} $(\gamma, \, \beta)=(3.0, \, 1.1)$, \textbf{(d)} $(\gamma, \, \beta)=(3.9, \, 0.1)$, \textbf{(e)} $(\gamma, \, \beta)=(3.9, \, 4.1)$), and the dynamical system (SIS (blue circles), MAK (yellow squares), MM (red triangles), ND (green diamonds), BD (purple hexagons)). Overall, the median node-wise MAE, $\tilde{\mathcal{L}}_\mathrm{mae}$, if $n^\mathrm{test}-n^\mathrm{obs}$ nodes remain unobserved (solid line) stays equivalent to the baseline value for a completely observed graph (dashed line) longer on very degree heterogeneous than on less degree heterogeneous graphs with clustering playing a minor role. However, the MAE tends to be higher in the very degree heterogeneous than the less degree heterogeneous setting. This means robustness to unobserved nodes depends on a complex interplay of the type of dynamical system and graph properties.}
    \label{fig:missing_median_mae_noiseless}
\end{figure*}

In Fig.~\ref{fig:missing_median_mae_noiseless}, we show how the median of the MAE, $\tilde{\mathcal{L}}_\mathrm{mae}$, increases as the number of unobserved nodes $n^\mathrm{test}-n^\mathrm{obs}$ increases. We consider test graphs of size $n^\mathrm{test}=8192$ with the same properties as the training graph, $(\gamma^\mathrm{test}, \, \beta^\mathrm{test})=(\gamma^\mathrm{train}, \, \beta^\mathrm{train})=(\gamma, \, \beta)$.

For all settings of the $\mathbb{S}^1$-model parameters, median MAE begins to rise later than the mean MAE, suggesting that the increase is potentially driven by a decrease in performance on a few nodes rather than a uniform increase in MAE at all nodes. This behavior is particularly noticeable for \model{}s trained on the SIS model and for all dynamical systems in the case of low degree heterogeneity ($(\gamma, \, \beta)=(3.9, \, 0.1)$, Fig.~\ref{fig:missing_median_mae_noiseless}\textbf{(d)}, $(\gamma, \, \beta)=(3.9, \, 4.1)$, Fig.~\ref{fig:missing_median_mae_noiseless}\textbf{(e)}). We also observe that the variability in MAE, as quantified by the interquartile range, tends to increase as soon as the MAE begins to rise for \model{}s trained on the SIS, MAK, and MM model.

\begin{figure*}[tb]
    \centering
    \includegraphics[width=\textwidth]{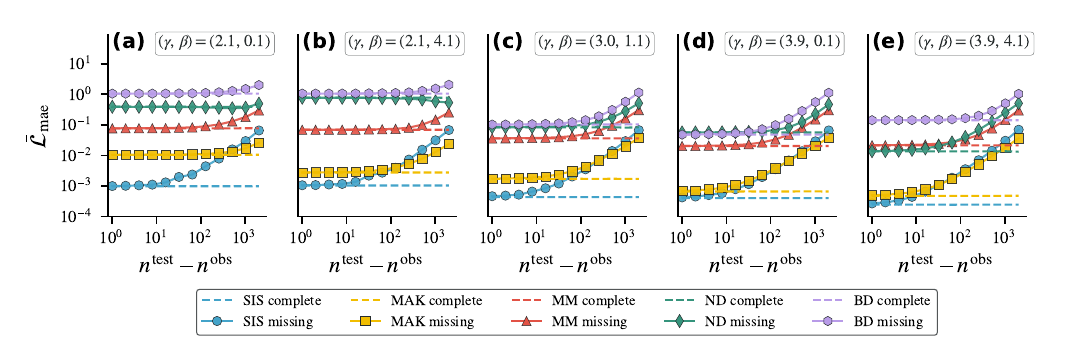}
    \caption{\textbf{Robustness of predictions to unobserved nodes in terms of mean MAE for noisy data.} The robustness of \model{}s trained on small graphs,  $n^\mathrm{train}=64$, to unobserved nodes in larger test graphs, $n^\mathrm{test}=8192$, with the same parameters as the training graph, $(\gamma^\mathrm{test}, \, \beta^\mathrm{test}) =(\gamma^\mathrm{train},\, \beta^\mathrm{train}) =(\gamma, \, \beta)$, depends on the number of observed nodes $n^\mathrm{obs}$, graph properties (\textbf{(a)} $(\gamma, \, \beta)=(2.1, \, 0.1)$, \textbf{(b)} $(\gamma, \, \beta)=(2.1, \, 4.1)$, \textbf{(c)} $(\gamma, \, \beta)=(3.0, \, 1.1)$, \textbf{(d)} $(\gamma, \, \beta)=(3.9, \, 0.1)$, \textbf{(e)} $(\gamma, \, \beta)=(3.9, \, 4.1)$), and the dynamical system (SIS (blue circles), MAK (yellow squares), MM (red triangles), ND (green diamonds), BD (purple hexagons)). Overall, the mean node-wise MAE, $\bar{\mathcal{L}}_\mathrm{mae}$, if $n^\mathrm{test}-n^\mathrm{obs}$ nodes remain unobserved (solid line) stays equivalent to the baseline value for a completely observed graph (dashed line) longer on very degree heterogeneous than on less degree heterogeneous graphs with clustering playing a minor role. However, the MAE tends to be higher in the very degree heterogeneous than the less degree heterogeneous setting. This means robustness to unobserved nodes depends on a complex interplay of the type of dynamical system and graph properties.} 
    \label{fig:missing_mean_mae_noisy}
\end{figure*}

\begin{figure*}[tb]
    \centering
    \includegraphics[width=\textwidth]{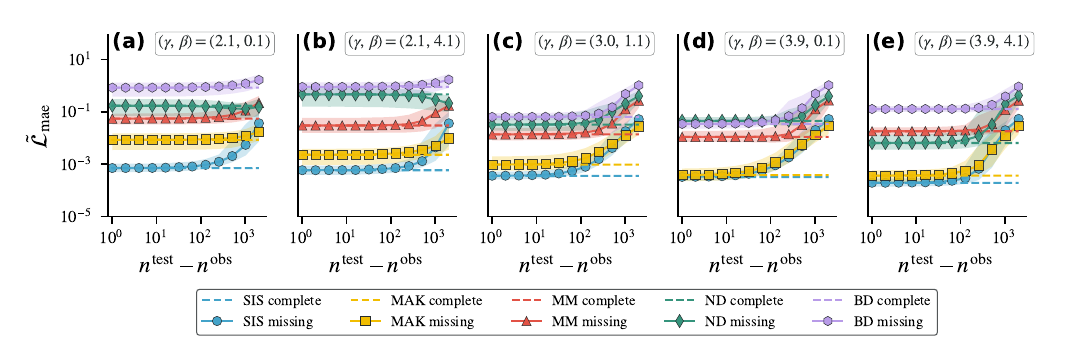}
    \caption{\textbf{Robustness of predictions to unobserved nodes in terms of median MAE for noisy data.} The robustness of \model{}s trained on small graphs,  $n^\mathrm{train}=64$, to unobserved nodes in larger test graphs, $n^\mathrm{test}=8192$, with the same parameters as the training graph, $(\gamma^\mathrm{test}, \, \beta^\mathrm{test}) =(\gamma^\mathrm{train},\, \beta^\mathrm{train}) =(\gamma, \, \beta)$, depends on the number of observed nodes $n^\mathrm{obs}$, graph properties (\textbf{(a)} $(\gamma, \, \beta)=(2.1, \, 0.1)$, \textbf{(b)} $(\gamma, \, \beta)=(2.1, \, 4.1)$, \textbf{(c)} $(\gamma, \, \beta)=(3.0, \, 1.1)$, \textbf{(d)} $(\gamma, \, \beta)=(3.9, \, 0.1)$, \textbf{(e)} $(\gamma, \, \beta)=(3.9, \, 4.1)$), and the dynamical system (SIS (blue circles), MAK (yellow squares), MM (red triangles), ND (green diamonds), BD (purple hexagons)). Overall, the median node-wise MAE, $\tilde{\mathcal{L}}_\mathrm{mae}$, if $n^\mathrm{test}-n^\mathrm{obs}$ nodes remain unobserved (solid line) stays equivalent to the baseline value for a completely observed graph (dashed line) longer on very degree heterogeneous than on less degree heterogeneous graphs with clustering playing a minor role. However, the MAE tends to be higher in the very degree heterogeneous than the less degree heterogeneous setting. This means robustness to unobserved nodes depends on a complex interplay of the type of dynamical system and graph properties.} 
    \label{fig:missing_median_mae_noisy}
\end{figure*}

If we compare the behavior of the MAE's mean and its quantiles in the case of training data without noise (Fig.~\ref{fig:missing_mean_mae_noiseless} and Fig.~\ref{fig:missing_median_mae_noiseless}, respectively) to their counterparts when independent Gaussian noise with standard deviation reported in Tab.~\ref{tab:dynamics} was added to the training data (Fig.~\ref{fig:missing_mean_mae_noisy} and Fig.~\ref{fig:missing_median_mae_noisy}, respectively), we find no major differences. 

\clearpage
\FloatBarrier
\section{Visualization of individual time series}
\label{appendix:timeseries}

To provide a more intuitive understanding of the quality of the \model{} predictions, we visualize time series and predictions for $n^\mathrm{test}_\mathrm{init}=1$ initial condition on $n_G^\mathrm{test}=1$ test graph with either the same size, $n^\mathrm{test}=64$, or much larger size, $n^\mathrm{test}=8192$, than the training graph. We use test graphs with the same $\mathbb{S}^1$-model parameters as the training graph, $(\gamma^\mathrm{test}, \, \beta^\mathrm{test})=(\gamma^\mathrm{train}, \, \beta^\mathrm{train})=(\gamma, \, \beta)$, and discuss both noise free training data and training data with additive Gaussian noise with standard deviation given in Tab.~\ref{tab:dynamics}.
To avoid cluttered visualizations and still show the full breadth of possible node-level performance, we sort time series by node-wise MAE and visualize every $q$th time series such that $n^\mathrm{test}/q=16$.

\begin{figure*}[tb]
    \centering
    \includegraphics[width=0.95\textwidth]{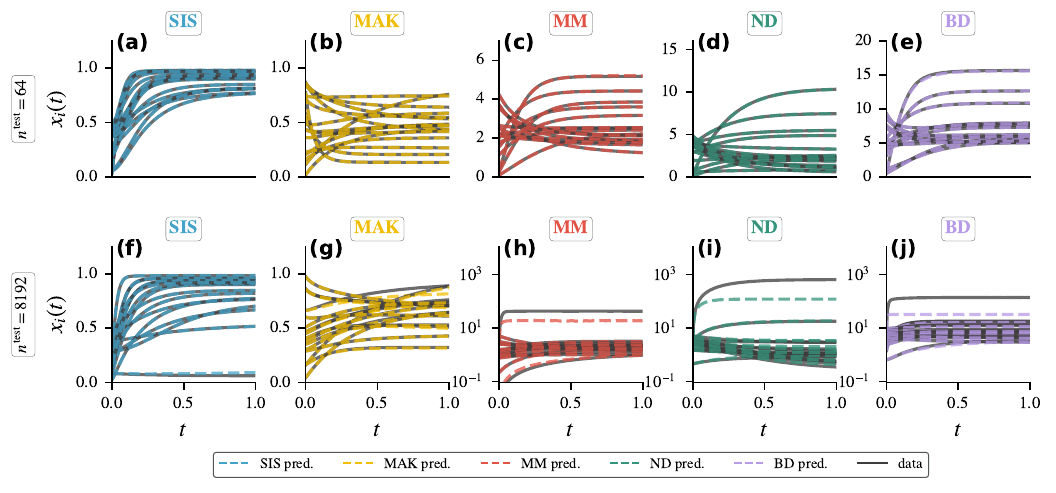}
    \caption{\textbf{Predictions on very degree heterogeneous graphs with weak clustering, $(\gamma, \, \beta)=(2.1, \, 0.1)$}. Predictions (dashed lines) and test data (solid line) agree well on graphs of the same size, $n^\mathrm{test}=64$, as the training graph for the \textbf{(a)} SIS, \textbf{(b)} MAK, \textbf{(c)} MM, \textbf{(d)} ND, and \textbf{(e)} BD model. On graphs larger, $n^\mathrm{test}=8192$, than the training graph, visual agreement is excellent for the \textbf{(f)} SIS model, mostly good for the \textbf{(g)} MAK and \textbf{(h)} MM model, less good for the \textbf{(i)} ND, and \textbf{(j)} BD model.}
    \label{fig:timeseries_2101}
\end{figure*}

\begin{figure*}[tb]
    \centering
    \includegraphics[width=0.95\textwidth]{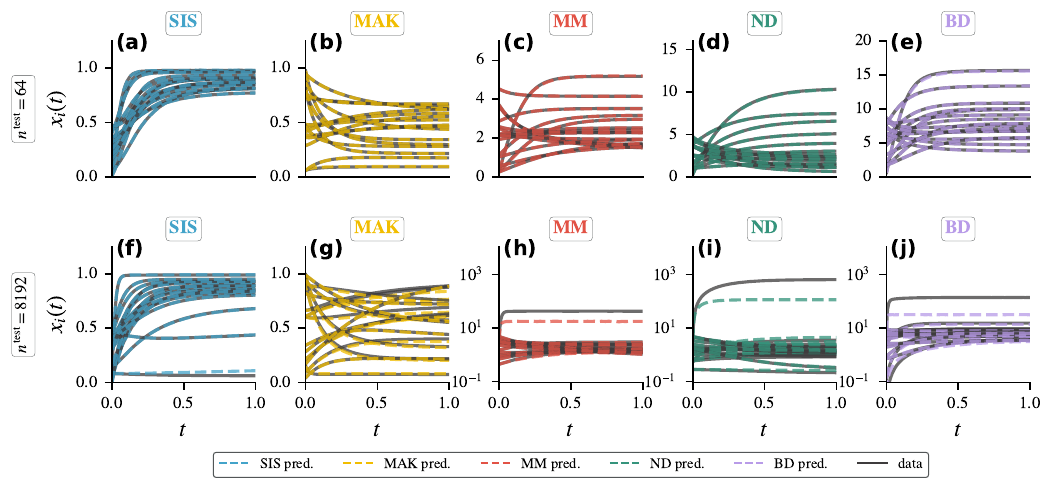}
    \caption{\textbf{Robustness to noise for predictions on very degree heterogeneous graphs with weak clustering, $(\gamma, \, \beta)=(2.1, \, 0.1)$}. Predictions (dashed lines) and test data (solid line) agree well on graphs of the same size, $n^\mathrm{test}=64$, as the training graph for the \textbf{(a)} SIS, \textbf{(b)} MAK, \textbf{(c)} MM, \textbf{(d)} ND, and \textbf{(e)} BD model. On graphs larger, $n^\mathrm{test}=8192$, than the training graph, visual agreement is excellent for the \textbf{(f)} SIS model, mostly good for the \textbf{(g)} MAK and \textbf{(h)} MM model, less good for the \textbf{(i)} ND, and \textbf{(j)} BD model---in agreement with the noiseless case Fig.~\ref{fig:timeseries_2101}.}
    \label{fig:timeseries_2101_noisy}
\end{figure*}

In the case of graphs with high degree heterogeneity and weak clustering (Fig.~\ref{fig:timeseries_2101}, $(\gamma, \, \beta)=(2.1, \, 0.1)$), we see excellent agreement between prediction and test data on test graphs of the same size as the training graph across dynamical systems (Fig~\ref{fig:timeseries_2101}\textbf{(a)}-\textbf{(e)}). If the size of the test graph increases we see predictions of different quality for different dynamical systems: 
For the \model{} trained on the SIS model (Fig.~\ref{fig:timeseries_2101}\textbf{(f)}) the visual agreement between test data and prediction is still high. We observe few deviations for \model{}s trained on the MAK (Fig.~\ref{fig:timeseries_2101}\textbf{(g)}). However, for high activity nodes we see substantial deviations of \model{} predictions and test data for the MM model (Fig.~\ref{fig:timeseries_2101}\textbf{(h)}), ND model (Fig.~\ref{fig:timeseries_2101}\textbf{(i)}) and BD model (Fig.~\ref{fig:timeseries_2101}\textbf{(j)}).
We observe similar patterns if training data is noisy (Fig.~\ref{fig:timeseries_2101_noisy}).

\begin{figure*}[tb]
    \centering
    \includegraphics[width=0.95\textwidth]{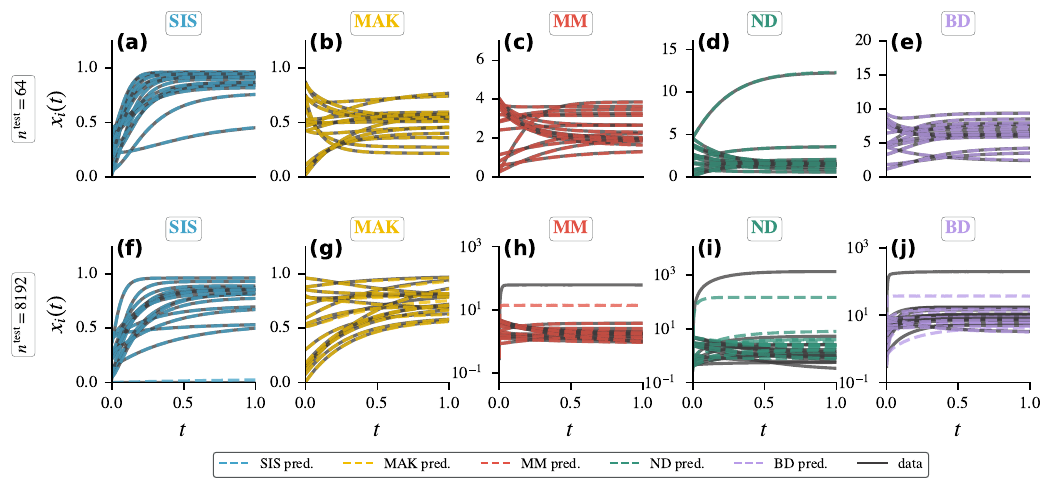}
    \caption{\textbf{Predictions on very degree heterogeneous graphs with strong clustering, $(\gamma, \, \beta)=(2.1, \, 4.1)$}. Predictions (dashed lines) and test data (solid line) agree well on graphs of the same size, $n^\mathrm{test}=64$, as the training graph for the \textbf{(a)} SIS, \textbf{(b)} MAK, \textbf{(c)} MM, \textbf{(d)} ND, and \textbf{(e)} BD model. On graphs larger, $n^\mathrm{test}=8192$, than the training graph, visual agreement is good for the \textbf{(f)} SIS model and \textbf{(g)} MAK model, but less good for the \textbf{(h)} MM model, \textbf{(i)} ND model, and \textbf{(j)} BD model.}
    \label{fig:timeseries_2141}
\end{figure*}

\begin{figure*}[tb]
    \centering
    \includegraphics[width=0.95\textwidth]{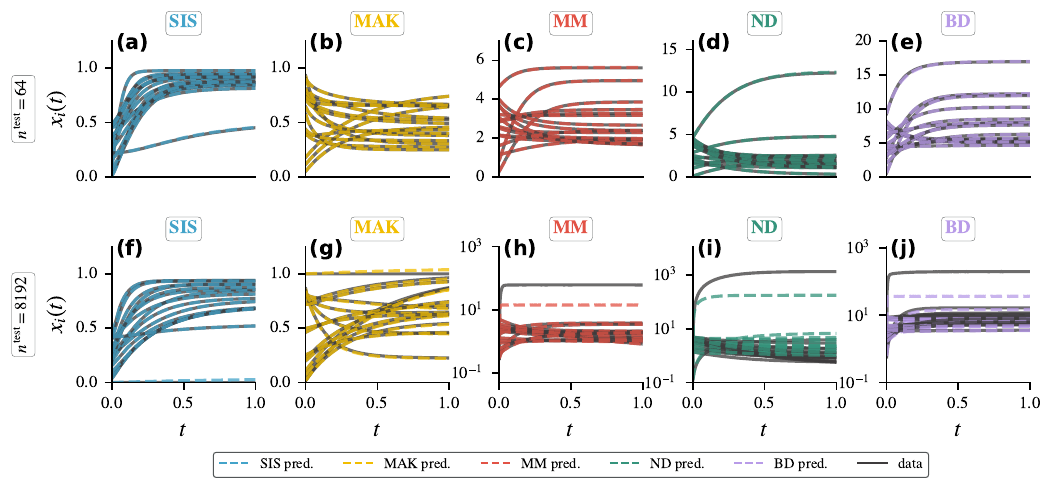}
    \caption{\textbf{Robustness to noise for predictions on very degree heterogeneous graphs with weak clustering, $(\gamma, \, \beta)=(2.1, \, 4.1)$}. Predictions (dashed lines) and test data (solid line) agree well on graphs of the same size, $n^\mathrm{test}=64$, as the training graph for the \textbf{(a)} SIS, \textbf{(b)} MAK, \textbf{(c)} MM, \textbf{(d)} ND, and \textbf{(e)} BD model. On graphs larger, $n^\mathrm{test}=8192$, than the training graph, visual agreement is good for the \textbf{(f)} SIS model and \textbf{(g)} MAK model, but less good for the \textbf{(h)} MM model, \textbf{(i)} ND model, and \textbf{(j)} BD model---in agreement with the noiseless case Fig.~\ref{fig:timeseries_2141}.}
    \label{fig:timeseries_2141_noisy}
\end{figure*}

If graphs have high degree heterogeneity and strong clustering (Fig.~\ref{fig:timeseries_2141}, $(\gamma, \, \beta)=(2.1, \, 4.1)$), we observe overall similar patterns as in the case of weak clustering on graphs of the same size as the training graph: Across dynamical systems (Fig~\ref{fig:timeseries_2141}\textbf{(a)}-\textbf{(e)}) visual agreement between prediction and test data is very good. We observe low discrepancy even on larger graphs for the \model{}s trained on the SIS (Fig.~\ref{fig:timeseries_2141}\textbf{(f)}), MAK (Fig.~\ref{fig:timeseries_2141}\textbf{(g)}). For \model{}s trained on the MM model (Fig.~\ref{fig:timeseries_2141}\textbf{(h)}),  ND model (Fig.~\ref{fig:timeseries_2141}\textbf{(i)}) and BD model (Fig.~\ref{fig:timeseries_2141}\textbf{(j)}) predictions and test data disagree, especially at high activity nodes.
These patterns persist in the setting with noisy training data (Fig.~\ref{fig:timeseries_2141_noisy}), i.e., noise neither remedies nor aggravates deviations.

\begin{figure*}[tb]
    \centering
    \includegraphics[width=0.95\textwidth]{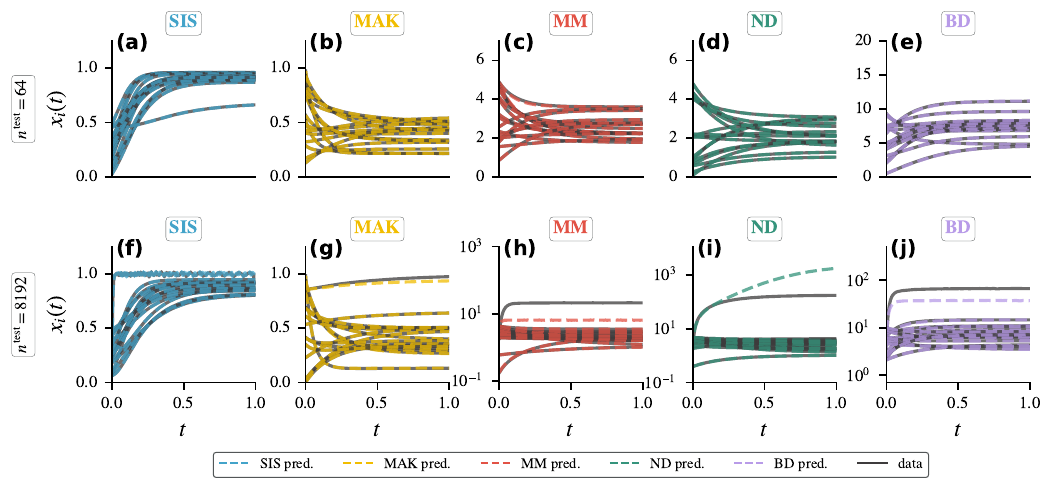}
    \caption{\textbf{Predictions on graphs with moderate degree heterogeneity and clustering, $(\gamma, \, \beta)=(3.0, \, 1.1)$}. Predictions (dashed lines) and test data (solid line) agree well on graphs of the same size, $n^\mathrm{test}=64$, as the training graph for the \textbf{(a)} SIS, \textbf{(b)} MAK, \textbf{(c)} MM, \textbf{(d)} ND, and \textbf{(e)} BD model. On graphs larger, $n^\mathrm{test}=8192$, than the training graph, visual agreement is good for the \textbf{(f)} SIS model and \textbf{(g)} MAK model, but less good for the \textbf{(h)} MM model, \textbf{(i)} ND model, and \textbf{(j)} BD model.}
    \label{fig:timeseries_3011}
\end{figure*}

\begin{figure*}[tb]
    \centering
    \includegraphics[width=0.95\textwidth]{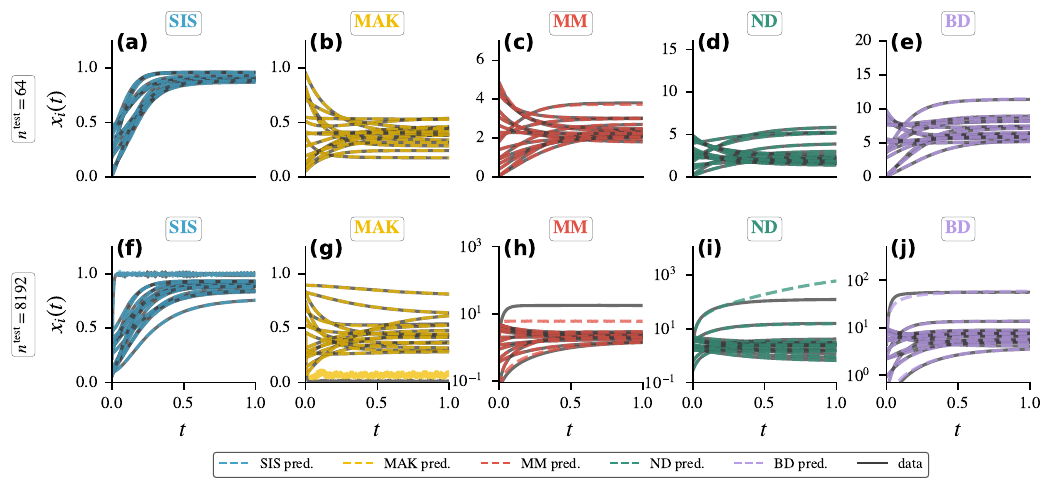}
    \caption{\textbf{Robustness to noise for predictions on graphs with moderate degree heterogeneity and clustering, $(\gamma, \, \beta)=(3.0, \, 1.1)$}. Predictions (dashed lines) and test data (solid line) agree well on graphs of the same size, $n^\mathrm{test}=64$, as the training graph for the \textbf{(a)} SIS, \textbf{(b)} MAK, \textbf{(c)} MM, \textbf{(d)} ND, and \textbf{(e)} BD model. On graphs larger, $n^\mathrm{test}=8192$, than the training graph, visual agreement is good for the \textbf{(f)} SIS model and \textbf{(g)} MAK model, but less good for the \textbf{(h)} MM model, \textbf{(i)} ND model, and \textbf{(j)} BD model---in agreement with the noiseless case Fig.~\ref{fig:timeseries_3011}.}
    \label{fig:timeseries_3011_noisy}
\end{figure*}

In the case of moderate degree heterogeneity and clustering (Fig.~\ref{fig:timeseries_3011}), $(\gamma, \, \beta)=(3.0, \, 1.1)$, there is good agreement between test data and prediction on graphs of the same size as the training graph (Fig.~\ref{fig:timeseries_3011}\textbf{(a)}-\textbf{(e)}). On graphs larger than the training graph, we see that predictions and test data agree for the SIS model (Fig.~\ref{fig:timeseries_3011}\textbf{(f)}) and MAK model (Fig.~\ref{fig:timeseries_3011}\textbf{(g)}), apart from minor deviations at very high activity. These deviations are worse for the MM model (Fig.~\ref{fig:timeseries_3011}\textbf{(h)}). For the ND model (Fig.~\ref{fig:timeseries_3011}\textbf{(i)}), and BD model (Fig.~\ref{fig:timeseries_3011}\textbf{(j)}) substantial deviations can occur, mostly if node activity is high. The same patterns arise if \model{}s are trained on noisy training data (Fig.~\ref{fig:timeseries_3011_noisy}). However, we notice additionally mismatch for very low activity nodes in the MAK model in the case of graphs that are much larger than the training graph (Fig.~\ref{fig:timeseries_3011_noisy}\textbf{(g)}).

\begin{figure*}[tb]
    \centering
    \includegraphics[width=0.95\textwidth]{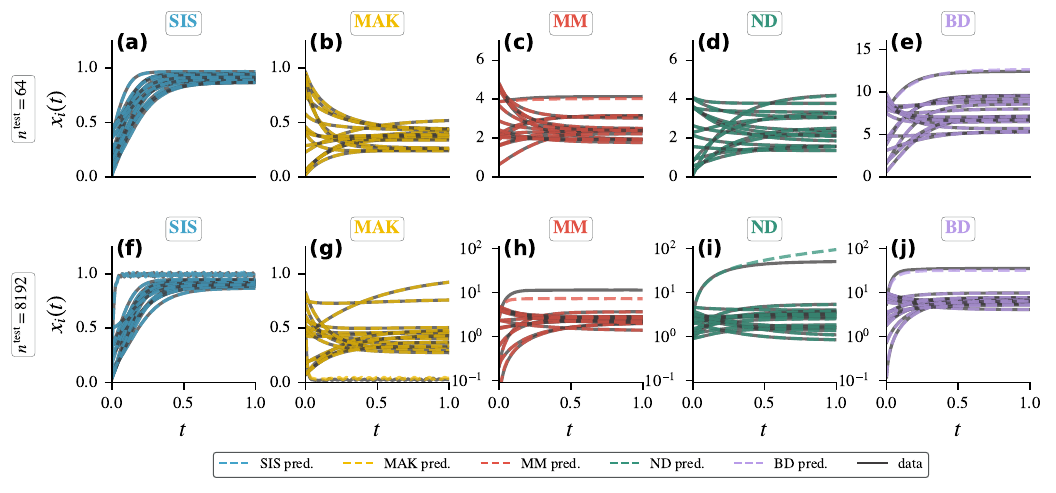}
    \caption{\textbf{Predictions on less degree heterogeneous graphs with weak clustering, $(\gamma, \, \beta)=(3.9, \, 0.1)$}. Predictions (dashed lines) and test data (solid line) agree well on graphs of the same size, $n^\mathrm{test}=64$, as the training graph for the \textbf{(a)} SIS, \textbf{(b)} MAK, \textbf{(c)} MM, \textbf{(d)} ND, and \textbf{(e)} BD model. On graphs larger, $n^\mathrm{test}=8192$, than the training graph, visual agreement is good for the \textbf{(f)} SIS model, \textbf{(g)} MAK model, and \textbf{(h)} MM model, but deviations are observed at high activity nodes for the \textbf{(i)} ND model and \textbf{(j)} BD model.}
    \label{fig:timeseries_3901}
\end{figure*}

\begin{figure*}[tb]
    \centering
    \includegraphics[width=0.95\textwidth]{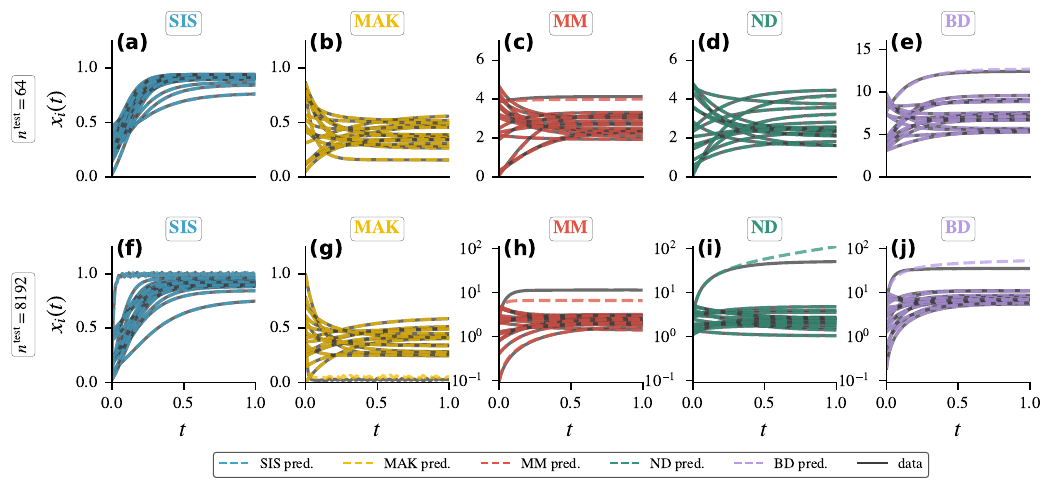}
    \caption{\textbf{Robustness to noise for predictions on less degree heterogeneous graphs with weak clustering, $(\gamma, \, \beta)=(3.9, \, 0.1)$}. Predictions (dashed lines) and test data (solid line) agree well on graphs of the same size, $n^\mathrm{test}=64$, as the training graph for the \textbf{(a)} SIS, \textbf{(b)} MAK, \textbf{(c)} MM, \textbf{(d)} ND, and \textbf{(e)} BD model. On graphs larger, $n^\mathrm{test}=8192$, than the training graph, visual agreement is good for the \textbf{(f)} SIS model and \textbf{(g)} MAK model, but less good for the \textbf{(h)} MM model, \textbf{(i)} ND model, and \textbf{(j)} BD model---in agreement with the noiseless case Fig.~\ref{fig:timeseries_3901}.}
    \label{fig:timeseries_3901_noisy}
\end{figure*}

If graphs are less degree heterogeneous and have weak clustering, $(\gamma, \, \beta) = (3.9, \, 0.1)$, predictions and test data exhibit very good visual agreement on small graphs across dynamical systems (Fig.~\ref{fig:timeseries_3901}\textbf{(a)}-\textbf{(e)}), which persists on larger graphs for the SIS model (Fig.~\ref{fig:timeseries_3901}\textbf{(f)}), MAK model (Fig.~\ref{fig:timeseries_3901}\textbf{(g)}), and MM model (Fig.~\ref{fig:timeseries_3901}\textbf{(h)}). While we still observe deviations on high activity nodes for the ND model (Fig.~\ref{fig:timeseries_3901}\textbf{(i)}) and BD model (Fig.~\ref{fig:timeseries_3901}\textbf{(j)}), they are less severe than on more degree heterogeneous graphs. While we make overall similar observations if training data is corrupted with Gaussian noise (Fig.~\ref{fig:timeseries_3901_noisy}), we note that deviations at very low and high activity nodes on large graphs are slightly stronger in the case of the SIS model (Fig.~\ref{fig:timeseries_3901_noisy}\textbf{(f)}) and MAK model (Fig.~\ref{fig:timeseries_3901_noisy}\textbf{(g)}).

\begin{figure*}[tb]
    \centering
    \includegraphics[width=0.95\textwidth]{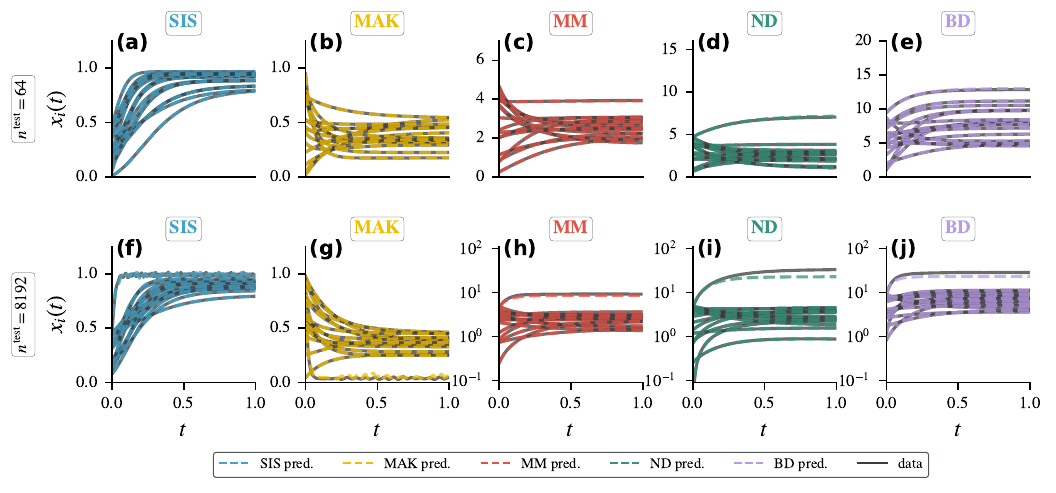}
    \caption{\textbf{Predictions on less degree heterogeneous graphs with strong clustering, $(\gamma, \, \beta)=(3.9, \, 4.1)$}. Predictions (dashed lines) and test data (solid line) agree well on graphs of the same size, $n^\mathrm{test}=64$, as the training graph for the \textbf{(a)} SIS, \textbf{(b)} MAK, \textbf{(c)} MM, \textbf{(d)} ND, and \textbf{(e)} BD model. On graphs larger, $n^\mathrm{test}=8192$, than the training graph, visual agreement is good for the \textbf{(f)} SIS model, \textbf{(g)} MAK model, and \textbf{(h)} MM model, but deviations are observed at high activity nodes for the \textbf{(i)} ND model and \textbf{(j)} BD model.}
    \label{fig:timeseries_3941}
\end{figure*}

\begin{figure*}[tb]
    \centering
    \includegraphics[width=0.95\textwidth]{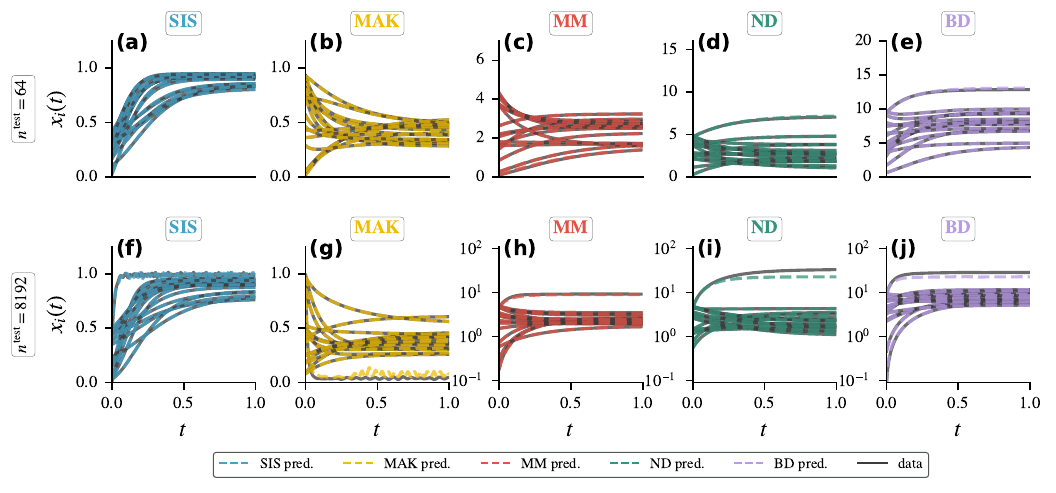}
    \caption{\textbf{Robustness to noise for predictions on less degree heterogeneous graphs with strong clustering, $(\gamma, \, \beta)=(3.9, \, 4.1)$}. Predictions (dashed lines) and test data (solid line) agree well on graphs of the same size, $n^\mathrm{test}=64$, as the training graph for the \textbf{(a)} SIS, \textbf{(b)} MAK, \textbf{(c)} MM, \textbf{(d)} ND, and \textbf{(e)} BD model. On graphs larger, $n^\mathrm{test}=8192$, than the training graph, visual agreement is good for the \textbf{(f)} SIS model and \textbf{(g)} MAK model, but less good for the \textbf{(h)} MM model, \textbf{(i)} ND model, and \textbf{(j)} BD model---in agreement with the noiseless case Fig.~\ref{fig:timeseries_3941}.}
    \label{fig:timeseries_3941_noisy}
\end{figure*}

Finally, for the case of less degree heterogeneous graphs with strong clustering, $(\gamma, \, \beta)=(3.9, \, 4.1)$, we find that test data and predictions are visually similar on small graphs irrespective of the dynamical system (Fig.~\ref{fig:timeseries_3941}\textbf{(a)}-\textbf{(e)}). If test graphs are much larger than the training graph, we find a good visual agreement between test data and predictions for the SIS model (Fig.~\ref{fig:timeseries_3941}\textbf{(f)}), MAK model (Fig.~\ref{fig:timeseries_3941}\textbf{(g)}), and MM model (Fig.~\ref{fig:timeseries_3941}\textbf{(h)}). For the ND model, predictions disagree at high activity nodes (Fig.~\ref{fig:timeseries_3941}\textbf{(i)}). The same is true for the BD model (Fig.~\ref{fig:timeseries_3941}\textbf{(j)}) but the mismatch is smaller. We observe the same patterns on noisy training data (Fig.~\ref{fig:timeseries_3941_noisy}). We find that in the noisy case deviations at very low activity nodes are larger for the MAK model on large graphs (Fig.~\ref{fig:timeseries_3941_noisy}\textbf{(g)}).

\clearpage
\FloatBarrier
\section{Visualization of the learned vector field}
\label{appendix:vectorfield}

Dynamical systems in BB form parameterize a high dimensional vector field in terms of three functions, $f, \, h^\mathrm{ego}, \, h^\mathrm{alt}$, between real numbers. As our \model{}s mirror this form, we can visually inspect the learned vector field and compare it to the analytical form. 

\subsection{Methods}

Recall that the interaction term in dynamical systems of BB form (Eq.~\eqref{eq:bb-dynamics}), as well as in our \model{}s (Eq.~\eqref{eq:nODE}) factorizes. However, the way the interaction term factorizes is not unique. This means even for a \model{} that is perfectly fit to data, it is not guaranteed that $h^\mathrm{ego}_\omega(x)=h^\mathrm{ego}(x)$ and $h^\mathrm{alt}_\omega(x)=h^\mathrm{alt}(x)$ but only that $h^\mathrm{ego}_\omega(x)h^\mathrm{alt}_\omega(x)=h^\mathrm{ego}(x)h^\mathrm{alt}(x)$. We thus introduce a scalar $\phi\neq 0$, and rely on numerical optimization via Brent's method as implemented in \texttt{scipy.optimize}~\cite{virtanen2020_scipy10fundamental} to find
\begin{equation}    
    \phi^\star
    =
    \underset{\phi \in\mathbb{R}, \, \phi\neq 0}{\arg\min}
    \sum_{x\in \mathcal{X}}
    \left(
    \frac{
     \frac{1}{\phi} h^\mathrm{ego}_\omega(x) - h^\mathrm{ego}(x)}{h^\mathrm{ego}(x)}
    \right)^2
    +
     \left(
     \frac{\phi h^\mathrm{alt}_\omega(x) - h^\mathrm{alt}(x)}{h^\mathrm{alt}(x)}
     \right)^2
\end{equation}
where $\mathcal{X}$ are $1000$ equally spaced points in the interval of initial conditions $[\tilde{x}_\mathrm{min}, \tilde{x}_\mathrm{max}]$ as detailed in Tab.~\ref{tab:dynamics}.

In the following, we visualize $\frac{1}{\phi^\star} h_\omega^\mathrm{ego}$ and $\phi^\star h_\omega^\mathrm{alt}$ but refer to these quantities as $h_\omega^\mathrm{alt}$ and $h_\omega^\mathrm{ego}$ respectively. We emphasize that this form is only used for the purpose of visualization, and not during quantitative analyses. It is also important to keep in mind when interpreting the figures in this Appendix that joint but opposite deviations in $h_\omega^\mathrm{ego}(x)$ and $h_\omega^\mathrm{alt}(x)$ from $h^\mathrm{ego}(x)$ and $h^\mathrm{alt}(x)$ respectively might cancel out when making predictions, because \model{} predictions depend on the product $h_\omega^\mathrm{ego}(x_i)h_\omega^\mathrm{alt}(x_j)$.

\subsection{Results}

\begin{figure*}[tb]
    \centering
    \includegraphics[width=0.825\textwidth]{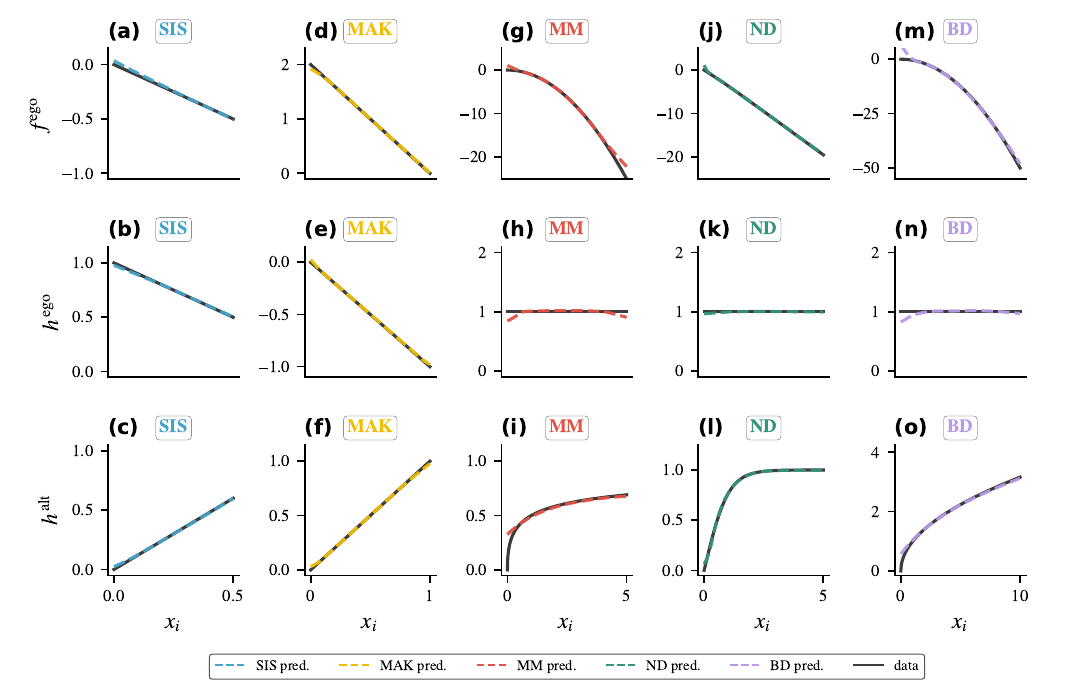}
    \caption{\textbf{Comparison of analytical and learned vector field, } $(\gamma^\mathrm{train}, \, \beta^\mathrm
    {train})=(2.1, \, 0.1)$. A comparison of the analytical (solid line) and learned (dashed line) form of the self-dynamics $f$ (upper row), ego contribution $h^\mathrm{ego}$ (middle row), and alter contribution $h^\mathrm{alt}$ (lower row) across dynamical systems (SIS \textbf{(a)}, \textbf{(b)}, \textbf{(c)}, MAK \textbf{(d)}, \textbf{(e)}, \textbf{(f)}, MM \textbf{(g)}, \textbf{(h)}, \textbf{(i)}, ND \textbf{(j)}, \textbf{(k)}, \textbf{(l)}, BD \textbf{(m)}, \textbf{(n)}, \textbf{(o)} model) shows only minor differences suggesting that \model{}s are well trained.}
    \label{fig:vec_2101}
\end{figure*}

\begin{figure*}[tb]
    \centering
    \includegraphics[width=0.825\textwidth]{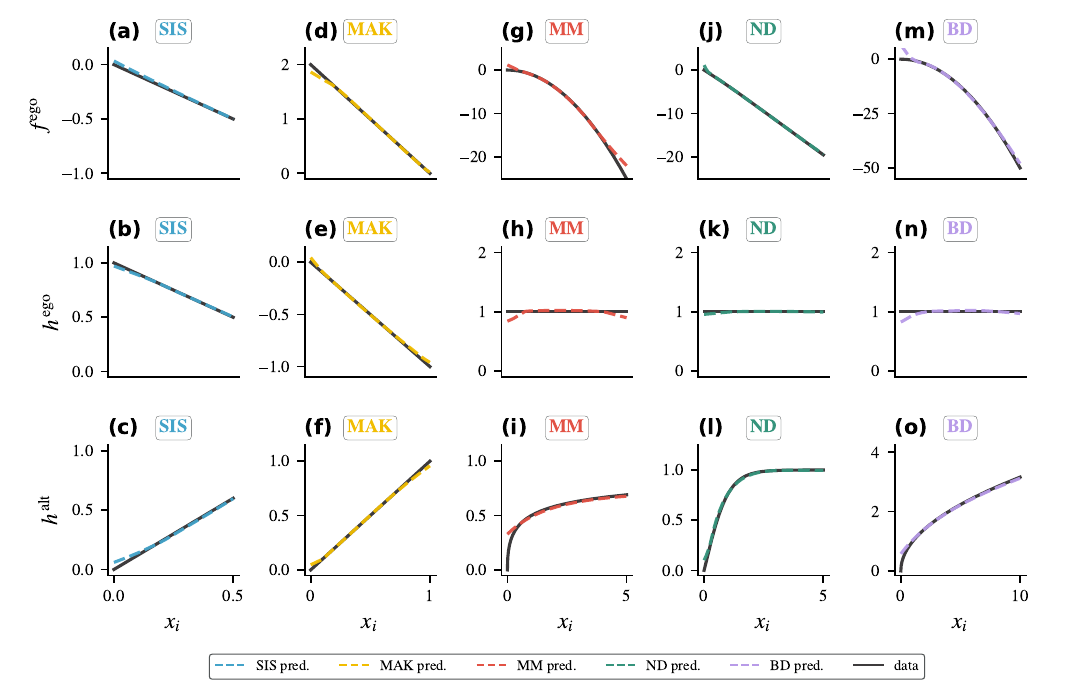}
    \caption{\textbf{Robustness of the comparison of analytical and learned vector field to noise, } $(\gamma^\mathrm{train}, \, \beta^\mathrm
    {train})=(2.1, \, 0.1)$. A comparison of the analytical (solid line) and learned (dashed line) form of the self-dynamics $f$ (upper row), ego contribution $h^\mathrm{ego}$ (middle row), and alter contribution $h^\mathrm{alt}$ (lower row) across dynamical systems (SIS \textbf{(a)}, \textbf{(b)}, \textbf{(c)}, MAK \textbf{(d)}, \textbf{(e)}, \textbf{(f)}, MM \textbf{(g)}, \textbf{(h)}, \textbf{(i)}, ND \textbf{(j)}, \textbf{(k)}, \textbf{(l)}, BD \textbf{(m)}, \textbf{(n)}, \textbf{(o)} model) shows only minor differences suggesting that \model{}s are well trained even in the presence of noise.}
    \label{fig:vec_2101_noisy}
\end{figure*}

In Fig.~\ref{fig:vec_2101}, we show the vector field for different dynamical systems and corresponding \model{}s for the case of training graphs that are very degree heterogeneous and have weak clustering, $(\gamma^\mathrm{train}, \, \beta^\mathrm{
train})=(2.1, \, 0.1)$. In case of the SIS model we observe good agreement between analytical and learned vector field for the self-dynamics $f$ (Fig.~\ref{fig:vec_2101}\textbf{(a)}), ego node's contribution $h^\mathrm
{ego}$ (Fig.~\ref{fig:vec_2101}\textbf{(b)}) and alter contribution $h^\mathrm{alt}$ (Fig.~\ref{fig:vec_2101}\textbf{(c)}) to the interaction term. The same is true in the case of the MAK model (Fig.~\ref{fig:vec_2101}\textbf{(d)}-\textbf{(f)}). For \model{}s trained on the MAK model (Fig.~\ref{fig:vec_2101}\textbf{(g)}-\textbf{(i)}), we notice slight deviations at the edges of the interval. In case of the ND model (Fig.~\ref{fig:vec_2101}\textbf{(j)}-\textbf{(l)}) agreement is overall good. For \model{}s trained on the BD model (Fig.~\ref{fig:vec_2101}\textbf{(m)}-\textbf{(o)}) minor differences are visible for small values of $x_i$.
Overall, these patterns also hold if training data is corrupted by independent Gaussian noise with standard deviation described in Tab.~\ref{tab:dynamics} (Fig.~\ref{fig:vec_2101_noisy}). Only in the case of \model{}s trained on the SIS model we observe slightly larger mismatch in $h^\mathrm{alt}$ (Fig.~\ref{fig:vec_2101_noisy}\textbf{(c)}).

\begin{figure*}[tb]
    \centering
    \includegraphics[width=0.825\textwidth]{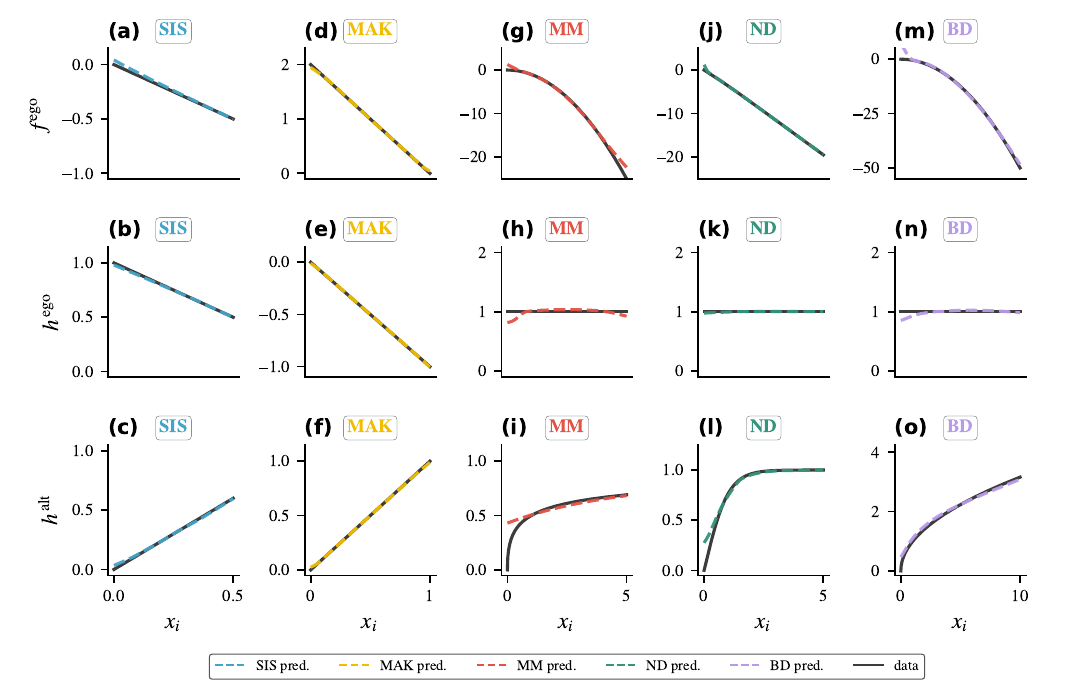}
    \caption{\textbf{Comparison of analytical and learned vector field, } $(\gamma^\mathrm{train}, \, \beta^\mathrm
    {train})=(2.1, \, 4.1)$. A comparison of the analytical (solid line) and learned (dashed line) form of the self-dynamics $f$ (upper row), ego contribution $h^\mathrm{ego}$ (middle row), and alter contribution $h^\mathrm{alt}$ (lower row) across dynamical systems (SIS \textbf{(a)}, \textbf{(b)}, \textbf{(c)}, MAK \textbf{(d)}, \textbf{(e)}, \textbf{(f)}, MM \textbf{(g)}, \textbf{(h)}, \textbf{(i)}, ND \textbf{(j)}, \textbf{(k)}, \textbf{(l)}, BD \textbf{(m)}, \textbf{(n)}, \textbf{(o)} model) shows only minor differences suggesting that \model{}s are well trained.}
    \label{fig:vec_2141}
\end{figure*}

\begin{figure*}[tb]
    \centering
    \includegraphics[width=0.825\textwidth]{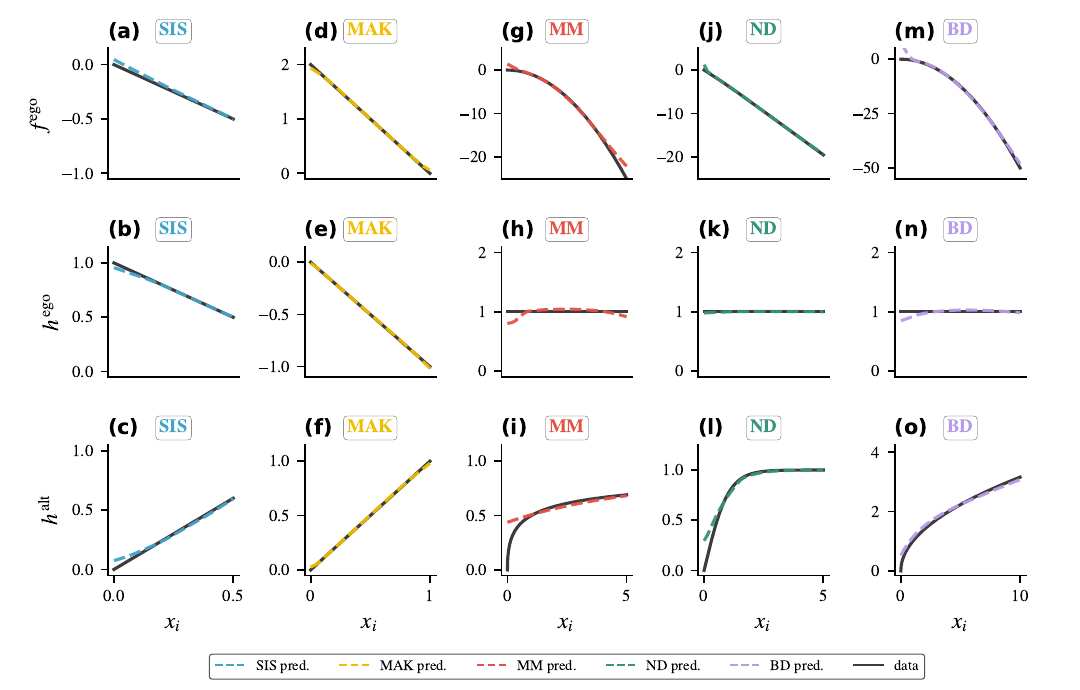}
    \caption{\textbf{Robustness of the comparison of analytical and learned vector field to noise, } $(\gamma^\mathrm{train}, \, \beta^\mathrm
    {train})=(2.1, \, 4.1)$. A comparison of the analytical (solid line) and learned (dashed line) form of the self-dynamics $f$ (upper row), ego contribution $h^\mathrm{ego}$ (middle row), and alter contribution $h^\mathrm{alt}$ (lower row) across dynamical systems (SIS \textbf{(a)}, \textbf{(b)}, \textbf{(c)}, MAK \textbf{(d)}, \textbf{(e)}, \textbf{(f)}, MM \textbf{(g)}, \textbf{(h)}, \textbf{(i)}, ND \textbf{(j)}, \textbf{(k)}, \textbf{(l)}, BD \textbf{(m)}, \textbf{(n)}, \textbf{(o)} model) shows only minor differences suggesting that \model{}s are well trained even in the presence of noise.}
    \label{fig:vec_2141_noisy}
\end{figure*}

In the case of \model{}s trained on very degree heterogeneous graphs with strong clustering, $(\gamma^\mathrm{train}, \, \beta^\mathrm{train})=(2.1, \, 4.1)$, our observation in both the case of noiseless training data (Fig.~\ref{fig:vec_2141}) and noisy training data (Fig.~\ref{fig:vec_2141_noisy}) mirror those on graphs with high degree heterogeneity and weak clustering. In the case of the MM model and the ND model deviations in $h^\mathrm{alt}$ (Fig.~\ref{fig:vec_2141}\textbf{(i)}, \textbf{(l)}) are slightly larger at small values of $x_i$ than in the case of weak clustering.

\begin{figure*}[tb]
    \centering
    \includegraphics[width=0.825\textwidth]{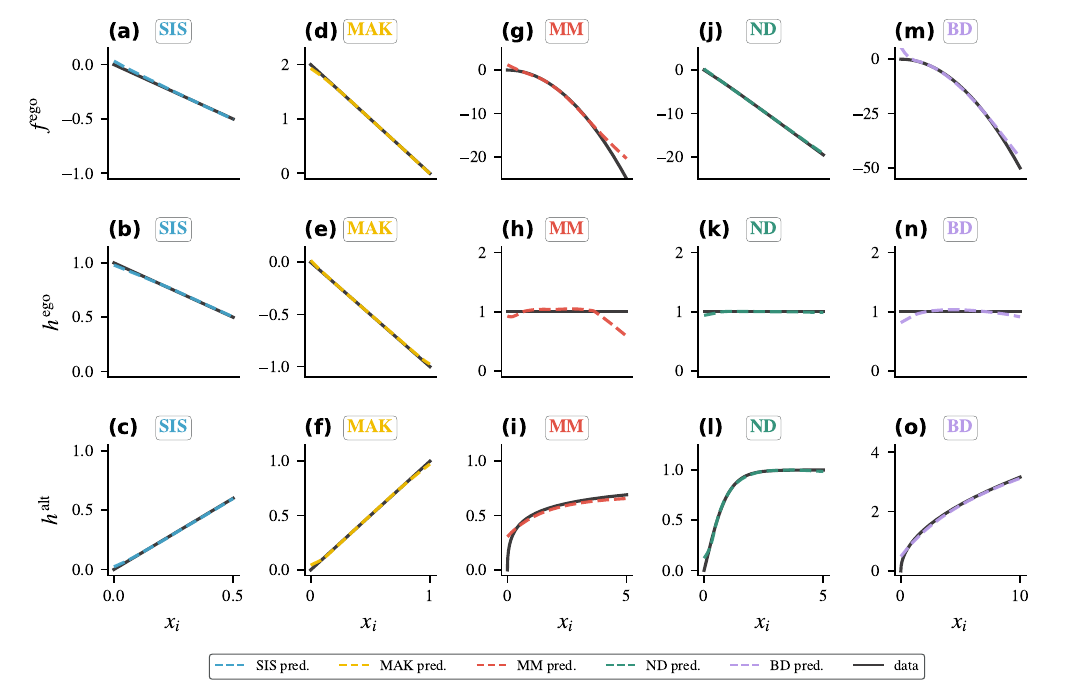}
    \caption{\textbf{Comparison of analytical and learned vector field, } $(\gamma^\mathrm{train}, \, \beta^\mathrm
    {train})=(3.0, \, 1.1)$. A comparison of the analytical (solid line) and learned (dashed line) form of the self-dynamics $f$ (upper row), ego contribution $h^\mathrm{ego}$ (middle row), and alter contribution $h^\mathrm{alt}$ (lower row) across dynamical systems (SIS \textbf{(a)}, \textbf{(b)}, \textbf{(c)}, MAK \textbf{(d)}, \textbf{(e)}, \textbf{(f)}, MM \textbf{(g)}, \textbf{(h)}, \textbf{(i)}, ND \textbf{(j)}, \textbf{(k)}, \textbf{(l)}, BD \textbf{(m)}, \textbf{(n)}, \textbf{(o)} model) shows only minor differences suggesting that \model{}s are well trained.}
    \label{fig:vec_3011}
\end{figure*}

\begin{figure*}[tb]
    \centering
    \includegraphics[width=0.825\textwidth]{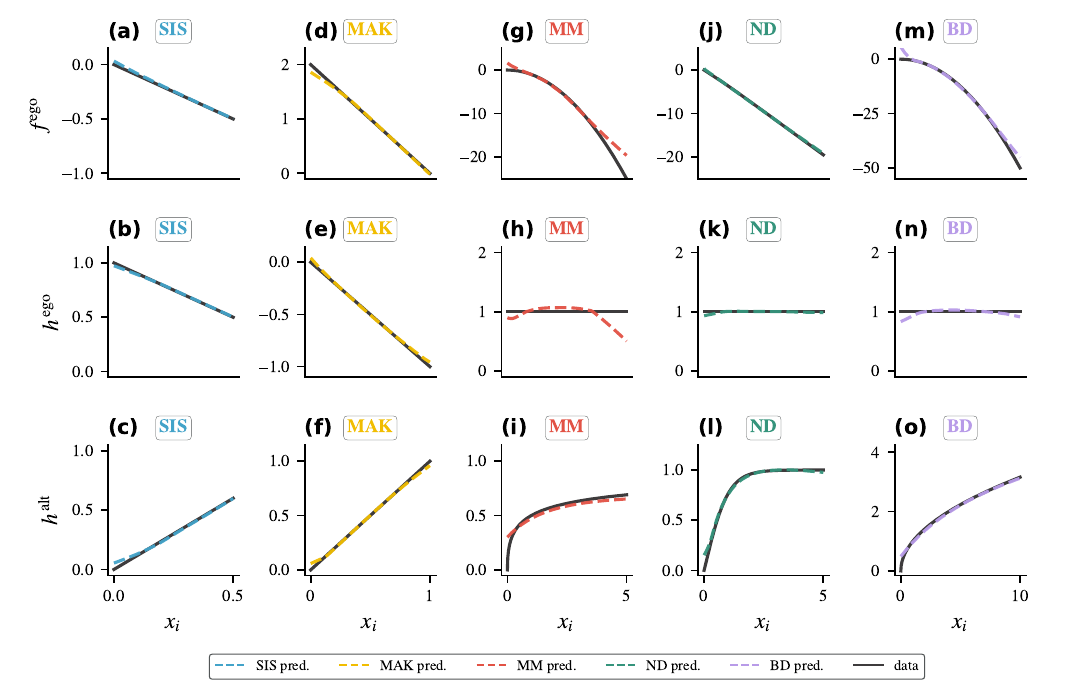}
    \caption{\textbf{Robustness of the comparison of analytical and learned vector field to noise, } $(\gamma^\mathrm{train}, \, \beta^\mathrm
    {train})=(3.0, \, 1.1)$. A comparison of the analytical (solid line) and learned (dashed line) form of the self-dynamics $f$ (upper row), ego contribution $h^\mathrm{ego}$ (middle row), and alter contribution $h^\mathrm{alt}$ (lower row) across dynamical systems (SIS \textbf{(a)}, \textbf{(b)}, \textbf{(c)}, MAK \textbf{(d)}, \textbf{(e)}, \textbf{(f)}, MM \textbf{(g)}, \textbf{(h)}, \textbf{(i)}, ND \textbf{(j)}, \textbf{(k)}, \textbf{(l)}, BD \textbf{(m)}, \textbf{(n)}, \textbf{(o)} model) shows only minor differences suggesting that \model{}s are well trained even in the presence of noise.}
    \label{fig:vec_3011_noisy}
\end{figure*}

If training graphs have intermediate degree heterogeneity and clustering, $(\gamma^\mathrm{train}, \, \beta^\mathrm{train})=(3.0, \, 1.1)$, we find good agreement of the learned and analytical vector field for most dynamical systems, irrespective of whether training data is noise free (Fig.~\ref{fig:vec_3011}) or noisy (Fig.~\ref{fig:vec_3011_noisy}). Only for \model{}s trained on the MM model do we see differences in $h^\mathrm{ego}$ at large values of $x_i$ in both cases (Fig.~\ref{fig:vec_3011}\textbf{(h)}, Fig.~\ref{fig:vec_3011_noisy}\textbf{(h)}).

\begin{figure*}[tb]
    \centering
    \includegraphics[width=0.825\textwidth]{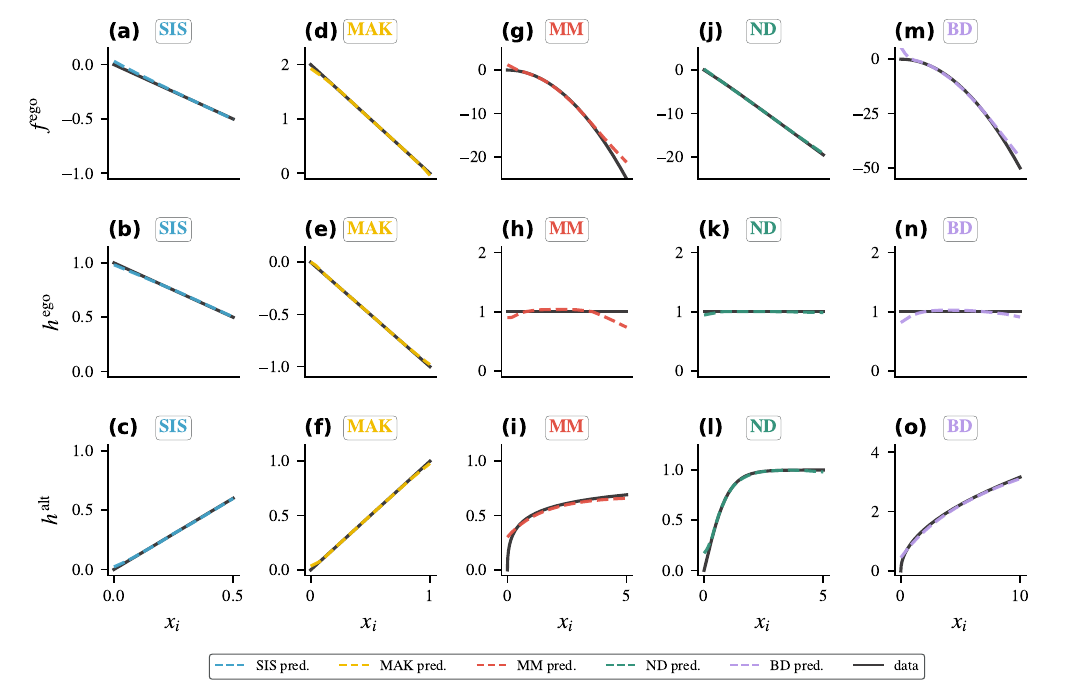}
    \caption{\textbf{Comparison of analytical and learned vector field, } $(\gamma^\mathrm{train}, \, \beta^\mathrm
    {train})=(3.9, \, 0.1)$. A comparison of the analytical (solid line) and learned (dashed line) form of the self-dynamics $f$ (upper row), ego contribution $h^\mathrm{ego}$ (middle row), and alter contribution $h^\mathrm{alt}$ (lower row) across dynamical systems (SIS \textbf{(a)}, \textbf{(b)}, \textbf{(c)}, MAK \textbf{(d)}, \textbf{(e)}, \textbf{(f)}, MM \textbf{(g)}, \textbf{(h)}, \textbf{(i)}, ND \textbf{(j)}, \textbf{(k)}, \textbf{(l)}, BD \textbf{(m)}, \textbf{(n)}, \textbf{(o)} model) shows only minor differences suggesting that \model{}s are well trained.}
    \label{fig:vec_3901}
\end{figure*}

\begin{figure*}[tb]
    \centering
    \includegraphics[width=0.825\textwidth]{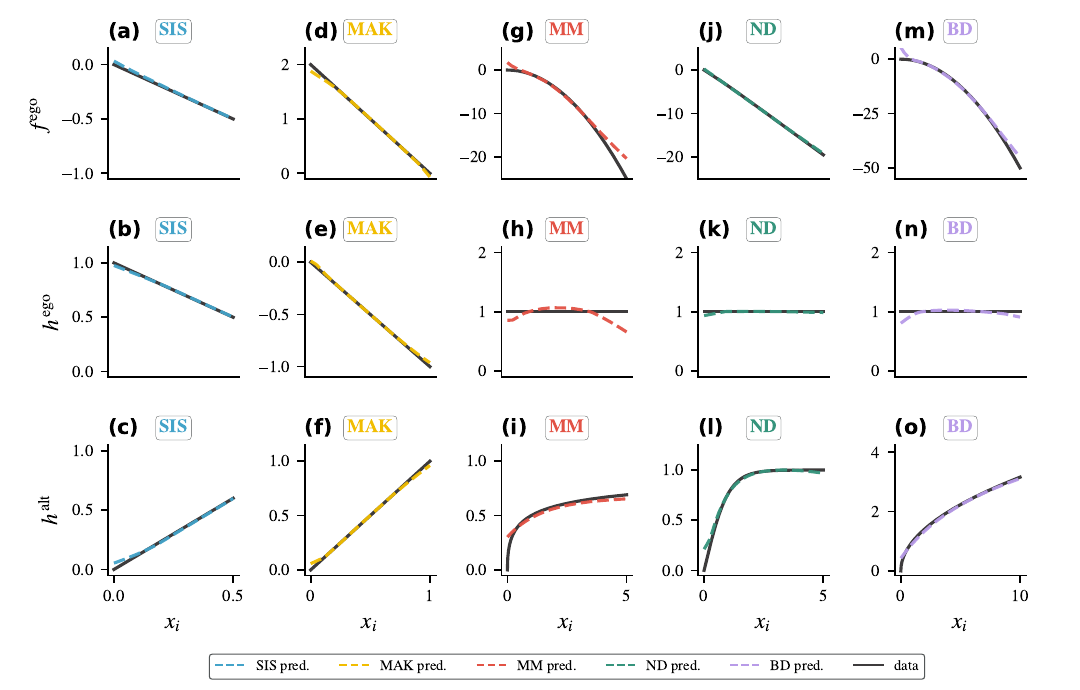}
    \caption{\textbf{Robustness of the comparison of analytical and learned vector field to noise, } $(\gamma^\mathrm{train}, \, \beta^\mathrm
    {train})=(3.9, \, 0.1)$. A comparison of the analytical (solid line) and learned (dashed line) form of the self-dynamics $f$ (upper row), ego contribution $h^\mathrm{ego}$ (middle row), and alter contribution $h^\mathrm{alt}$ (lower row) across dynamical systems (SIS \textbf{(a)}, \textbf{(b)}, \textbf{(c)}, MAK \textbf{(d)}, \textbf{(e)}, \textbf{(f)}, MM \textbf{(g)}, \textbf{(h)}, \textbf{(i)}, ND \textbf{(j)}, \textbf{(k)}, \textbf{(l)}, BD \textbf{(m)}, \textbf{(n)}, \textbf{(o)} model) shows only minor differences suggesting that \model{}s are well trained even in the presence of noise.}
    \label{fig:vec_3901_noisy}
\end{figure*}

For training graphs that are less degree heterogeneous and have weak clustering, $(\gamma^\mathrm{train}, \, \beta^\mathrm{train})=(3.9, \, 0.1)$ we make an analogous observation: There are only minor differences between the learned and analytical vector field on both noise-free (Fig.~\ref{fig:vec_3901}) and noisy (Fig.~\ref{fig:vec_3901_noisy}) training data for most dynamical systems. However, in the case of the MM model, deviations occur at high values of $x_i$ for $h^\mathrm{ego}$.

\begin{figure*}[tb]
    \centering
    \includegraphics[width=0.825\textwidth]{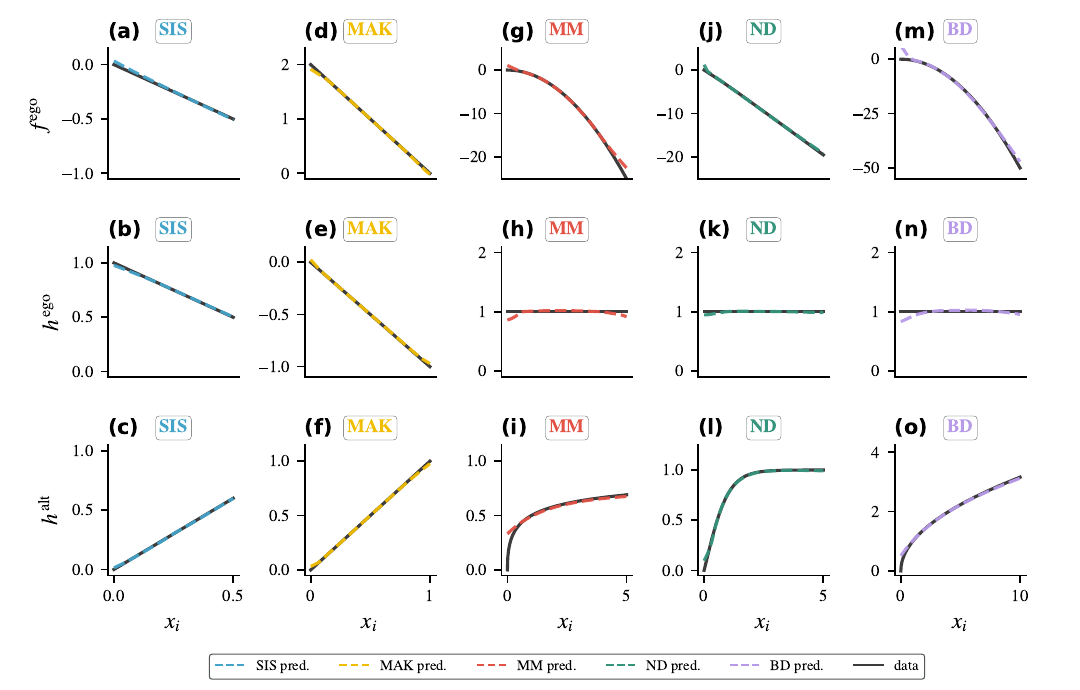}
    \caption{\textbf{Comparison of analytical and learned vector field, } $(\gamma^\mathrm{train}, \, \beta^\mathrm
    {train})=(3.9, \, 4.1)$. A comparison of the analytical (solid line) and learned (dashed line) form of the self-dynamics $f$ (upper row), ego contribution $h^\mathrm{ego}$ (middle row), and alter contribution $h^\mathrm{alt}$ (lower row) across dynamical systems (SIS \textbf{(a)}, \textbf{(b)}, \textbf{(c)}, MAK \textbf{(d)}, \textbf{(e)}, \textbf{(f)}, MM \textbf{(g)}, \textbf{(h)}, \textbf{(i)}, ND \textbf{(j)}, \textbf{(k)}, \textbf{(l)}, BD \textbf{(m)}, \textbf{(n)}, \textbf{(o)} model) shows only minor differences suggesting that \model{}s are well trained.}
    \label{fig:vec_3941}
\end{figure*}

\begin{figure*}[tb]
    \centering
    \includegraphics[width=0.825\textwidth]{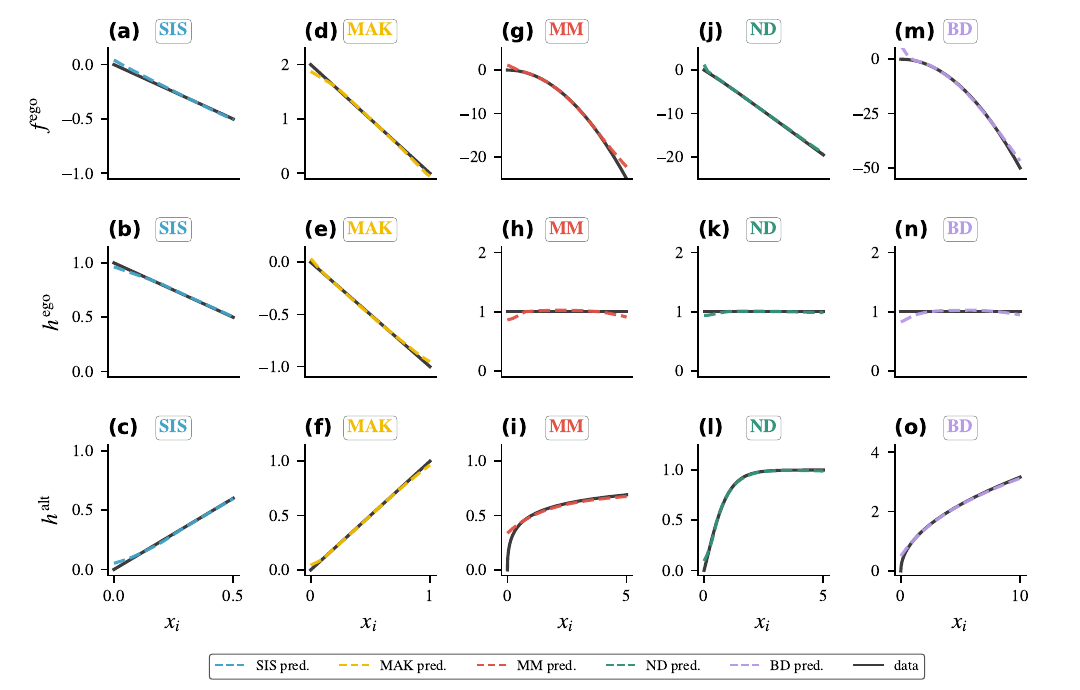}
    \caption{\textbf{Robustness of the comparison of analytical and learned vector field to noise, } $(\gamma^\mathrm{train}, \, \beta^\mathrm
    {train})=(3.9, \, 4.1)$. A comparison of the analytical (solid line) and learned (dashed line) form of the self-dynamics $f$ (upper row), ego contribution $h^\mathrm{ego}$ (middle row), and alter contribution $h^\mathrm{alt}$ (lower row) across dynamical systems (SIS \textbf{(a)}, \textbf{(b)}, \textbf{(c)}, MAK \textbf{(d)}, \textbf{(e)}, \textbf{(f)}, MM \textbf{(g)}, \textbf{(h)}, \textbf{(i)}, ND \textbf{(j)}, \textbf{(k)}, \textbf{(l)}, BD \textbf{(m)}, \textbf{(n)}, \textbf{(o)} model) shows only minor differences suggesting that \model{}s are well trained even in the presence of noise.}
    \label{fig:vec_3941_noisy}
\end{figure*}

Finally, we turn to less degree heterogeneous graphs with strong clustering, $(\gamma^\mathrm{train}, \, \beta^\mathrm{train})=(3.9, \, 4.1)$, where we observe that the analytical and learned vector fields agree well across dynamical systems and independent of whether training data is free of noise (Fig.~\ref{fig:vec_3941}) or contains additive noise (Fig.~\ref{fig:vec_3941_noisy}).

In summary, these findings suggest that the \model{}s where able to extract good approximations of the vector field from training data. They also provide further evidence for the robustness of our findings to noise in the training data.

\end{document}